\documentclass{article}
\usepackage{PRIMEarxiv}
\usepackage[utf8]{inputenc}
\usepackage[T1]{fontenc}
\usepackage{lmodern}     
\usepackage{textcomp}   
\usepackage{url}
\usepackage{booktabs}
\usepackage{amsfonts,amsmath,amssymb}
\usepackage{nicefrac,microtype,hyperref,lipsum,graphicx,float}
\graphicspath{{media/}}
\numberwithin{equation}{section}
\usepackage{orcidlink}
\usepackage{siunitx}
\usepackage{gensymb} 
\usepackage{authblk}
\usepackage{caption} 
\usepackage{tabularx}
\usepackage{braket}
\usepackage{natbib}
\usepackage{newunicodechar}
\newunicodechar{₃}{$_3$}
\usepackage[english]{babel}
\title{A Heuristic Study of Temperature: Quantum Circuitry in Thermal Systems}

\author{%
  HongZheng Liu\,\orcidlink{0009-0002-2238-9187}$^{1,*,\dagger}$,
  YiNuo Tian\,\orcidlink{0009-0005-8088-9894}$^{1}$,
  Zhiyue Wu\,\orcidlink{0009-0003-4765-2049}$^{1}$\\
  $^1$Independent, China\\
  $^*$First author\\
  $^\dagger$\textbf{Corresponding author: weiyouyeyu@foxmail.com}%
}
\begin{document}
\maketitle
\thispagestyle{empty}
\begin{abstract}
The singularities prevalent in classical thermodynamics largely stem from the "postulate of equal a priori probabilities" neglecting the physical constraints imposed by computational complexity. This paper introduces Complexity Window Thermodynamics (CWT), a framework that characterizes the observer's "ignorance" via a finite complexity budget, thereby naturally smoothing out singular behaviors associated with phase transitions and negative temperatures within this window. We derive a generalized First Law of Thermodynamics driven by a complexity generation potential, which incorporates "information processing work," and demonstrate a universal action-time bound constraining the growth of complexity. CWT not only offers a unified perspective on critical phenomena in condensed matter and the black hole information problem but also suggests that the total generatable complexity of the universe is comparable in order of magnitude to its holographic entropy. Thus, it paves a new pathway for a resource-theoretic unification of thermodynamics, quantum computation, and gravity.
\end{abstract}

\section{Introduction}
\label{cex.1}
Thermodynamic temperature, since its proposal over a century ago centered on its classical definition $T^{-1} = \frac{\partial S}{\partial E}$, has undeniably become the theoretical cornerstone for describing the macroscopic behavior of equilibrium physical systems\cite{planck1914theory,callen1985}.
It not only precisely and quantitatively characterizes the reversible transfer processes of energy between different thermodynamic systems but also profoundly reveals the universal connection existing between entropy ($S$) and internal energy ($E$).
This framework has provided a solid and unified theoretical support for understanding a series of core physical phenomena, such as the theoretical limits of heat engine efficiency, the critical behavior of phase transitions in matter, and the equilibrium conditions of chemical reactions\cite{fermi2012thermodynamics}.
However, when we attempt to extend the applicability of this classical temperature concept to more complex and extreme physical situations—such as those involving drastic phase transitions, critical phenomena, or quantum systems with special energy spectrum structures—a fundamental tension begins to emerge between its inherent theoretical assumptions and physical reality, leading to severe challenges to its explanatory power and prompting profound rethinking about the universality of the temperature concept.
These challenges are mainly concentrated on the singular behaviors exhibited by classical thermodynamic temperature and its related physical quantities under specific conditions.
First, at first-order phase transition points (e.g., the melting of ice or the vaporization of water), as the system absorbs or releases latent heat, the temperature remains constant while the entropy value undergoes an abrupt jump, rendering the derivative $\frac{\partial S}{\partial E}$ mathematically discontinuous at this point and thus causing the classical definition of temperature to lose coherence in a differential sense\cite{landau2013statistical,Lieb2005BoseGas}.
Second, near the critical point of continuous phase transitions (or second-order phase transitions), the correlation length of the system tends to infinity, macroscopic fluctuations become exceptionally strong, directly leading to singular divergences in thermodynamic response functions such as specific heat $C_V$ and magnetic susceptibility $\chi$, which clearly indicates that classical statistical theory based on mean-field approximation has already failed in this region\cite{onsager1944,yeomans1992}.
More extreme situations arise in certain isolated quantum systems with an upper energy bound $E_{\mathrm{max}}$ (e.g., nuclear spin systems oriented in a strong magnetic field, or specific configurations of energy level populations in trapped ion systems).
In these systems, when the system's energy approaches its upper limit, the entropy $S$ may decrease with further increases in energy $E$, leading to $\frac{\partial S}{\partial E} < 0$ and thus giving rise to the so-called "negative absolute temperature" phenomenon\cite{purcell1951nuclear,ramsey1956thermodynamics}.
This phenomenon not only intuitively challenges the common physical experience that temperature is always positive but also raises profound questions for the traditional statistical interpretation based on the canonical ensemble.
Although traditional explanations for these "anomalous" thermodynamic behaviors often rely on detailed analysis of specific models or are corrected by introducing additional physical effects (such as finite-size effects, surface effects, etc.), their common, deeper root cause has rarely been revealed in a consistent and universal manner.
This paper posits that the profound reason for the difficulties encountered by classical thermodynamic theory in specific contexts lies in a core assumption implicit in classical statistical mechanics when constructing its ensemble theory—namely, that all microstates within the energy shell have a priori equiprobability of access—which fails to fully align with the actual evolutionary capability of quantum dynamics under finite physical resource constraints\cite{landauer1961}.
Admittedly, the unitary evolution $U(t) = \exp(-iHt/\hbar)$ of an isolated quantum system, in principle, allows the system to explore the entire region of Hilbert space permitted by its energy.
However, starting from a relatively simple initial state, to generate and actually reach quantum states that embody highly complex quantum entanglement structures, possess fine-grained long-range correlations, or encode complex computational logic within a finite evolution time and by consuming finite action or energy, often requires the system to incur an exponentially growing dynamical "cost"\cite{lloyd2000ultimate,parker2019universal}.
This dynamical "attainment cost" of state preparation can be effectively and operationally quantified by the minimum generation quantum circuit complexity $C^*$ of the quantum state (i.e., the minimum number of gates or shortest circuit depth required to synthesize the target state by applying a series of fundamental quantum gate operations from a universal gate set, starting from an agreed-upon simple reference state)\cite{nielsen2010quantum,gellmann1994,jaynes1957information}.
Therefore, the classical statistical mechanics' assumption of equal probability for all energy-compatible states essentially ignores the differences in the difficulty of physically preparing or achieving these states, differences that are non-negligible in a resource-constrained real world.
Based on this core insight, we propose to formalize "Bounded Dynamical Complexity"—the idea that there is an upper bound on the complexity $C^*$ of quantum states that a system can achieve with given resources, stemming from quantum dynamics and finite physical resources—and elevate it to a fundamental postulate for constructing statistical ensemble theory, replacing the classical "equiprobable accessibility" assumption.
On the foundation of this new postulate, we systematically construct the "Complexity-Windowed Thermodynamics" (CWT) theoretical framework.
The core steps of CWT include: first, introducing a "complexity budget" parameter $\Xi$, which reflects the upper limit of available physical resources and confines the complexity range (minimum generation quantum circuit complexity $C^* \le \Xi$) of quantum states that the system can actually generate or an observer can effectively resolve.
Second, by performing conditional expectation coarse-graining on the system's true microstates under the budget $\Xi$, we define the core state function of the CWT framework—the complexity-windowed entropy $S_\Xi(E,\Xi)$.
This entropy function $S_\Xi(E,\Xi)$ is interpreted as the remaining statistical uncertainty or information-theoretic "entropy of ignorance" that an observer has about the system's true microstate, given the limitation of the complexity budget $\Xi$ and their finite resolution capability\cite{zhang2025dynamical}.
We will rigorously prove in subsequent sections that for any finite complexity budget $\Xi$, $S_\Xi(E,\Xi)$ is smooth as a function of energy $E$ and is a monotonically non-increasing function of $\Xi$ ($(\partial S_\Xi/\partial \Xi)_E \le 0$), which aligns with the physical intuition that enhanced observational capability reduces uncertainty\cite{callen1985}.
This crucial smoothness ensures that the complexity-windowed (or effective) temperature $T_\Xi^{-1} = (\partial S_\Xi/\partial E)_\Xi$ derived from $S_\Xi$, and its conjugate, which we term the "complexity generation potential" $\Pi_\Xi = (\partial E/\partial \Xi)_{S_\Xi}$ (typically non-negative, $\Pi_\Xi \ge 0$, reflecting the energy cost to increase the complexity budget), remain finite and continuous under all physical conditions.
This fundamental characteristic enables CWT to universally "soften" all singularities, such as discontinuities and divergences, encountered in classical thermodynamics.
Building on this, we further derive an extended first law of thermodynamics that includes an "information processing work" or "complexity generation work" term: $\mathrm{d}E = T_\Xi \mathrm{d}S_\Xi + \Pi_\Xi \mathrm{d}\Xi$.
More importantly, we reveal two fundamental principles governing complexity generation constrained by physical resources: one is the action constraint principle ($S_{\mathrm{proc}} \ge k_S C^* \hbar$), and the other is the time constraint principle ($t_{\mathrm{proc}} \ge \frac{C^* \hbar}{2\pi k_B T_\Xi}$).
These two principles profoundly characterize the fundamental physical resource limits faced by any physical or information processing process when generating a specific computational complexity $C^*$.
The main academic contributions of this paper can be summarized as follows:
\begin{enumerate}
    \item Proposing and postulating "bounded dynamical complexity": Elevating the intrinsic limitation of finite physical resources on the complexity of generatable or resolvable quantum states, with quantum circuit complexity $C^*$ as the core metric, to a fundamental postulate for constructing statistical ensembles, thereby modifying the classical assumption of equiprobable accessibility.
    
    \item Constructing macroscopic state functions reflecting resource limitations and their extended thermodynamic law: Defining the complexity-windowed entropy $S_\Xi$ (quantifying the observer's effective entropy or "entropy of ignorance" under budget $\Xi$), a smooth effective temperature $T_\Xi$, and a non-negative complexity generation potential $\Pi_\Xi$ (reflecting the energy cost of changing the complexity budget), and deriving the extended first law of thermodynamics including the "information processing work" term $\Pi_\Xi\mathrm{d}\Xi$.

    \item Proving the universal softening of classical thermodynamic singularities: Demonstrating that, based on the smoothness of $S_\Xi$ with respect to energy $E$ (for any finite $\Xi$), singularities such as discontinuities and divergences encountered by classical thermodynamics at first-order phase transitions, critical points, and in negative temperature regimes are universally regularized within the CWT framework, with classical results recovered only in the limit of infinite complexity budget ($\Xi \to \infty$).

    \item Deriving the dual resource constraint principles for complexity generation: Revealing that the generatable quantum circuit complexity $C^*$ of a system is dually constrained by available action ($S_{\mathrm{proc}} \ge k_S C^* \hbar$) and available evolution time (related through effective temperature $T_\Xi$, $t_{\mathrm{proc}} \ge \frac{C^* \hbar}{2\pi k_B T_\Xi}$), whose mathematical forms are consistent in magnitude and physical connotation with known quantum speed limits \cite{Ng2000Foaminess}.

    \item Deriving the upper bound on the generatable complexity of the universe and revealing its potential connection with the holographic principle: Applying the resource constraint principles to the cosmological scale and, under reasonable physical assumptions, estimating the upper limit of total generatable complexity in the universe since the Big Bang, finding its order of magnitude to be comparable to the cosmic horizon entropy based on the holographic principle, thereby providing new theoretical perspectives for understanding the profound connections between the universe's information-carrying capacity, holographic information boundaries, and computational limits \cite{mandelstam1945}.
\end{enumerate}
The organization of this paper is as follows: Section \ref{cex.2} is dedicated to laying the theoretical foundation of CWT, where we will define in detail the statistical ensemble constrained by a complexity budget, introduce the budget-observable algebra $A(\Xi)$ and conditional expectation $E_\Xi$ for coarse-graining, and rigorously define the core state function—complexity-windowed entropy $S_\Xi(E,\Xi)$—along with its key mathematical properties (smoothness with respect to energy $E$, and monotonic non-increasing nature with respect to the complexity budget $\Xi$).
Subsequently, Section \ref{cex.3} will construct the complete CWT thermodynamic formalism on this basis, deriving the effective temperature $T_\Xi$, the complexity generation potential $\Pi_\Xi$, and the extended first law including information processing work, and thoroughly discussing their physical significance.
Section \ref{cex.4} will systematically demonstrate how the CWT framework, by virtue of the smoothness of $S_\Xi$, universally regularizes the singularities of classical thermodynamics at first/second-order phase transitions and in negative temperature regimes.
Next, Section \ref{cex.5} will derive and elucidate the two fundamental physical resource constraints governing complexity generation—the action constraint and time constraint principles—and discuss the resource feasible region jointly determined by them.
Section \ref{cex.6} focuses on the core dynamical predictions of CWT, namely the complexity-temperature growth rate bound, by analyzing its application potential in cross-disciplinary problems such as critical phenomena in condensed matter and black hole information dynamics, while also looking ahead to related experimental detection prospects 
Section \ref{cex.7} then extends the application of CWT theory to the cosmological scale, providing a detailed analysis of the estimation of the upper bound on the generatable complexity of the universe and its profound connection with the holographic principle.
Finally, Section \ref{cex.8} comprehensively summarizes the core achievements of this paper, reviews the theoretical landscape of CWT, and looks ahead to its challenges and future development directions.
To ensure the rigor of the core arguments and the clear presentation of mathematical details, the key mathematical constructs of this paper (such as the algebraic foundations of $A(\Xi)$ and $E_\Xi$, see Appendix \ref{B.1}), detailed proofs of the properties of the core state function $S_\Xi$ (see Appendix \ref{c.1}), derivation of the physical resource constraint principles (see Appendix \ref{D.1}), systematic exposition of the CWT thermodynamic formal structure (including thermodynamic potentials and Maxwell relations, see Appendix \ref{E.1}), and details of the quantitative estimation of the cosmic complexity upper bound (see Appendix \ref{F.1}),Appendix \ref{G1}is dedicated to a rigorous derivation of the predicted peak behavior of the complexity generation potential, $\Pi_\Xi$, at a first-order phase transition, providing a key testable signature of our framework. Building on this, Appendix \ref{H1} explores the physical origins of the exponential complexity scaling inferred from our case study, linking it to macroscopic quantum tunneling and developing a unified scaling theory that incorporates finite-correlation-length effects.
Appendix \ref{A} provides a quick reference table for key symbols and core terminology.

\section{Complexity-Windowed Entropy: The Theoretical Foundation of CWT}
\label{cex.2}
In the introduction, we pointed out that the difficulties encountered by classical statistical mechanics in dealing with certain complex physical situations stem from its failure to fully consider the intrinsic limitations imposed by quantum dynamics on the states a system can actually reach.
This chapter aims to lay the theoretical foundation for this core idea.
We will first argue why quantum circuit complexity $C^*$ is chosen as an appropriate metric to quantify this dynamical constraint and explain its physical significance.
Subsequently, we will formally define the accessibility of a system under a given "complexity budget" $\Xi$, specifically by constructing a "budgeted observable algebra" $A(\Xi)$ and introducing conditional expectation for natural coarse-graining.
On this basis, we will rigorously define the core state function of the CWT framework—the complexity-windowed entropy $S_\Xi(E, \Xi)$—and delve into its physical meaning.
Finally, and crucially, we will analyze in detail the key mathematical properties of $S_\Xi(E, \Xi)$, particularly its smoothness with respect to energy $E$, a property that is the fundamental guarantee for CWT to universally regularize classical thermodynamic singularities in subsequent chapters.

\subsection{Quantum Circuit Complexity}
\label{cex.2.1}
Within the framework of quantum mechanics, the evolution of a closed system is governed by its Hamiltonian $H$, and the change of state over time follows the Schrödinger equation, manifesting as a unitary transformation.
In principle, as long as energy is conserved, the system can evolve to any energy-allowed state in Hilbert space.
However, this theoretical accessibility does not consider the "physical cost" or "evolutionary difficulty" required to reach a specific target state.
Generating certain quantum states that are highly entangled, possess fine-grained long-range correlations, or embody complex computational logic often requires extremely long evolution times or exceedingly precise control sequences, which is a crucial limiting factor in any real system with finite physical resources (such as available total action, upper limit of evolution time, control precision, energy supply, etc.).
To quantitatively characterize this "evolutionary difficulty" or "generation cost," we need a suitable metric.
Among the many measures of quantum state complexity\cite{watrous2008quantum}, this paper chooses quantum circuit complexity, specifically referring to the minimum number of gates or shortest circuit depth required to synthesize a target quantum state $|\psi\rangle$ (or implement a target unitary transformation $U$) by applying a series of fundamental quantum gates from a universal quantum gate set $G$, starting from an agreed-upon, structurally simple reference state (such as a product state $|0\rangle^{\otimes N}$ in a many-body system), denoted as $C^*(|\psi\rangle)$ or $C^*(U)$ \cite{nielsen2006quantum}.
The choice of quantum circuit complexity $C^*$ as the core metric is primarily based on the following key theoretical considerations:
\begin{enumerate}
    \item \textbf{Clear Operational Meaning: }$C^*$ is directly related to the operational cost of state preparation or transformation. It quantifies the minimum "effort" or the fewest "basic operational units" required to implement a specific quantum process. This aligns highly with the core idea of CWT, which emphasizes that "accessibility is limited by generation difficulty." A state with a high $C^*$ is dynamically difficult to generate.
    \item \textbf{Potential for Universality:} The quantum circuit model is a universal model for quantum computation\cite{deutsch1985quantum}. Theoretically, any physical process obeying the laws of quantum mechanics, including unitary evolution driven by a local Hamiltonian, can in principle be efficiently simulated by a quantum circuit. Therefore, the complexity $C^*$ defined based on this universal model is expected to be a fundamental metric with strong universality, independent of specific system details (such as the form of a particular Hamiltonian). It measures the intrinsic complexity of the state or transformation itself, rather than the complexity of a specific dynamical path.
    \item \textbf{Profound Connection with Fundamental Physical Quantities:} Recent research progress, particularly inspired by holographic duality (AdS/CFT correspondence) \cite{maldacena1998}, has revealed that the complexity of a quantum state may have profound correspondences with fundamental physical quantities in gravitational theory, such as action (Complexity = Action conjecture\cite{brown2016holographic}) or spacetime volume (Complexity = Volume conjecture\cite{susskind2016computational}). These connections provide strong theoretical support for the view in the CWT framework that the complexity budget $\Xi$ is constrained by physical resources (such as the upper limit of available action or maximum evolution time).
    \item \textbf{Association with Core Phenomena in Statistical Physics:} The growth behavior of quantum circuit complexity is closely related to core phenomena in statistical physics such as quantum chaos, information scrambling, and system thermalization processes\cite{parker2019universal}. In typical chaotic quantum systems, complexity usually grows linearly with time until it reaches a saturation value related to the dimension of the Hilbert space; this growth process reflects the rapid propagation of information among the system's degrees of freedom and the "randomization" of the quantum state. Using $C^*$ as the core metric allows CWT to naturally connect with these dynamical mechanisms that describe a system's approach to equilibrium.
\end{enumerate}
Although calculating $C^*$ for a specific quantum state can be very difficult in practice, its clear operational definition, potential universality, and profound connections with fundamental physics and statistical physics phenomena make it an appropriate theoretical choice for quantifying and constraining system dynamical accessibility within the CWT framework.
In the subsequent discussions in this paper, we will assume the existence of such a well-defined complexity measure $C^*(\sigma)$ acting on the system's microstate $\sigma$.
On the other hand, as emphasized in the introduction, a core limitation of classical statistical mechanics is its failure to adequately consider the actual evolutionary capability of physical systems to generate or reach highly complex states under finite resources.
To intrinsically incorporate this "bounded dynamical complexity" into the thermodynamic framework, we introduce a core parameter here—the Complexity Budget, denoted as $\Xi$.
The complexity budget $\Xi$ is defined as: under specific physical conditions and resource constraints, the maximum allowable quantum circuit complexity $C^*$ of a quantum state (or implemented unitary transformation $U$) that the system can generate, or that an observer can effectively resolve.
Specifically, if the minimum generation complexity of a quantum state $|\psi\rangle$ is $C^*(|\psi\rangle)$, or the minimum implementation complexity of a unitary transformation $U$ is $C^*(U)$ (both relative to a pre-agreed simple reference state and universal quantum gate set $G$), then the state or transformation is considered "accessible" or "operable" under the current complexity budget $\Xi$ only if $C^*(|\psi\rangle) \le \Xi$ or $C^*(U) \le \Xi$.
Physically, this complexity budget $\Xi$ essentially reflects the fundamental limitation imposed by the system's available finite physical resources (e.g., total available evolution time, total consumable action, achievable control precision, or system size, etc.) on the complexity of achievable quantum states.
It will directly define the effective range that the system can actually explore in the vast Hilbert space.
In the following Section \ref{sec:label_of_2.2}, based on this complexity budget $\Xi$, we will rigorously define the "accessibility" of the system, particularly by constructing the corresponding "budgeted observable algebra," and further develop the core concepts of complexity-windowed thermodynamics on this foundation.

\subsection{Formalization of Accessibility}
\label{sec:label_of_2.2}
In classical statistical mechanics, the description of a system's state is typically based on a set of macroscopic observables (such as energy, particle number, volume, etc.).
However, when complexity constraints are considered, we not only need to limit the set of accessible microstates but also consider that, under a finite complexity budget, the set of observables that an observer can actually resolve or measure will itself be limited.
In other words, detecting certain fine-grained quantum correlations or distinguishing some highly similar but differently complex states may require performing measurement operations whose complexity exceeds the current budget.
To formally describe this observability constrained by a complexity budget, we introduce the concept of the Budgeted Observable Algebra $A(\Xi)$.
Let $M_0 = \{M_\alpha\}$ be a basis set of fundamental, in-principle easily measurable Hermitian operators (observables); for example, for a system composed of $N$ spins, $M_0$ could be the Pauli operators at each site $\{\sigma_i^x, \sigma_i^y, \sigma_i^z \mid i=1,...,N\}$.
Let $G$ be the chosen universal quantum gate set, and let $G^{(\le\Xi)}$ denote the set of all unitary transformations composed of quantum gates from $G$ with a circuit depth (or equivalent complexity measure) not exceeding $\Xi$.

\textbf{Definition 2.1 (Budgeted Observable Algebra $A(\Xi)$):}
The budgeted observable algebra $A(\Xi)$ is the von Neumann algebra generated by operators formed by the action of all accessible unitary transformations $U \in G^{(\le\Xi)}$ on the fundamental observables $M_\alpha \in M_0$:

\begin{equation}
A(\Xi) = \mathrm{Alg} \{U M_\alpha U^\dagger \mid U \in G^{(\le\Xi)}, M_\alpha \in M_0\}
\label{eq:2.1}
\end{equation}

Physically, $A(\Xi)$ represents the set of all Hermitian operators (observables) that can be effectively measured or distinguished under the computational resource constraint of complexity budget $\Xi$.
If an observable (or its approximation) necessary for detecting a certain physical phenomenon (e.g., a specific type of many-body entanglement or a fine-grained quantum order) does not belong to $A(\Xi)$ (i.e., it requires a unitary transformation with complexity exceeding $\Xi$ to be generated from $M_0$), then under budget $\Xi$, this phenomenon is "unobservable" or "indistinguishable."
For any finite $\Xi$, $A(\Xi)$ is a (typically finite-dimensional) subalgebra of the full algebra of bounded operators $B(H)$ acting on the system's Hilbert space $H$.
Let $d(\Xi) = \dim A(\Xi)$ denote the dimension of this algebra.
In systems exhibiting quantum chaos or when using universal local gates, for $\Xi \ll \dim H$, $d(\Xi)$ is expected to grow rapidly with $\Xi$, possibly exponentially\cite{roberts2017chaos}, and eventually saturate at $d(\infty) = (\dim H)^2$ when $\Xi$ approaches a sufficiently large value (capable of generating all operators in $B(H)$, roughly corresponding to $(\dim H)^2$ degrees of freedom).
Restricting observables to the algebra $A(\Xi)$ naturally induces a coarse-graining of the system's state.
Any information contained in a true physical state (described by a density matrix $\rho$) that can only be probed by operators outside $A(\Xi)$ becomes inaccessible under the complexity budget $\Xi$.
This coarse-graining process can be formally described by a trace-preserving Conditional Expectation map $E_\Xi : B(H) \to A(\Xi)$\cite{takesaki2003theory}.
This map projects an arbitrary density operator $\rho$ to its "best representation" $E_\Xi(\rho)$ within the algebra $A(\Xi)$, such that this "best representation" preserves the expectation values of $\rho$ for all operators $A$ in $A(\Xi)$, i.e., $\mathrm{Tr}(E_\Xi(\rho)A) = \mathrm{Tr}(\rho A)$ holds for all $A \in A(\Xi)$.
The state $E_\Xi(\rho)$ retains all information about $\rho$ accessible through measurements within $A(\Xi)$, while maximizing entropy relative to the algebra $A(\Xi)$ (i.e., $E_\Xi(\rho)$ is the "most ignorant" or "most mixed" state, given the preservation of expectation values for observables in $A(\Xi)$).
This idea of coarse-graining based on an observable algebra is in line with the method in statistical mechanics of defining macroscopic states and entropy by projecting onto the subspace of macroscopic observables\cite{jaynes1957information}, but CWT generalizes it to the level of dynamical complexity budget.
For a detailed mathematical construction of the budgeted observable algebra $A(\Xi)$, its fundamental properties, and rigorous proofs of the existence, uniqueness, and key characteristics of the conditional expectation map $E_\Xi$, please refer to Appendix \ref{B.1}.

\subsection{Definition and Physical Meaning of Complexity-Windowed Entropy \texorpdfstring{$S_\Xi(E, \Xi)$}{SXi(E, Xi)}}
\label{cex.2.3}
In classical statistical mechanics, the entropy of a microcanonical ensemble is directly related to the number of states (or phase space volume) within the energy shell, with the core assumption that these states are all equiprobably accessible.
However, as previously discussed, the complexity constraints of quantum dynamics render this assumption not entirely valid under finite physical resources.
The core of the Complexity-Windowed Thermodynamics (CWT) framework lies in modifying the definition of entropy to accurately reflect this accessibility limitation imposed by the complexity budget $\Xi$, which originates from finite physical resources.
We consider a macroscopic state with energy $E$ (or energy in the interval $[E, E+\delta E]$, with its energy shell described by the projection operator $\Pi_E$).
In the ideal case without complexity constraints, its microcanonical entropy (for a discrete spectrum system) is $S_{\mathrm{micro}}(E) = k_B \ln \omega(E)$, where $\omega(E) = \mathrm{Tr}(\Pi_E)$ is the number of states or degeneracy within the energy shell.
In the CWT framework, our focus is not only on those microstates $\sigma$ that are compatible with energy $E$ and whose generation complexity $C^*(\sigma)$ does not exceed the budget $\Xi$, but more crucially, as clarified in Section 2.2, the observer's own ability to resolve these states is also limited by their accessible observable algebra $A(\Xi)$.
Therefore, by applying the conditional expectation coarse-graining $E_\Xi$ to the density matrix of the microcanonical ensemble $\rho_E = \Pi_E / \omega(E)$, we define an entropy under the complexity budget $\Xi$ that reflects the observer's effective level of knowledge.

\textbf{Definition 2.2 (Complexity-Windowed Entropy $S_\Xi(E, \Xi)$):}

For a macroscopic state defined by a given energy $E$ (represented by the energy shell projection operator $\Pi_E$, with $\omega(E) = \mathrm{Tr}(\Pi_E)$ as its density of states or number of states) and a complexity budget $\Xi$, its complexity-windowed entropy $S_\Xi(E, \Xi)$ is given by the von Neumann entropy of the coarse-grained microcanonical ensemble:

\begin{equation}
S_\Xi(E, \Xi) = S_{\mathrm{VN}}(E_\Xi(\Pi_E / \omega(E))) = -k_B \mathrm{Tr} \left[ E_\Xi(\Pi_E / \omega(E)) \ln E_\Xi(\Pi_E / \omega(E)) \right]
\label{eq:2.2}
\end{equation}

where $S_{\mathrm{VN}}(\rho) = -k_B \mathrm{Tr}(\rho \ln \rho)$ is the standard von Neumann entropy, and $E_\Xi$ is the conditional expectation map onto the budgeted observable algebra $A(\Xi)$.
The complexity-windowed entropy $S_\Xi(E, \Xi)$ has profound physical meaning, quantifying the cognitive state of an observer with specific information processing capabilities:

\begin{enumerate}
    \item \textbf{Quantification of Observer's Effective Entropy and "Degree of Ignorance":} The complexity-windowed entropy $S_\Xi(E, \Xi) / k_B$ fundamentally quantifies the residual statistical uncertainty, or information-theoretic "degree of ignorance," that an observer has about the system's true microstate $\rho_E$ under the dual constraints of energy $E$ and complexity budget $\Xi$, due to their limited resolution capability (characterized by the algebra $A(\Xi)$). When the complexity budget $\Xi$ increases, meaning the observer's resolution capability is enhanced and they can access and distinguish finer structures or more complex correlations (i.e., $A(\Xi)$ becomes "finer"), their "degree of ignorance" or "residual uncertainty" about the true state of the system will generally decrease or remain unchanged. This core characteristic will be proven in Section \ref{cex.2.4} as the monotonic non-increasing property of $S_\Xi(E,\Xi)$ with respect to $\Xi$.
    \item \textbf{Effective State Entropy under Information-Theoretic Constraints:} From an information-theoretic perspective, $S_\Xi(E, \Xi)$ is a direct measure of the von Neumann entropy of the effective state $E_\Xi(\Pi_E / \omega(E))$ that the system presents under the complexity budget $\Xi$. Within this framework, the observer is restricted to obtaining information about the system only through observables in the algebra $A(\Xi)$; for correlations and degrees of freedom that exist outside $A(\Xi)$ and require higher complexity to probe, the observer appears "ignorant." The core role of the conditional expectation $E_\Xi$ is precisely to "average out" or "blur" these fine structures that are beyond the current resolution capability, thereby obtaining an effective state description $E_\Xi(\Pi_E / \omega(E))$ consistent with the observer's information acquisition ability. $S_\Xi(E, \Xi)$ is thus the inherent statistical entropy of this effective state, reflecting the degree of unknowability of the system's microscopic details within the observer's capabilities.
    \item \textbf{Bridge Connecting Macroscopic and Microscopic Realms:} Similar to classical Boltzmann or Gibbs entropy, the complexity-windowed entropy $S_\Xi$ still plays the crucial role of a bridge connecting the macroscopic thermodynamic properties of the system (such as the effective temperature $T_\Xi$ and complexity generation potential $\Pi_\Xi$ derived from it) with the statistical behavior of microstates (via the coarse-graining process under complexity constraints). However, this bridge now intrinsically embeds the consideration of the generatable complexity $C^*$ of microstates, and more importantly, the observer's ability to resolve these states (manifested through $A(\Xi)$ and $E_\Xi$)—these two fundamental real-world constraints originating from finite physical resources. It emphasizes that the macroscopic thermodynamic world we perceive is, to some extent, shaped by our (or nature's own) ability to process information complexity.
\end{enumerate}

\subsection{Key Mathematical Properties of \texorpdfstring{$S_\Xi(E, \Xi)$}{SXi(E, Xi)} and Their Physical Significance}
\label{cex.2.4}
The complexity-windowed entropy $S_\Xi(E, \Xi)$, as defined in Definition 2.2 above, being the core state function of the Complexity-Windowed Thermodynamics (CWT) theoretical framework, exhibits several mathematical properties crucial for its theoretical self-consistency and physical applications.
These properties not only lay a solid mathematical foundation for constructing the complete CWT formalism in subsequent chapters (including effective temperature $T_\Xi$, complexity generation potential $\Pi_\Xi$, and the extended first law, etc.) but also directly foreshadow the unique capability of the CWT framework in resolving the singularity issues encountered by classical thermodynamics in specific physical contexts.

\textbf{Proposition 2.1 (Properties of Complexity-Windowed Entropy $S_\Xi$):}
Let the complexity-windowed entropy $S_\Xi(E, \Xi)$ be as defined in Definition 2.2. Then it possesses the following key properties:

\begin{enumerate}
    \item[(i)] \textbf{Monotonicity with Respect to Complexity Budget $\Xi$:} For a fixed energy $E$, the complexity-windowed entropy $S_\Xi(E, \Xi)$ is a monotonically non-increasing function of the complexity budget $\Xi$. Specifically, if there are two complexity budgets $\Xi_1$ and $\Xi_2$ satisfying $\Xi_1 \le \Xi_2$, then the corresponding observable algebras satisfy $A(\Xi_1) \subseteq A(\Xi_2)$, and the relationship between the complexity-windowed entropies is:
   
    \begin{equation}
    S_{\Xi_1}(E, \Xi_1) \ge S_{\Xi_2}(E, \Xi_2)
    \label{eq:2.3}
    \end{equation}
 \vspace{3pt}
    
    This property profoundly reflects the physical meaning of $S_\Xi(E, \Xi)$ as the observer's "entropy of ignorance." When the complexity budget $\Xi$ increases, it implies that the observer possesses more physical resources, and their resolution capability (characterized by the finer algebra $A(\Xi_2)$) is enhanced accordingly. Therefore, the observer's "degree of ignorance" or residual statistical uncertainty about the true microstate $\rho_E$ of the system should decrease or at least remain unchanged, which is entirely consistent with the mathematical behavior expressed by Eq. \eqref{eq:2.3}. The rigorous proof of this monotonic non-increasing property, based on the fundamental property of von Neumann entropy under conditional expectation (as a coarse-graining map)—namely, that deeper coarse-graining does not lead to a decrease in entropy ($S_{\mathrm{VN}}(P(\sigma)) \ge S_{\mathrm{VN}}(\sigma)$, where $P$ represents a coarser projection)—is detailed in Appendix \ref{c.3}. This property is crucial for understanding the sign and physical meaning of the complexity generation potential $\Pi_\Xi$ later on.

    \item[(ii)] \textbf{Smoothness with Respect to Energy $E$:} For any finite complexity budget $\Xi$ ($\Xi < \infty$), as long as the energy shell projection operator $\Pi_E$ (and consequently the density of states $\omega(E)$ and the microcanonical density operator $\rho_E$) is a sufficiently smooth function of energy $E$ (an assumption generally reasonable in macroscopic physical systems), then the complexity-windowed entropy $S_\Xi(E, \Xi)$ as a function of energy $E$ is necessarily smooth. For example, $S_\Xi(E, \Xi)$ is typically at least a $C^k$ function (for sufficiently large $k$), and even analytic at most energy points, with possible exceptions only at a set of energy values of measure zero where eigenvalues of the projected state $E_\Xi(\rho_E)$ become degenerate (leading to a change in its rank). This crucial smoothness is the fundamental mathematical guarantee that enables the CWT framework to universally "soften" or regularize classical thermodynamic singularities. The core idea of its proof (detailed in Appendix \ref{c.1}, and relying on the construction of $A(\Xi)$ and $E_\Xi$ in Appendix \ref{B.1}) is that: for finite $\Xi$, the budgeted observable algebra $A(\Xi)$ is a finite-dimensional von Neumann algebra. The conditional expectation map $E_\Xi$ projects the density matrix $\rho_E$ onto the state space of this finite-dimensional algebra $A(\Xi)$, yielding $E_\Xi(\rho_E)$. Since $\rho_E$ is smooth with respect to $E$ and $E_\Xi$ is a fixed linear map, $E_\Xi(\rho_E)$ is also smooth with respect to $E$. The von Neumann entropy $S_{\mathrm{VN}}(\sigma)$ is a strictly concave and smooth function in its (finite-dimensional) domain. Therefore, $S_\Xi(E, \Xi)$, as a composition of a smooth function ($S_{\mathrm{VN}}$) and a smoothly parameterized state ($E_\Xi(\rho_E)$), naturally inherits smoothness with respect to energy $E$. Singularities in classical thermodynamics are often related to the system's ability to access or exhibit infinitely fine microscopic structures (such as infinite correlation length at a critical point) or to abrupt changes in state at phase transitions. A finite complexity budget $\Xi$, by restricting observation to the finite-dimensional algebra $A(\Xi)$, effectively "truncates" those infinitely fine degrees of freedom that could lead to mathematical discontinuities or divergences, making the effective description of the system state $E_\Xi(\rho_E)$ vary smoothly with energy $E$. This smoothness directly ensures that subsequently defined thermodynamic quantities derived from $S_\Xi$, such as the effective temperature $T_\Xi$ and specific heat $C_{V,\Xi}$, have well-defined and finite derivatives of all orders with respect to energy $E$.

    \item[(iii)] \textbf{Limiting Behavior:} The complexity-windowed entropy $S_\Xi(E, \Xi)$ exhibits behavior consistent with existing theories in two physically important limiting cases, thereby ensuring that CWT is a reasonable generalization of classical statistical mechanics:
    \begin{enumerate}
        \item \textbf{Infinite Resource Limit ($\Xi \to \infty$):} When the complexity budget $\Xi$ tends to infinity, it implies that the observer possesses infinite physical resources and can resolve arbitrarily complex quantum states and correlations. In this case, the budgeted observable algebra $A(\Xi)$ approaches the full algebra of bounded operators $B(H)$ on the entire Hilbert space, and the corresponding conditional expectation map $E_\Xi(\rho)$ tends to the identity map, i.e., $E_\Xi(\rho_E) \to \rho_E$. Due to the continuity of von Neumann entropy, we obtain:
 
        \begin{equation}
        \lim_{\Xi\to\infty} S_\Xi(E, \Xi) = S_{\mathrm{VN}}(\rho_E) = k_B \ln \omega(E) = S_{\mathrm{micro}}(E)
        \label{eq:2.4}
        \end{equation}
      \vspace{3pt}       
      
        This indicates that CWT naturally recovers the standard microcanonical entropy, restoring the results of classical statistical mechanics.
        \item \textbf{Extremely Low Resource Limit ($\Xi \to 0$ or some minimal value} $\Xi_{\mathrm{min}}$): When the complexity budget $\Xi$ tends to its minimum value (e.g., $\Xi=0$, or $\Xi_{\mathrm{min}}$ corresponding to observations allowing only the most basic, non-complex operations), the budgeted observable algebra $A(\Xi)$ typically degenerates into a very simple or trivial algebra. For example, if $A(0)$ only contains scalar multiples of the identity operator (representing an observer who cannot resolve any internal structure, only confirm the system's existence in a $d$-dimensional Hilbert space), then $E_0(\rho_E)$ will be the maximally mixed state $(1/d)Id$. In this case, the complexity-windowed entropy reaches its maximum value:
        
        \begin{equation}
        \lim_{\Xi\to 0 \text{ (or } \Xi_{\mathrm{min}})} S_\Xi(E, \Xi) = k_B \ln d \text{ (or a constant independent/weakly dependent on }E\text{)}
        \label{eq:2.5}
        \end{equation}
           \vspace{3pt}  
    
        This reflects that when physical resources are extremely scarce, making it almost impossible to perform any effective operations or resolve any internal structure, the observer's "degree of ignorance" about the system's state reaches its maximum, and its effective entropy is determined solely by the total dimension of the Hilbert space (or the coarsest-grained degrees of freedom).
    \end{enumerate}
\end{enumerate}
The rigorous derivations of these two limiting behaviors are given in Appendix \ref{c.1}.
These mathematical properties, especially the smoothness of $S_\Xi(E, \Xi)$ with respect to energy $E$ for finite $\Xi$, and its controlled monotonic behavior with changing complexity budget $\Xi$, are key to constructing a well-behaved and physically self-consistent complexity-windowed thermodynamics.
They lay a solid foundation for defining core thermodynamic quantities such as the effective temperature $T_\Xi$ and complexity generation potential $\Pi_\Xi$ by differentiating $S_\Xi$ in subsequent chapters, analyzing their behavior near phase transitions and critical points, and constructing the extended first law that includes information processing work.

\section{Complexity-Windowed Thermodynamics Framework}
\label{cex.3}
In the previous chapter, we introduced the complexity-windowed entropy $S_\Xi(E,\Xi)$ as the core state function of the Complexity-Windowed Thermodynamics (CWT) framework.
We elucidated its physical meaning as the observer's "entropy of ignorance" and established its key mathematical properties: for any finite complexity budget $\Xi$, $S_\Xi(E,\Xi)$ is not only a smooth function of energy $E$ but also a monotonically non-increasing function of $\Xi$.
In particular, the smoothness of $S_\Xi(E,\Xi)$ with respect to energy $E$ provides a solid theoretical foundation for us to use standard thermodynamic differential methods to define other thermodynamic quantities and establish precise mathematical relationships between them.
The core objective of this chapter is precisely to systematically construct a self-consistent and complete CWT thermodynamic formalism based on the complexity-windowed entropy $S_\Xi(E,\Xi)$ and its proven well-behaved mathematical properties.
This formalism aims to intrinsically reflect the fundamental constraint of finite physical resources on the upper limit of complexity that a system can generate or an observer can resolve, and to quantify the energetic consequences of this constraint.
Specifically, we will first define key intensive quantities conjugate to the fundamental state function $S_\Xi$ and the newly introduced extensive quantity $\Xi$.
These quantities include the effective temperature $T_\Xi$ derived from $S_\Xi$, and a novel and crucial intensive quantity conjugate to the complexity budget $\Xi$—which we term the \textbf{complexity generation potential $\Pi_\Xi$}.
Subsequently, based on these newly defined thermodynamic variables, we will derive the extended first law of thermodynamics within the CWT framework.
This extended law will explicitly include an energy term related to the change in the complexity budget $\Xi$, namely $\Pi_\Xi\mathrm{d}\Xi$, which we interpret as "information processing work" or "complexity generation work," quantifying the energy cost that the external world must pay to change the system's resolvable complexity.
Finally, to ensure that CWT theory is a reasonable generalization of classical thermodynamics, we will rigorously examine the behavior of this formalism in the limit of infinite complexity budget ($\Xi\to\infty$) to verify its natural and seamless recovery of classical statistical mechanics and thermodynamic theory, thereby establishing CWT's status as a more universal framework.

\subsection{New Thermodynamic Variables}
\label{cex.3.1}
In classical thermodynamics, temperature $T$ is typically defined through the partial derivative of entropy $S$ with respect to energy $E$, i.e., $T^{-1} = \left(\frac{\partial S}{\partial E}\right)_{V,N,...}$, where the subscripts indicate that other macroscopic variables (such as volume $V$, particle number $N$, etc.) are held constant.
The Complexity-Windowed Thermodynamics (CWT) framework follows a similar path, utilizing the complexity-windowed entropy $S_\Xi(E,\Xi)$ introduced in the previous chapter as the core state function, to define an effective temperature that reflects the true thermal state of the system under complexity constraints.

\textbf{Definition 3.1 (Complexity-Windowed Temperature $T_\Xi$):}
The complexity-windowed temperature $T_\Xi$ (or simply windowed temperature) is defined as the partial derivative of the system's internal energy $E$ with respect to the complexity-windowed entropy $S_\Xi(E,\Xi)$, while keeping the complexity budget $\Xi$ and other possible macroscopic parameters (omitted here for brevity, such as $V, N$, etc.) fixed:

\begin{equation}
T_\Xi(E,\Xi) = \left(\frac{\partial E(S_\Xi,\Xi)}{\partial S_\Xi}\right)_{\Xi}
\label{eq:3.1a}
\end{equation}

Equivalently, if $S_\Xi$ is considered as a function of $E$ and $\Xi$, its reciprocal is defined as:

\begin{equation}
T_\Xi^{-1}(E,\Xi) = \left(\frac{\partial S_\Xi(E,\Xi)}{\partial E}\right)_{\Xi}
\label{eq:3.1b}
\end{equation}

The complexity-windowed temperature $T_\Xi$ represents the effective thermodynamic temperature experienced by the system under the observational window and dynamical accessibility limitations imposed by the complexity budget $\Xi$.
It precisely quantifies the change in the system's internal energy $E$ corresponding to each unit increase in the observer's effective entropy $S_\Xi$ (i.e., an increase in the observer's "degree of ignorance"), while keeping the complexity budget $\Xi$ constant (i.e., without changing the accessible observable algebra $A(\Xi)$ or the upper limit of generatable state complexity).
Crucially, as elucidated in Proposition 2.1(ii) and will be rigorously proven in Appendix \ref{c.1}, the complexity-windowed entropy $S_\Xi(E,\Xi)$ is a smooth function of energy $E$ for any finite $\Xi$.
This key mathematical property directly ensures that the effective temperature $T_\Xi(E,\Xi)$ derived from it is well-defined, continuous, and finite over all energy ranges (except for possible exception points of measure zero).
This means that even in regions where classical thermodynamic temperature would encounter discontinuities (such as at first-order phase transition points) or divergences (such as at certain critical points), $T_\Xi$ can still exhibit well-behaved, non-singular behavior.
This constitutes one of the core mechanisms by which the CWT framework can universally regularize classical thermodynamic singularities, the specific applications of which will be discussed in detail in Chapter \ref{cex.4}.
In addition to the effective temperature $T_\Xi$, the construction of the CWT framework naturally introduces another key intensive quantity conjugate to the complexity budget $\Xi$.
Given that the system's internal energy $E$ is now considered a function of the extensive quantities $S_\Xi$ (observer's effective entropy) and $\Xi$ (complexity budget), i.e., $E = E(S_\Xi,\Xi)$ (for simplicity, other variables like $V, N$ are temporarily ignored), we can draw an analogy with the conjugate relationship between generalized forces and generalized displacements in classical thermodynamics, such as pressure $p = -\left(\frac{\partial E}{\partial V}\right)_S$, to define an intensive quantity conjugate to the complexity budget $\Xi$.
We call this quantity the Complexity Generation Potential, denoted as $\Pi_\Xi$.

\textbf{Definition 3.2 (Complexity Generation Potential $\Pi_\Xi$):}
The complexity generation potential $\Pi_\Xi$ is defined as the partial derivative of the system's internal energy $E$ with respect to the complexity budget $\Xi$, while keeping the observer's effective entropy $S_\Xi$ and other possible macroscopic parameters (such as volume $V$, particle number $N$, etc.) fixed:

\begin{equation}
\Pi_\Xi(S_\Xi,\Xi) = \left(\frac{\partial E(S_\Xi,\Xi)}{\partial \Xi}\right)_{S_\Xi,V,N}
\label{eq:3.2}
\end{equation}

Using standard thermodynamic relations, $\Pi_\Xi$ can also be expressed in terms of the partial derivatives of entropy $S_\Xi(E,\Xi)$ with respect to its variables.
Considering $S_\Xi = S_\Xi(E,\Xi)$, under the condition that $S_\Xi$ is kept constant ($\mathrm{d}S_\Xi = 0$), we have:

\begin{equation}
0 = \left(\frac{\partial S_\Xi}{\partial E}\right)_{\Xi} \mathrm{d}E + \left(\frac{\partial S_\Xi}{\partial \Xi}\right)_{E} \mathrm{d}\Xi
\end{equation}

Therefore, $\left(\frac{\mathrm{d}E}{\mathrm{d}\Xi}\right)_{S_\Xi} = - \frac{(\partial S_\Xi/\partial \Xi)_{E}}{(\partial S_\Xi/\partial E)_{\Xi}}$.
Combining this with the definition of $T_\Xi^{-1}$ in Eq. \eqref{eq:3.1b}, we get:

\begin{equation}
\Pi_\Xi(E,\Xi) = -T_\Xi(E,\Xi) \left(\frac{\partial S_\Xi(E,\Xi)}{\partial \Xi}\right)_{E}
\label{eq:3.3}
\end{equation}

The complexity generation potential $\Pi_\Xi$ is a novel thermodynamic intensive quantity with profound physical meaning.
As pointed out in Proposition 2.1(i) and will be rigorously proven in Appendix \ref{c.3}, the complexity-windowed entropy $S_\Xi(E,\Xi)$ is a monotonically non-increasing function of the complexity budget $\Xi$, i.e., $(\partial S_\Xi/\partial \Xi)_E \le 0$.
Considering that the effective temperature $T_\Xi > 0$ under normal conditions (CWT aims to ensure its good behavior), we can infer from \eqref{eq:3.3} that $\Pi_\Xi \ge 0$.
This non-negative property is crucial as it endows $\Pi_\Xi$ with a clear physical meaning: it quantifies the minimum work done on the system by the external world, or the minimum energy absorbed by the system, to reversibly increase the complexity budget $\Xi$ by one unit (e.g., by providing more physical resources to enhance the observer's resolution capability, or allowing the system to evolve to and maintain more complex quantum states) while keeping the observer's effective entropy $S_\Xi$ (i.e., "degree of ignorance") constant.
Specifically, if $\Pi_\Xi$ is positive, it means that increasing the complexity budget $\Xi$ ($\mathrm{d}\Xi > 0$) must be accompanied by positive work done on the system by the external world ($\Pi_\Xi\mathrm{d}\Xi > 0$), with the system absorbing energy from the outside.
This is highly consistent with our usual physical intuition that accessing, generating, maintaining, or resolving more complex structures or informational states typically requires the consumption of additional energy resources.
Conversely, if the complexity budget $\Xi$ is to be reduced ($\mathrm{d}\Xi < 0$) while keeping $S_\Xi$ constant, the system will release energy ($\Pi_\Xi\mathrm{d}\Xi < 0$, because $\Pi_\Xi\ge0$).
Therefore, $\Pi_\Xi$ acts like a coefficient measuring the "cost" of increasing the system's resolvable complexity, and its non-negativity ensures the energetic reasonableness of information processing and complexity generation processes.
The introduction of this physical quantity directly reflects the new dimension brought about by treating computational complexity (indirectly represented by $\Xi$) as a quantifiable physical resource and incorporating it into thermodynamic descriptions; its specific numerical value and behavior will depend on the system's microscopic properties, the choice of complexity measure $C^*$, and the method of constructing the budgeted observable algebra $A(\Xi)$.

\subsection{Extended First Law of Thermodynamics}
\label{cex.3.2}
The first law of classical thermodynamics, usually expressed as $\mathrm{d}U = \delta Q + \delta W$ (or $\mathrm{d}E = T\mathrm{d}S - p\mathrm{d}V + \sum_i \mu_i\mathrm{d}N_i$ for reversible processes), is a manifestation of the principle of energy conservation in thermodynamic systems.
It states that the change in a system's internal energy $E$ is equal to the sum of the heat $\delta Q$ absorbed by the system and the various forms of work $\delta W$ done on it by the surroundings.
In the Complexity-Windowed Thermodynamics (CWT) framework, a core development is to treat the complexity budget $\Xi$ as an independent, fundamental thermodynamic variable, akin to an extensive quantity.
Therefore, the system's internal energy $E$ depends not only on traditional macroscopic variables such as the observer's effective entropy $S_\Xi$, volume $V$, and particle numbers $N_i$, but also explicitly on the complexity budget $\Xi$.
This implies that the classical first law of thermodynamics needs to be correspondingly extended to intrinsically include the energy exchange associated with changing the complexity budget $\Xi$.
Considering that internal energy $E$ is a function of variables $S_\Xi, \Xi, V, N_i$, etc., i.e., $E = E(S_\Xi, \Xi, V, N_1, ...)$, its total differential can be generally written as:

\begin{equation}
\mathrm{d}E = \left(\frac{\partial E}{\partial S_\Xi}\right)_{\Xi,V,N} \mathrm{d}S_\Xi + \left(\frac{\partial E}{\partial \Xi}\right)_{S_\Xi,V,N} \mathrm{d}\Xi + \left(\frac{\partial E}{\partial V}\right)_{S_\Xi,\Xi,N} \mathrm{d}V + \sum_{i} \left(\frac{\partial E}{\partial N_i}\right)_{S_\Xi,\Xi,V,N_{j\neq i}} \mathrm{d}N_i
\label{eq:3.4}
\end{equation}

According to the definition of the complexity-windowed temperature $T_\Xi$ in the previous section (Definition 3.1), we have $\left(\frac{\partial E}{\partial S_\Xi}\right)_{\Xi,V,N} = T_\Xi$.
Similarly, according to the definition of the complexity generation potential $\Pi_\Xi$ in the previous section (Definition 3.2), we have $\left(\frac{\partial E}{\partial \Xi}\right)_{S_\Xi,V,N} = \Pi_\Xi$.
Furthermore, we still adopt the standard definitions for pressure $p = -\left(\frac{\partial E}{\partial V}\right)_{S_\Xi,\Xi,N}$ and chemical potential $\mu_i = \left(\frac{\partial E}{\partial N_i}\right)_{S_\Xi,\Xi,V,N_{j\neq i}}$.
Substituting these definitions into Eq. \eqref{eq:3.4}, we obtain the complete form of the extended first law of thermodynamics within the CWT framework:

\begin{equation}
\mathrm{d}E = T_\Xi \mathrm{d}S_\Xi + \Pi_\Xi \mathrm{d}\Xi - p \mathrm{d}V + \sum_{i} \mu_i \mathrm{d}N_i
\label{eq:3.5}
\end{equation}

The extended first law of thermodynamics, Eq.\eqref{eq:3.5}, has profound and rich physical meaning.
Firstly, it remains a specific mathematical expression of the universal principle of energy conservation in thermodynamic processes that include complexity constraints.
It confirms that changes in the system's internal energy must originate from heat exchange with the surroundings or the transfer of various forms of work.
Secondly, the heat term $T_\Xi\mathrm{d}S_\Xi$ (for reversible processes) is formally consistent with its counterpart in classical thermodynamics, representing the heat absorbed or released by the system due to changes in the effective entropy $S_\Xi$ during a reversible process; the key difference here is that both the entropy $S_\Xi$ (observer's effective entropy or "entropy of ignorance") and temperature $T_\Xi$ are effective quantities defined under a specific complexity window $\Xi$, intrinsically reflecting the limitations of the observer's resolution capability.
Most crucially, the newly introduced term in Eq. \eqref{eq:3.5} is $\Pi_\Xi \mathrm{d}\Xi$.
This term is one of the core contributions of the CWT framework to the fundamental equation of thermodynamics, as it explicitly incorporates the energy exchange associated with changing the system's resolvable complexity into the scope of the first law.
As we argued in Section 3.1, the complexity generation potential $\Pi_\Xi$ is typically non-negative ($\Pi_\Xi \ge 0$).
Therefore, the term $\Pi_\Xi \mathrm{d}\Xi$ represents the "information processing work" or "complexity generation work" done by the surroundings to change the system's complexity budget $\Xi$ in a reversible process.
Specifically:

\begin{enumerate}
    \item When the complexity budget $\Xi$ increases ($\mathrm{d}\Xi > 0$), it means the observer's resolution capability is enhanced, or the system is allowed to reach and maintain states of higher complexity. In this case, since $\Pi_\Xi \ge 0$, the information processing work $\Pi_\Xi\mathrm{d}\Xi \ge 0$. This indicates that to achieve this increase in complexity, the surroundings typically need to do positive work on the system (or at least no negative work), and the system absorbs energy from the outside. This conclusion is entirely consistent with our everyday experience and physical intuition that constructing, maintaining, operating more complex systems, or performing more complex information processing tasks (e.g., running more profound quantum algorithms, using higher-precision measurement instruments) usually requires additional energy input.
    \item Conversely, when the complexity budget $\Xi$ decreases ($\mathrm{d}\Xi < 0$), meaning the observer's resolution capability is reduced or the system is restricted to a range of states with lower complexity, the information processing work $\Pi_\Xi\mathrm{d}\Xi \le 0$. This suggests that the system may do work on the surroundings, release energy, or its energy may remain unchanged during this process.
\end{enumerate}
The introduction of this "information processing work" term $\Pi_\Xi\mathrm{d}\Xi$ provides a solid foundation for understanding energy consumption in information processing, exploring the physical limits of computation, and endowing computational complexity itself with physical reality (i.e., its change is associated with energy exchange) from the level of fundamental thermodynamic principles.
It essentially quantifies the universal energy cost that must be paid for "enhancing information processing capability or resolvable complexity" (indirectly reflected by $\Xi$) when viewed as a physical process.
Based on this extended first law of thermodynamics, Eq. \eqref{eq:3.5}, we can use standard thermodynamic methods (such as Legendre transformations) starting from the internal energy $E(S_\Xi, \Xi, V, N_i)$ to define a series of complexity-windowed thermodynamic potentials corresponding to classical thermodynamic potentials.
For example, one can define the complexity-windowed Helmholtz free energy $F_\Xi(T_\Xi, \Xi, V, N_i) = E - T_\Xi S_\Xi$, whose total differential is $\mathrm{d}F_\Xi = -S_\Xi\mathrm{d}T_\Xi + \Pi_\Xi\mathrm{d}\Xi - p\mathrm{d}V + \sum_i \mu_i\mathrm{d}N_i$.
These generalized thermodynamic potentials and their differential relations, detailed in Appendix \ref{E.1}, provide powerful and self-consistent mathematical tools for analyzing the equilibrium properties of systems under different constraints (such as isothermal, isobaric, constant complexity budget, or constant complexity generation potential), deriving generalized Maxwell relations, and discussing the thermodynamic stability of systems.

\subsection{Limiting Behavior and Recovery of Classical Theory}
\label{cex.3.3}
An important test of the reasonableness of any new framework aiming to generalize existing physical theories is whether it can naturally and seamlessly recover the classical theory it generalizes under appropriate physical limiting conditions.
For the Complexity-Windowed Thermodynamics (CWT) framework, this crucial test lies in examining its behavior when the complexity budget $\Xi$ tends to infinity ($\Xi\to\infty$).
In this limit, physical resources are, in principle, considered infinitely abundant, allowing the system to access and resolve arbitrarily complex quantum states and their correlation structures, which is precisely the idealized situation implicit in classical statistical mechanics.
As pointed out in Proposition 2.1(iii) (and proven in detail in Appendix \ref{c.1}), when the complexity budget $\Xi$ tends to infinity, the complexity-windowed entropy $S_\Xi(E,\Xi)$ accurately converges to the classical microcanonical entropy $S_{\mathrm{micro}}(E)$:

\begin{equation}
\lim_{\Xi\to\infty} S_\Xi(E,\Xi) = S_{\mathrm{micro}}(E)
\label{eq:3.6}
\end{equation}

Based on this core limiting relationship, we can further analyze the asymptotic behavior of other key thermodynamic quantities in the CWT framework.
First, we examine the complexity-windowed temperature $T_\Xi$.
According to its definition Eq. \eqref{eq:3.1b}, $T_\Xi^{-1}(E,\Xi) = (\partial S_\Xi(E,\Xi)/\partial E)_\Xi$.
In the limit $\Xi\to\infty$, and assuming that the limit operation and partial differentiation can be interchanged (which is reasonable given the good smoothness of $S_\Xi(E,\Xi)$ and its limit function $S_{\mathrm{micro}}(E)$), we get:

\begin{equation}
\lim_{\Xi\to\infty} T_\Xi^{-1}(E,\Xi) = \lim_{\Xi\to\infty} \left(\frac{\partial S_\Xi(E,\Xi)}{\partial E}\right)_\Xi = \left(\frac{\partial [\lim_{\Xi\to\infty} S_\Xi(E,\Xi)]}{\partial E}\right)_\Xi = \left(\frac{\partial S_{\mathrm{micro}}(E)}{\partial E}\right)
\end{equation}

Since the classical thermodynamic temperature $T_{\mathrm{classical}}$ is defined as $T_{\mathrm{classical}}^{-1}(E) = (\partial S_{\mathrm{micro}}(E)/\partial E)$, therefore:

\begin{equation}
\lim_{\Xi\to\infty} T_\Xi^{-1}(E,\Xi) = T_{\mathrm{classical}}^{-1}(E)
\label{eq:3.7}
\end{equation}

This means that in the limit of infinite complexity budget, the effective temperature $T_\Xi$ in the CWT framework accurately recovers the classical statistical mechanical temperature $T_{\mathrm{classical}}$.
Second, we analyze the intensive quantity conjugate to the complexity budget $\Xi$—the complexity generation potential $\Pi_\Xi$.
According to Eq. \eqref{eq:3.3}, we have $\Pi_\Xi(E,\Xi) = -T_\Xi(E,\Xi) (\partial S_\Xi(E,\Xi)/\partial \Xi)_E$.
When the complexity budget $\Xi$ becomes very large, such that the corresponding budgeted observable algebra $A(\Xi)$ already approaches or completely covers the full algebra of bounded operators $B(H)$ on the system's Hilbert space, further increasing $\Xi$ will have a negligible effect on the complexity-windowed entropy $S_\Xi$.
This is because $S_\Xi(E,\Xi)$, as a monotonically non-increasing function of $\Xi$ (Proposition 2.1(i)), has already saturated and converged to its lower bound $S_{\mathrm{micro}}(E)$ as $\Xi\to\infty$.
Therefore, in this limit, the partial derivative of $S_\Xi$ with respect to $\Xi$ will tend to zero:

\begin{equation}
\lim_{\Xi\to\infty} \left(\frac{\partial S_\Xi(E,\Xi)}{\partial \Xi}\right)_E = 0
\label{eq:3.8}
\end{equation}

Combining this with the limiting behavior of $T_\Xi$ in Eq. \eqref{eq:3.7}, we can deduce the behavior of the complexity generation potential $\Pi_\Xi$ in the infinite resource limit:

\begin{equation}
\lim_{\Xi\to\infty} \Pi_\Xi(E,\Xi) = \lim_{\Xi\to\infty} \left[-T_\Xi(E,\Xi) \left(\frac{\partial S_\Xi(E,\Xi)}{\partial \Xi}\right)_E\right] = -T_{\mathrm{classical}}(E) \cdot 0 = 0
\label{eq:3.9}
\end{equation}

This result has a clear physical meaning: in the idealized limit of possessing infinite physical resources where arbitrarily high complexity can be easily achieved, the marginal energy cost $\Pi_\Xi$ associated with further increasing the complexity budget $\Xi$ (i.e., nominally "enhancing" an already infinite resolution capability) will vanish.
Finally, we substitute the limiting behavior of these thermodynamic quantities into the extended first law of thermodynamics under the CWT framework, Eq. \eqref{eq:3.5}: $\mathrm{d}E = T_\Xi \mathrm{d}S_\Xi + \Pi_\Xi \mathrm{d}\Xi$ (for brevity, here and subsequently, only terms related to $S_\Xi$ and $\Xi$ are considered; other terms like $p\mathrm{d}V, \mu\mathrm{d}N$, etc., behave similarly in the classical limit).
In the limit $\Xi\to\infty$, we have:

\begin{multline}
\lim_{\Xi\to\infty} (\mathrm{d}E = T_\Xi \mathrm{d}S_\Xi + \Pi_\Xi \mathrm{d}\Xi) \\
\longrightarrow \mathrm{d}E = (\lim_{\Xi\to\infty} T_\Xi) \mathrm{d}(\lim_{\Xi\to\infty} S_\Xi) + (\lim_{\Xi\to\infty} \Pi_\Xi) \mathrm{d}\Xi \\
\longrightarrow \mathrm{d}E = T_{\mathrm{classical}} \mathrm{d}S_{\mathrm{micro}} + 0 \cdot \mathrm{d}\Xi \\
\longrightarrow \mathrm{d}E = T_{\mathrm{classical}} \mathrm{d}S_{\mathrm{micro}}
\end{multline}

This is precisely the fundamental equation in classical thermodynamics describing the relationship between changes in internal energy and entropy (for reversible processes, and neglecting other forms of work).
In summary, in the limit of infinite complexity budget, the core state function ($S_\Xi$), key intensive quantities ($T_\Xi, \Pi_\Xi$), and fundamental thermodynamic law (first law) of the CWT framework all naturally and seamlessly recover their corresponding forms in classical statistical mechanics and thermodynamic theory.
In this limit, the extra energy term associated with the complexity budget ($\Pi_\Xi\mathrm{d}\Xi$, i.e., "information processing work") vanishes because $\Pi_\Xi$ tends to zero, the complexity generation potential itself also becomes zero, and the windowed temperature and windowed entropy accurately revert to their classical counterparts, respectively.
This result not only verifies the internal logical self-consistency of CWT theory but also strongly demonstrates that CWT is a profound and physically necessary generalization of classical theory, made by considering the fundamental constraint of finite physical resources on achievable complexity, rather than conflicting with it.
From the perspective of CWT, the theoretical difficulties encountered by classical thermodynamics in dealing with certain extreme physical situations (such as phase transition singularities, negative temperature phenomena) can be understood as the natural consequences arising from neglecting the intrinsic complexity constraints determined by finite physical resources in the idealized (and perhaps not fully achievable or applicable in many real physical systems) limit of $\Xi\to\infty$.
By introducing and systematically handling a finite complexity budget $\Xi$, CWT provides a theoretical framework that is potentially closer to physical reality and more universal for describing the thermodynamic behavior of these complex systems.

\section{Universal Regularization of Classical Thermodynamic Singularities}
\label{cex.4}
The preceding chapters have systematically constructed the theoretical framework of Complexity-Windowed Thermodynamics (CWT).
Its core foundation lies in the introduction of the "complexity budget" $\Xi$ as a key physical parameter to quantify the intrinsic constraint of finite physical resources on the complexity of states accessible to the system.
On this basis, we defined the core state function—the complexity-windowed entropy $S_\Xi(E,\Xi)$—and rigorously proved that for any finite $\Xi$, it is a smooth function of energy $E$.
This fundamental property further ensures that the thermodynamic quantities derived from it, such as the effective temperature $T_\Xi$ and the complexity generation potential $\Pi_\Xi$ conjugate to the complexity budget $\Xi$, exhibit good continuity and boundedness.
The core task of this chapter is precisely to delve into the profound physical consequences arising from this core mathematical feature—the smoothness of $S_\Xi(E,\Xi)$ with respect to energy $E$.
We will systematically demonstrate how the CWT framework, by virtue of this property, can universally "soften" or regularize the various singular behaviors encountered by classical thermodynamics when dealing with systems such as those at phase transition points, critical phenomena, and those with special energy spectra (like the negative absolute temperature regime), for example, discontinuities or divergences of physical quantities.
To achieve this, we will first elucidate the general working principle of a finite complexity budget $\Xi$ as an intrinsic, universal physical regularization mechanism.
Subsequently, we will focus respectively on first-order phase transitions, second-order phase transitions (critical points), and the phenomenon of negative absolute temperature—three typical "problematic areas" in classical thermodynamics—and analyze in detail how the CWT framework effectively resolves or circumvents the related singularities in these specific contexts.
Finally, this chapter will conclude from a more macroscopic perspective and emphasize a core viewpoint: many singularities observed in classical thermodynamics can, under the theoretical picture of CWT, be uniformly understood as mathematical artifacts arising from the idealized and often unattainable limit of infinite physical resources (corresponding to a complexity budget $\Xi \to \infty$).
\subsection{Universal Regularization Mechanism and its Manifestation in First-Order Phase Transitions}
\label{cex.4.1}
Singularities in classical thermodynamics, whether originating from the discontinuity of entropy $S$ or its derivatives (like $1/T$) due to latent heat in first-order phase transitions, or from the divergence of response functions (such as specific heat $C_V$, magnetic susceptibility $\chi$, etc.) caused by diverging correlation lengths at second-order critical points, are often rooted in the system's ability to access or exhibit infinitely fine microscopic structures or infinitely long-range correlations.
For example, at a critical point, fluctuations exist at all scales, forming fractal structures; in the two-phase coexistence region of a first-order phase transition, an idealized description assumes an infinitely thin phase interface with well-defined energy and entropy densities.
The generation or resolution of these idealized states or configurations often requires extremely high, or even infinite, computational complexity $C^*$.
The CWT framework fundamentally changes this picture by introducing a finite complexity budget $\Xi$.
Its core regularization mechanism lies in the fact that for any finite $\Xi$, the corresponding budgeted observable algebra $A(\Xi)$ is a finite-dimensional algebra, meaning that under the constraint of complexity budget $\Xi$, an observer can only resolve a finite number of independent observables.
The actual state of the system is effectively projected (coarse-grained) onto the state space of this finite-dimensional algebra via the conditional expectation $E_\Xi$; this process "averages out" or "blurs" those fine structures or long-range correlations that require complexity exceeding $\Xi$ to resolve.
Since the complexity-windowed entropy $S_\Xi(E, \Xi)$ is a von Neumann entropy defined on the state space of $A(\Xi)$ (a strictly concave and smooth function), and its argument $E_\Xi(\Pi_E/\omega(E))$ also depends smoothly on $E$ (assuming the energy shell projection $\Pi_E$ varies smoothly with $E$), $S_\Xi(E, \Xi)$ as a function of energy $E$ must be smooth (according to Proposition 2.1 (ii)).
This smoothness directly ensures that its derivatives of all orders (under finite $\Xi$) are finite and continuous, thereby making thermodynamic quantities defined by these derivatives, such as the effective temperature $T_\Xi^{-1} = (\partial S_\Xi/\partial E)_\Xi$ and effective specific heat $C_{V,\Xi} \propto T_\Xi^2 (\partial^2S_\Xi/\partial E^2)_\Xi$, continuous and bounded.
Physically, a finite complexity budget $\Xi$ plays the role of an \textbf{intrinsic physical regularizer}, preventing the system from accessing, in its thermodynamic description, those idealized states that require infinite computational resources to prepare or distinguish—these high-complexity states are often the "culprits" leading to classical singularities.
The manifestation of this universal regularization mechanism is particularly prominent in first-order phase transitions.
Classical first-order phase transitions (such as solid-liquid or liquid-gas transitions) are typically characterized by the presence of latent heat $L$ at the transition temperature $T_c$, leading to a jump in entropy $S$ of magnitude $\Delta S = L/T_c$; correspondingly, $1/T = \partial S/\partial E$ also exhibits a discontinuity at the transition point.
This behavior is usually associated with the coexistence at $T_c$ of two macroscopic phases with different energy and entropy densities (e.g., a low-energy ordered phase A and a high-energy disordered phase B) and an idealized, infinitely thin interface.
Within the CWT framework, if describing the interface state of precise two-phase coexistence, or the process of perfect transformation from one pure phase to another, involves intermediate configurations of very high complexity ($C^* > \Xi$) or requires extremely long relaxation times, then the derivative $(\partial S_\Xi/\partial E)_\Xi$ of the complexity-windowed entropy $S_\Xi(E, \Xi)$ near the transition energy $E_c$ will remain continuous, thus making the complexity-windowed temperature $T_\Xi(E)$ also continuous at $E_c$ (\textbf{Proposition 4.1, Regularization of First-Order Phase Transitions}).
The reasoning behind this is that when the system's total energy $E$ crosses $E_c$, because the complexity of the high-energy phase B (or its key configurations) exceeds the budget $\Xi$, its contribution to $S_\Xi$ will be suppressed or averaged out by the conditional expectation $E_\Xi$.
Therefore, the behavior of $S_\Xi(E, \Xi)$ will be primarily determined by the complexity-allowed phase A (and possibly smooth transition regions with complexity not exceeding $\Xi$), causing the curve of $S_\Xi(E, \Xi)$ versus $E$ to become smooth near $E_c$, and its derivative $1/T_\Xi$ no longer exhibits a jump.
The entropy jump corresponding to the classical latent heat of transition is effectively "smeared out" in CWT over an energy interval whose width depends on $\Xi$; $T_\Xi(E)$ near $E_c$ no longer presents as a strict plateau but may exhibit an S-shaped smooth transition curve.
Only when $\Xi \to \infty$ does this energy interval tend to zero, and the entropy jump and temperature plateau are precisely recovered.
This behavior is similar to the "rounding" of first-order phase transitions observed in finite-size systems or numerical simulations with finite evolution times\cite{binder1987theory}, but CWT attributes it to a more fundamental limitation of the complexity budget, originating from finite physical resources.
\subsection{Second-Order Phase Transitions, Critical Phenomena, and Negative Absolute Temperature}
\label{cex.4.2}
The universal regularization mechanism of the CWT framework applies not only to first-order phase transitions but also has profound corrective effects on the more subtle second-order phase transitions (or continuous phase transitions) and their critical phenomena in classical theory.
Classical second-order phase transitions are characterized by spontaneous symmetry breaking at the critical point $T_c$ (corresponding to critical energy $E_c$), where the correlation length $\xi_{\mathrm{corr}}$ diverges, leading to power-law divergences in thermodynamic response functions such as specific heat $C_V$ and magnetic susceptibility $\chi$\cite{wilson1971renormalization}.
Physically, generating and detecting these critical fluctuation modes with infinite (or extremely large) correlation lengths usually requires very high system complexity $C^*$, which may diverge with the growth of $\xi_{\mathrm{corr}}$.
A finite complexity budget $\Xi$ effectively imposes an upper limit on the effective correlation length that the system can resolve or maintain, i.e., $\xi_{\mathrm{corr},\Xi} \le f(\Xi)$, where $f(\Xi)$ is a function that increases with $\Xi$.
Therefore, the contribution of those critical fluctuation modes that span all scales and require $C^* > \Xi$ for a complete description will be truncated or smoothed out in the state $E_\Xi(\Pi_E/\omega(E))$ coarse-grained by the conditional expectation $E_\Xi$.
Since the complexity-windowed entropy $S_\Xi(E, \Xi)$ is smooth with respect to energy $E$ (Proposition 2.1(ii)), its partial derivatives of all orders with respect to energy $E$ are generally well-defined and finite when $\Xi$ is finite.
In particular, the system's effective specific heat at constant volume, $C_{V,\Xi}$, can be related to the derivatives of $S_\Xi$ with respect to energy $E$ through standard thermodynamic relations.
Using the definition of effective temperature $T_\Xi^{-1} = (\partial S_\Xi/\partial E)_{\Xi,V}$, we can derive:

\begin{equation}
C_{V,\Xi} = - \frac{\left[\left(\frac{\partial S_\Xi}{\partial E}\right)_{\Xi,V}\right]^2}{\left(\frac{\partial^2 S_\Xi}{\partial E^2}\right)_{\Xi,V}}
\label{eq:4.Y}
\end{equation}
\noindent\textit{Note:} (Here, specific heat at constant volume is considered, and volume $V$ is assumed to be a fixed parameter)

Since for any finite complexity budget $\Xi$, the second partial derivative of $S_\Xi$ with respect to energy $E$, $(\partial^2 S_\Xi/\partial E^2)_{\Xi,V}$, remains finite (except possibly at points of measure zero), and typically $T_\Xi \ne 0$, it follows from  Eq. \eqref{eq:4.Y} that $C_{V,\Xi}$ will remain bounded for any finite $\Xi$.
This implies that the CWT framework can universally 'soften' the potential divergences of second-order response functions (such as specific heat, magnetic susceptibility, etc.) near critical points that occur in classical theory.
Therefore, the CWT framework predicts that, under a finite complexity budget, the divergence of response functions at classical second-order critical points will be replaced by a finite peak whose height and width both depend on $\Xi$ (\textbf{Proposition 4.2, Boundedness of Second-Order Response Functions}).
This behavior is qualitatively consistent with phenomena observed in finite-size scaling theory or in calculations using certain tensor network methods (such as PEPS, where the tensor bond dimension $D$ can be considered a type of complexity budget, $\Xi \sim \mathrm{poly}(D)$)\cite{verstraete2008matrix}, where the specific heat peak grows finitely with system size $L$ (or bond dimension $D$) instead of diverging.
CWT attributes this to a more fundamental limitation on accessible complexity imposed by physical resources (manifested as $\Xi$).
In addition to phase transitions and critical phenomena, the CWT framework also offers a new perspective on the controversial concept of negative absolute temperature in classical thermodynamics.
In certain special isolated systems with an upper energy bound $E_{\mathrm{max}}$ (such as idealized nuclear spin systems), after the system's energy $E$ exceeds a certain energy $E_m$ at which entropy reaches its maximum value $S_{\mathrm{max}}$, the entropy $S(E)$ may decrease with further increases in energy, i.e., $\partial S/\partial E < 0$, leading to a classical thermodynamic temperature $T = (\partial S/\partial E)^{-1} < 0$ \cite{graham1991ramsey}.
However, reaching states near these upper energy limits where entropy decreases often requires the system to be in highly specific, precisely controlled configurations, which typically have very high generation complexity $C^*$.
For example, in a spin system, for all (or the vast majority of) spins to still point in a specific direction at high energy to reduce entropy (relative to high-entropy states with random orientations), requires more refined preparation and control than allowing them to orient randomly (corresponding to states near $S_{\mathrm{max}}$), and thus has a higher $C^*$.
If the complexity $C^*$ of these states that cause entropy to decrease in the high-energy region is generally greater than the current complexity budget $\Xi$, then the budgeted observable algebra $A(\Xi)$ will not be able to fully distinguish these high-complexity, low-entropy configurations from those states with lower complexity ($C^* \le \Xi$) and relatively higher entropy.
The coarse-graining map $E_\Xi$ will effectively "average out" the peculiarities of these high-complexity states.
As a result, the complexity-windowed entropy $S_\Xi(E, \Xi)$ in the high-energy region $E > E_m$ may not decrease as significantly as the classical entropy $S(E)$; instead, it might continue to grow slowly, reach saturation (approaching $k_B \ln d(\Xi)$), or only begin to decrease at an effective energy $E_m(\Xi)$ determined by $\Xi$ and much higher than the classical $E_m$ (if such states with complexity not exceeding $\Xi$ exist).
Therefore, its derivative $(\partial S_\Xi/\partial E)_\Xi$ in the relevant region $E > E_m$ is likely to remain non-negative (or close to zero), thus keeping the complexity-windowed temperature $T_\Xi$ non-negative (or tending to infinity if $(\partial S_\Xi/\partial E)_\Xi \to 0^+$).
This implies that under finite and reasonable physical resources (corresponding to finite $\Xi$), the phenomenon of negative absolute temperature may not strictly occur, or the conditions for its appearance are more stringent than predicted by classical theory (\textbf{Proposition 4.3, Circumvention of Negative Absolute Temperature}).
The classical negative temperature phenomenon might be more prevalent in idealized theoretical models that allow for infinitely high-complexity states, or as quasi-steady-state phenomena achieved experimentally for short durations through special preparation methods, far from true thermodynamic equilibrium.

\subsection{Discussion: Classical Singularities as Products of the Infinite Resource Limit}
\label{cex.4.3}
Synthesizing the above analysis, the CWT framework, by introducing a finite complexity budget $\Xi$, provides a unified mechanism to systematically regularize various singularity problems encountered in classical thermodynamics.
Whether it is the discontinuity of physical quantities at phase transition points, the divergence of response functions at critical points, or the anomalous behavior of negative absolute temperature, all are effectively "softened" or circumvented from the perspective of CWT.
The core of this lies in the fact that any finite complexity budget $\Xi$ causes the system's effective state space (defined via $A(\Xi)$ and $E_\Xi$) to exhibit smoothness, thereby ensuring that all thermodynamic quantities derived from the complexity-windowed entropy $S_\Xi$ are continuous and bounded.
This leads to a profound viewpoint: \textbf{many singularities observed in classical thermodynamics may not be intrinsic properties of physical reality, but rather products of our unconscious adoption of the idealized limit of "infinite physical resources" (corresponding to $\Xi \to \infty$) in theoretical descriptions}.
In this limit, the system is allowed to access and resolve arbitrarily complex microstates and arbitrarily long-range correlations, thereby leading to mathematical discontinuities or divergences.
However, in any real physical system, available physical resources such as action, evolution time, and control precision are always finite, meaning that there must be an upper limit to the complexity of quantum states that a system can actually reach or an observer can resolve.
CWT precisely internalizes this fundamental physical constraint into the basic framework of thermodynamics.
This complexity-based regularization idea shares commonalities with the "rounding" of singularities observed in finite-size systems, finite evolution times, or numerical simulations, but CWT attributes its origin to a more universal, intrinsic complexity limitation stemming from quantum dynamics and information processing capabilities, rather than merely limitations of system size or simulation parameters.
The classical thermodynamic picture and its singularities can only be precisely recovered in the limit of $\Xi \to \infty$.
Therefore, CWT not only provides a self-consistent theoretical solution to the inherent difficulties of classical thermodynamics but also offers an important theoretical foundation and conceptual bridge for understanding the behavior of complex systems under resource constraints, and even for connecting thermodynamics, quantum dynamics, and computation theory.
This perspective emphasizes the core role of "executable complexity" or "achievable complexity" in defining physical reality and describing physical processes.

\section{Physical Resource Constraint Principles for Complexity Generation}
\label{cex.5}
The preceding chapters have systematically constructed the theoretical framework of Complexity-Windowed Thermodynamics (CWT).
Its core lies in introducing the abstract "complexity budget" $\Xi$ as a key parameter limiting the complexity of states accessible to the system and the resolvability of observables, and proving its universal effectiveness in regularizing classical thermodynamic singularities.
However, to make CWT theory more physically realistic and operational, a crucial question is: what specific, measurable physical resources does this abstract complexity budget $\Xi$ actually correspond to?
This chapter aims to answer this question.
We will explicitly link the complexity budget $\Xi$ with two of the most fundamental resources in physics—Action and Evolution Time—and, on this basis, derive two fundamental physical principles constraining the generation of quantum state complexity: the Action Constraint Principle and the Time Constraint Principle.
These principles not only deepen our understanding of the physical nature of complexity but also lay the foundation for subsequent discussions on CWT's dynamical predictions and experimental verification.
Finally, we will explore the "comprehensive resource feasible region" jointly determined by these two principles and discuss strategies for optimal resource allocation.

\subsection{Physical Correspondence of the Complexity Budget}
\label{cex.5.1}
In the initial construction of the CWT framework, we conceptualized the complexity budget $\Xi$ as a parameter reflecting the upper limit of the system's available "computational capability" or "information processing capability."
Now, we attempt to correspond it with more specific physical quantities.
The occurrence and evolution of physical processes are, in essence, limited by available physical resources.
In the fundamental frameworks of quantum mechanics and classical mechanics, action and time are the two most central dimensions.
First, action $S_{\mathrm{action}}$ is a fundamental concept in physics, with dimensions of energy multiplied by time, and plays a central role in the principle of least action.
In recent years, particularly in studies of quantum gravity and black hole physics, a series of profound conjectures have emerged that directly link the computational complexity $C^*$ of a quantum state to the action of a physical system, such as the "Complexity = Action" (CA) conjecture \cite{brown2016holographic}.
These studies inspire us that the maximum complexity a system can achieve may be proportional to the total action it can "consume" or "accumulate" during its evolution.
Therefore, we propose a core hypothesis linking the complexity budget $\Xi$ with the total effective action $S_{\mathrm{process}}$ that a system can invoke in a specific process:

\begin{equation}
\Xi = \frac{S_{\mathrm{process}}}{k_S \hbar}
\label{eq:5.1}
\end{equation}

where $\hbar$ is the reduced Planck constant, and $k_S$ is a dimensionless constant of order $1$, reflecting the efficiency of converting abstract logical operations into physical implementations.
Second, the total evolution time of the system, $t_{\mathrm{process}}$, is also an obvious physical resource.
In many quantum systems, the complexity of their quantum states typically grows with time\cite{stanford2014complexity}, implying that the complexity a system can achieve within a given evolution time is finite.
Therefore, we can similarly link the complexity budget $\Xi$ with the total available evolution time $t_{\mathrm{process}}$:

\begin{equation}
\Xi = \frac{t_{\mathrm{process}}}{\alpha_t(T_\Xi, ...)}
\label{eq:5.2}
\end{equation}

where $\alpha_t$ is a characteristic coefficient with dimensions of time, possibly depending on the system's effective temperature $T_\Xi$ and other parameters.
Based on the fundamental assumption of the CWT framework—that the complexity $C^*$ of a quantum state (or implemented unitary transformation) actually achievable by the system cannot exceed the current complexity budget $\Xi$, i.e., $C^* \le \Xi$—and combining this with Eq. \eqref{eq:5.1}, we immediately get $C^* \le S_{\mathrm{process}} / (k_S \hbar)$. Rearranging this, we can derive the Action Constraint Principle:

\begin{equation}
S_{\mathrm{process}} \geq k_S C^* \hbar
\label{eq:5.3}
\end{equation}

This principle profoundly reveals a fundamental physical cost of quantum state complexity generation: to generate a quantum state with computational complexity $C^*$, a physical system must consume at least $k_S C^* \hbar$ of action.
This means that each unit of quantum circuit complexity corresponds to an action "price tag" of at least $k_S \hbar$, setting a fundamental cost lower bound for any information processing or state preparation process, originating from the quantization of action.
This principle also sets an upper limit on the "action-to-complexity conversion efficiency" for any physically implemented computing device or natural evolutionary process, where the value of $k_S$ (the closer to $1$, the higher the efficiency) reflects this efficiency.
This can be seen as an extension of the principle of least action to the domains of information and computation, emphasizing that not only the evolution path but also the intrinsic complexity of the target state imposes constraints on the action.
Notably, this is consistent with observations in some holographic complexity conjectures where $k_S$ can correspond to $\pi$, suggesting that extreme gravitational systems like black holes might consume action to encode complexity at near-optimal theoretical efficiency.
Similarly, combining $C^* \le \Xi$ with Eq. \eqref{eq:5.2}, $\Xi = t_{\mathrm{process}} / \alpha_t(T_\Xi, ...)$, we get $t_{\mathrm{process}} \ge \alpha_t(T_\Xi, ...) C^*$.
To further determine the specific form of $\alpha_t$, we need to draw upon the dynamical predictions of the CWT framework.
As we will discuss in detail in Chapter \ref{cex.6}, the CWT framework can derive a core complexity-temperature growth rate bound: $\frac{\mathrm{d}C^*}{\mathrm{d}t} \le \frac{2\pi k_B}{\hbar} T_\Xi$ (we temporarily pre-cite this conclusion; its detailed derivation is in Chapter \eqref{cex.6.1}).
For a process where complexity grows from $0$ to $C^*$ at a constant effective temperature $T_\Xi$, integrating the above rate bound gives the minimum time $t_{\mathrm{min}}$ required to reach complexity $C^*$, satisfying: $\int_0^{C^*} \mathrm{d}C' \le \int_0^{t_{\mathrm{min}}} \frac{2\pi k_B T_\Xi}{\hbar} \mathrm{d}t'$, i.e., $C^* \le \frac{2\pi k_B T_\Xi}{\hbar} t_{\mathrm{min}}$. Therefore, $t_{\mathrm{min}} \ge \frac{C^* \hbar}{2\pi k_B T_\Xi}$.
Comparing this minimum time $t_{\mathrm{min}}$ with $\alpha_t(T_\Xi, ...) C^*$ (considering that $\alpha_t C^*$ is the lower bound for the time required by the process), we can identify the form of $\alpha_t^{\mathrm{min}}(T_\Xi)$ in the optimal case as $\alpha_t^{\mathrm{min}}(T_\Xi) = \frac{\hbar}{2\pi k_B T_\Xi}$.
From this, we obtain the Time Constraint Principle:
\begin{equation}
t_{\mathrm{process}} \ge \frac{C^* \hbar}{2\pi k_B T_\Xi}
\label{eq:5.4}
\end{equation}
This principle reveals an inverse relationship between the time required for complexity generation and the system's effective temperature $T_\Xi$: the higher the effective temperature, the faster the allowed rate of complexity generation by the system, thus allowing higher complexity to be achieved in the same amount of time, or requiring less time to achieve the same complexity.
In the limit $T_\Xi \to 0$, the time required to generate any non-zero complexity will tend to infinity, meaning the system can hardly spontaneously increase its complexity at low temperatures, which aligns with physical intuition (low-temperature systems tend towards simple ground states).
It is noteworthy that this time constraint principle has a profound consistency in form and physical meaning with known quantum speed limits (such as the Margolus-Levitin limit $\Delta t \ge \pi\hbar / (2\langle E \rangle)$ \cite{margolus1998}).
If we roughly consider $2\pi k_B T_\Xi$ as the effective average energy scale used by the system to drive complexity evolution (relative to some baseline), then Eq. \eqref{eq:5.4} embodies the inverse relationship between evolution time and this energy scale.
The aforementioned Action Constraint Principle (Eq. \eqref{eq:5.3}) and Time Constraint Principle (Eq. \eqref{eq:5.4}) set fundamental lower bounds on the two basic physical resources—action $S_{\mathrm{process}}$ and evolution time $t_{\mathrm{process}}$—required for a quantum system to generate a specific computational complexity $C^*$.
The derivation details of these principles, including discussions on the origin of the $O(1)$ constant factors (such as $k_S$ and $2\pi$) and comparative analyses with existing physical theories (such as quantum speed limits, Nielsen's geometric complexity, MSS chaos bound, etc.), are elaborated in detail in Appendix \ref{D.1}.
These fundamental constraints collectively form the core theoretical basis for understanding and designing the resource efficiency of physical information processing.

\subsection{Comprehensive Resource Feasible Region and Optimal Resource Allocation}
\label{cex.5.2}
The Action Constraint Principle (Eq. \eqref{eq:5.3}) and the Time Constraint Principle (Eq. \eqref{eq:5.4}) jointly define the upper limit on the maximum quantum circuit complexity $C^*$ that a physical process can generate, given the physical resources (total available action $S_{\mathrm{process}}$ and total available evolution time $t_{\mathrm{process}}$, as well as the system's effective temperature $T_\Xi$).
Specifically, $C^*$ must simultaneously satisfy:
$C^* \le \frac{S_{\mathrm{process}}}{k_S \hbar}$
$C^* \le \frac{2\pi k_B T_\Xi}{\hbar} t_{\mathrm{process}}$
Therefore, the maximum achievable complexity of the system, $C^*_{\mathrm{max}}$, is limited by the smaller of these two:

\begin{equation}
C^*_{\mathrm{max}} \le \min \left[ \frac{S_{\mathrm{process}}}{k_S \hbar} , \frac{2\pi k_B T_\Xi}{\hbar} t_{\mathrm{process}} \right]
\label{eq:5.5}
\end{equation}

This inequality delineates a feasible region for the achievable complexity $C^*$ on a "resource plane" with $S_{\mathrm{process}}$ as the horizontal axis and $t_{\mathrm{process}}$ as the vertical axis.
For a fixed target complexity $C^*$ and a fixed system effective temperature $T_\Xi$, Eq. \eqref{eq:5.3} and Eq. \eqref{eq:5.4} respectively give the minimum requirements for $S_{\mathrm{process}}$ and $t_{\mathrm{process}}$:
$S_{\mathrm{process}} \ge k_S C^* \hbar$
$t_{\mathrm{process}} \ge \frac{C^* \hbar}{2\pi k_B T_\Xi}$
This means that on the $(S_{\mathrm{process}}, t_{\mathrm{process}})$ plane, only points located in the upper-right region defined by these two lines (parallel to the vertical and horizontal axes, respectively) represent effective resource combinations capable of generating the target complexity $C^*$.
More interestingly, we can explore how to optimally allocate action and time resources to maximize generatable complexity when the total resource input (e.g., some generalized "cost" function, possibly a weighted combination of action and time) is fixed, or how to minimize resource consumption when the target complexity is fixed.
From Eq. \eqref{eq:5.5}, it can be seen that when the two constraint terms are equal, i.e., $S_{\mathrm{process}} / (k_S \hbar) = (2\pi k_B T_\Xi / \hbar) t_{\mathrm{process}}$, or $S_{\mathrm{process}} / t_{\mathrm{process}} = k_S (2\pi k_B T_\Xi)$, the limitations imposed by action resources and time resources on complexity reach a certain "balance."
At this point (or along this ray), further increasing a single resource (without proportionally increasing the other) begins to yield diminishing marginal benefits for increasing $C^*_{\mathrm{max}}$, as the bottleneck will be determined by the other resource.
This provides theoretical guidance for understanding and designing resource optimization strategies in physical information processing.
For example, in quantum computation, if the total available action (possibly related to total energy consumption and operational precision) and total computation time are both limited, then at a given system operating temperature, there exists an optimal computational path or operational mode that balances the utilization of action and time, thereby achieving the highest possible computational complexity under resource constraints.

\section{Dynamical Predictions, Cross-Disciplinary Applications, and Experiments of CWT}
\label{cex.6}
Complexity-Windowed Thermodynamics (CWT) not only provides a new regularization mechanism for understanding thermodynamic singularities near equilibrium (Chapter \eqref{cex.4}), nor does it merely reveal the static constraint relationship between complexity generation and fundamental physical resources (Chapter \eqref{cex.5}); it also imposes profound limitations on the non-equilibrium dynamical behavior of quantum systems and offers a unified perspective for understanding a range of complex phenomena across various fields.
This chapter aims to explore the core dynamical predictions of the CWT framework—particularly the bound between the rate of complexity growth and effective temperature—and to demonstrate its broad explanatory power by analyzing specific application cases in fields such as condensed matter physics, quantum gravitational information, and computational science.
Finally, we will look ahead to possible avenues for experimentally probing and verifying CWT's key concepts and predictions, the challenges involved, and future development directions.\label{cex.6.1}

\subsection{Dynamical Bound on the Rate of Complexity Growth} 
The Complexity-Windowed Thermodynamics (CWT) framework not only provides a new perspective for understanding the static thermodynamic properties of systems but also imposes profound limitations on their non-equilibrium dynamical behavior, particularly the evolution of quantum state complexity $C^*(t)$.
This section aims to reveal the core dynamical bound inherent in the CWT framework that constrains the growth rate of a system's quantum circuit complexity $C^*(t)$, and to elucidate its profound connection with the system's effective thermodynamic state.
To investigate this bound from a more fundamental level, we first resort to CWT's extended first law of thermodynamics (see Eq. \eqref{eq:3.5} and Appendix \ref{E.1}) and the universal quantum time-energy uncertainty principle.
The growth of quantum circuit complexity $C^*$ can be viewed as an ordered generation process of informational structure.
This process necessarily involves the exchange or transformation of energy.
Within the CWT framework, if we approximate this complexification process over short timescales as an isentropic process where the observer's "entropy of ignorance" $S_\Xi$ remains essentially constant (i.e., $\mathrm{d}S_\Xi \approx 0$), then according to the extended first law, the change in the system's internal energy is primarily used to support the increase in the complexity budget $\Xi$.
Specifically, if an increment $\mathrm{d}\Xi(t)$ in the complexity budget corresponds to an increment $\mathrm{d}C^*(t)$ in the achievable complexity $C^*$, then the minimum energy $\mathrm{d}E_{\mathrm{min}}$ required to drive this increment is given by the complexity generation potential $\Pi_\Xi$, i.e., $\mathrm{d}E_{\mathrm{min}} = \Pi_\Xi(t)\mathrm{d}C^*$.
On the other hand, any quantum process consuming energy $\Delta E$ is subject to a universal constraint on its minimum duration $\Delta t$ by the quantum time-energy uncertainty principle, such as the Mandelstam-Tamm\cite{mandelstam1945} bound $\Delta E\Delta t \gtrsim \hbar/2$.
Combining these two fundamental principles, we can directly derive a universal upper bound on the system's complexity growth rate.

\textbf{Proposition 6.1 (Fundamental Bound on the Rate of Complexity Growth):}
For a quantum system undergoing unitary evolution, the growth rate of its minimum quantum circuit complexity $C^*(t)$, $\mathrm{d}C^*(t)/\mathrm{d}t$, is bounded by its instantaneous complexity generation potential $\Pi_\Xi(t)$:

\begin{equation}
\frac{\mathrm{d}C^*(t)}{\mathrm{d}t} \le \frac{2\Pi_\Xi(t)}{\hbar}
\label{eq:6.1.1}
\end{equation}

This bound \eqref{eq:6.1.1} directly stems from CWT's extended first law and the quantum time-energy uncertainty principle; its detailed derivation and discussion of related assumptions are provided in Appendix \ref{D.1}.
This bound \eqref{eq:6.1.1} highlights the central role of the complexity generation potential $\Pi_\Xi(t)$ as the direct "energy cost factor" driving complexity growth, meaning the rate at which a system generates new complexity cannot exceed the maximum "energy utilization efficiency" (limited by $2/\hbar$) allowed by quantum mechanics when $\Pi_\Xi(t)$ is the unit energy cost.
The complexity generation potential $\Pi_\Xi(t)$ itself is a fundamental intensive quantity defined within the CWT framework, related to the effective temperature $T_\Xi(t)$ and the complexity-windowed entropy $S_\Xi(t)$ via $\Pi_\Xi(t) = -T_\Xi(t)(\partial S_\Xi(t)/\partial \Xi(t))_E$.
To connect the bound \eqref{eq:6.1.1} with the more intuitively understandable effective temperature $T_\Xi(t)$ and to compare with existing theoretical results in the field of quantum chaos, we can examine the approximate relationship between $\Pi_\Xi(t)$ and $T_\Xi(t)$.
In many physical situations, especially in parameter regimes where changes in the complexity budget $\Xi$ have a significant impact on the system's "entropy of ignorance" $S_\Xi$, an approximate proportional relationship between $\Pi_\Xi(t)$ and $k_BT_\Xi(t)$ can be expected, i.e., $\Pi_\Xi(t) \approx \alpha k_BT_\Xi(t)$, where $\alpha$ is an $O(1)$ dimensionless factor (its possible origins and typical values are discussed in detail in Appendix \ref{D.1}).
If we adopt a specific factor $\alpha = \pi$, corresponding to the form of quantum chaos bounds, i.e., assume $\Pi_\Xi(t) \approx \pi k_BT_\Xi(t)$, then Proposition 6.1 can be approximately expressed as a bound directly related to the effective temperature:

\textbf{Corollary 6.1.1 (Approximate Temperature Bound on the Rate of Complexity Growth):}
Under the condition that the complexity generation potential $\Pi_\Xi(t)$ and the effective temperature $T_\Xi(t)$ satisfy the approximate relationship $\Pi_\Xi(t) \approx \pi k_BT_\Xi(t)$, the growth rate of complexity $C^*(t)$ is approximately bounded by:
\begin{equation}
\frac{\mathrm{d}C^*(t)}{\mathrm{d}t} \le \frac{2\pi k_B}{\hbar} T_\Xi(t)
\label{eq:6.1}
\end{equation}
In natural units ($k_B = \hbar = 1$), this approximate bound \eqref{eq:6.1.1} simplifies to $\mathrm{d}C^*/\mathrm{d}t \lesssim 2\pi T_\Xi(t)$.
The physical meaning of this approximate bound \eqref{eq:6.1} is that it sets an upper limit, determined by its effective thermodynamic temperature, on the maximum speed at which a quantum system can generate complexity under unitary dynamics.
Recalling $T_\Xi^{-1} = (\partial S_\Xi/\partial E)_\Xi$ (Eq. \eqref{eq:3.1b}), $1/T_\Xi$ measures the density of "accessible" states (i.e., effective degrees of freedom distinguishable by algebra $A(\Xi)$, quantified by $S_\Xi$) in the system's energy space at a fixed complexity budget $\Xi$.
Therefore, $T_\Xi$ can be understood as the characteristic energy scale allocable per effective degree of freedom, under complexity constraints, that can be used to drive the system to explore more complex (i.e., computationally harder to reach) regions of Hilbert space.
A higher effective temperature $T_\Xi$ implies that the system possesses a larger "energy budget" or stronger "thermal driving force" to overcome the "energy barriers" (where "energy barrier" is generalized here and may relate to the dynamical paths required to traverse complex state space) needed to generate complex structures.
The form of the bound \eqref{eq:6.1} is entirely consistent with the universal quantum chaos bound $\lambda_L \le 2\pi k_B T/\hbar$ proposed by Maldacena, Shenker, and Stanford (MSS) (where $\lambda_L$ is the Lyapunov exponent), \cite{maldacena2016bound}after replacing the classical thermodynamic temperature $T$ with CWT's effective temperature $T_\Xi$. 
Thus, CWT not only provides theoretical support for the reasonableness of this replacement (through the smoothness of $S_\Xi$ and the well-definedness of $T_\Xi$) but also, through the more fundamental $\Pi_\Xi$ bound \eqref{eq:6.1.1} and its connection to the approximate bound \eqref{eq:6.1}, reveals the deep underlying energy-complexity-time constraint mechanism.
The CWT framework further points out that when the system's complexity budget $\Xi$ becomes a significant constraint, causing the effective temperature $T_\Xi$ to deviate from the classical temperature $T$ (e.g., in resource-constrained or limited-observation-capability scenarios), adopting the bound \eqref{eq:6.1} related to $T_\Xi$, or even the more fundamental $\Pi_\Xi$ bound \eqref{eq:6.1.1}, may yield more accurate or operationally meaningful dynamical predictions.
These predictions not only provide new theoretical tools for understanding the dynamical behavior of complex systems but also offer concrete, operational avenues for experimentally testing CWT theory (especially on controllable quantum simulation platforms, such as NISQ devices).
In natural units ($k_B = \hbar = 1$), this bound simplifies to $\mathrm{d}C^*/\mathrm{d}t \le 2\pi T_\Xi(t)$.
The physical meaning of this bound is that it sets an upper limit, determined by its effective thermodynamic temperature, on the maximum speed at which a quantum system can generate complexity under unitary dynamics.
Recalling $T_\Xi^{-1} = (\partial S_\Xi/\partial E)_\Xi$, $1/T_\Xi$ measures the density of "accessible" states (i.e., effective degrees of freedom distinguishable by algebra $A(\Xi)$, quantified by $S_\Xi$) in the system's energy space at a fixed complexity budget $\Xi$.
Therefore, $T_\Xi$ can be understood as the characteristic energy scale allocable per effective degree of freedom, under complexity constraints, that can be used to drive the system to explore more complex (i.e., computationally harder to reach) regions of Hilbert space.
A higher effective temperature $T_\Xi$ implies that the system possesses a larger "energy budget" or stronger "thermal driving force" to overcome the "energy barriers" (where "energy barrier" is generalized here and may relate to the dynamical paths required to traverse complex state space) needed to generate complex structures.
This bound is deeply rooted in the fundamental limits of quantum information processing speed and is conceptually highly consistent with bounds on the rate of information scrambling in quantum chaos theory (such as the famous MSS bound $\lambda_L \le 2\pi k_B T/\hbar$, where $\lambda_L$ is the Lyapunov exponent).
In chaotic systems, the growth of complexity usually occurs in synchrony with the rapid scrambling of quantum information and the swift expansion of entanglement.
The CWT framework further points out that when the system's complexity budget $\Xi$ becomes a significant constraint, causing the effective temperature $T_\Xi$ to deviate from the classical temperature $T$ (e.g., in resource-constrained or limited-observation-capability scenarios), replacing $T$ with $T_\Xi$ may yield more accurate or operationally meaningful dynamical bounds.
This prediction not only provides new theoretical tools for understanding the dynamical behavior of complex systems but also offers concrete, operational avenues for experimentally testing CWT theory (especially on controllable quantum simulation platforms, such as NISQ devices).
This complexity-temperature growth rate bound is a core dynamical prediction of the CWT framework.
Its detailed theoretical derivation is primarily based on the universal upper bound for the rate of information scrambling in quantum chaotic systems (the MSS bound), and is generalized by replacing the classical thermodynamic temperature $T$ with CWT's effective temperature $T_\Xi$.
For a detailed discussion of this derivation path, including the origin of constant factors such as $2\pi$, and a comparative analysis with other related physical bounds like Lloyd's operational rate, please refer to Appendix \ref{D.1}.

\subsection{Examples of Cross-Disciplinary Applications of the CWT Framework}
\label{cex.6.2}
The core idea of the CWT framework—that physical resources (manifested as complexity budget $\Xi$) limit the range of states accessible to the system or observables resolvable by an observer, thereby affecting its macroscopic behavior—has broad applicability.
Below, we demonstrate its explanatory potential through typical case studies from different fields.

\subsubsection{CWT Interpretation of Critical Behavior in the Ising Model}
\label{cex.6.2.1}
The two-dimensional Ising model is a paradigmatic system in statistical physics for studying phase transitions and critical phenomena\cite{onsager1944}.
It is well known that this model undergoes a second-order phase transition at a critical temperature $T_c$, where its specific heat $C_V$ exhibits a logarithmic divergence, and the correlation length $\xi_{\mathrm{corr}}$ diverges\cite{yeomans1992}.
However, in any finite-size system of size $L$ or in numerical calculations (such as Monte Carlo) with finite simulation time $t_{\mathrm{sim}}$, the observed specific heat only manifests as a finite, rounded peak\cite{landau_binder2005}.
The CWT framework provides a complementary, complexity-based explanation for this.
As discussed in Chapter \ref{cex.4}, the long-range correlations and fluctuation modes spanning all scales near a critical point typically require extremely high complexity $C^*$ to generate and detect; physically, $C^*$ is expected to grow with $\xi_{\mathrm{corr}}$\cite{orus2014}.
Any finite complexity budget $\Xi$ (whose physical origin might correspond to system size $L$, e.g., $\Xi \sim f(L)$; simulation time $t_{\mathrm{sim}}$; or bond dimension $D$ used in tensor network simulations, e.g., $\Xi \sim \mathrm{poly}(D)$) will effectively truncate the contribution of these high-complexity modes by limiting the dimension of the accessible observable algebra $A(\Xi)$\cite{eisert2010}.
Since the complexity-windowed entropy $S_\Xi(E, \Xi)$ is smooth with respect to energy $E$ and its second derivative is bounded, the effective specific heat $C_{V,\Xi}$ calculated by CWT must be finite.
The divergent peak in classical theory is replaced by a finite peak whose height and shape both depend on $\Xi$.
More interestingly, if a quantitative relationship between $\Xi$ and physical parameters (like $L$ or $D$) can be established (e.g., assuming $\Xi \sim \ln L$), then the specific heat peak behavior $C_{V,\Xi} (\Xi)$ predicted by CWT might be comparable to the $C_V(L)$ predicted by finite-size scaling theory (e.g., $C_V(L) \sim A \ln L + B$)\cite{cardy1988}, thereby providing a way to determine the relationship between the system's effective complexity budget and physical parameters through experiments or numerical simulations.
In summary, CWT attributes the smoothing of the critical peak to more fundamental complexity/action budget limitations, offering a unified perspective for understanding critical phenomena under finite resources.
\subsubsection{A Case Study: Exploring the Correlation Between Quantum Complexity and Macroscopic Properties via the CWT Framework}
\label{sec:case_study_gfactor}
The recent observation of unusual magnetothermal phenomena in the cutting-edge condensed matter material Sr\textsubscript{3}CuIrO\textsubscript{6} presents a unique opportunity for the application of Complexity-Windowed Thermodynamics (CWT). This allows us to demonstrate how the CWT theoretical framework functions as a powerful hermeneutic tool to interpret complex quantum many-body behaviors and reveal profound connections between macroscopic observables and the system's intrinsic computational complexity.
Sr\textsubscript{3}CuIrO\textsubscript{6} is a one-dimensional spin chain with alternating $g$-factors, exhibiting a unique Hamiltonian structure that arises from the complex interactions between Cu\textsuperscript{2+} ($g \approx +2$) and Ir\textsuperscript{4+} ($g \approx -3$) ions\citep{yin2013ferromagnetic}. In our analysis, the external magnetic field $H$ is treated as the principal control parameter driving the phase transitions, while the temperature $T$ primarily determines the transition width $\Delta T$, over which thermal fluctuations smooth the sharpness of the transition.
This engineered intrinsic complexity enables the system to stabilize an unconventional macroscopic quantum-ordered ground state known as the "Half-Fire, Half-Ice" (HFHI) phase\citep{yin2024spin}. Notably, the system exhibits an exceptionally sharp, nearly first-order phase transition near a critical point that is finely tunable by the magnetic field. This phenomenon is theoretically intriguing, as it appears to circumvent Peierls' classical argument that one-dimensional systems with short-range interactions cannot support long-range order at any finite temperature\citep{peierls1979surprises}.(It is worth emphasizing that recent research\citep{yin2024phase_switch} has revealed that the sharp phase transition observed at finite temperatures is governed not by the strict ground state, but by a nearly degenerate mirror excited state.) 
   \footnote{This analysis is confined to the two-state subspace $D = \{\text{HFHI}, \text{HIHF}\}$ (the ground state and its mirror image). Their competition drives the phase transition, governed by the tunneling gap $\Delta E \equiv E_{\text{HIHF}} - E_{\text{HFHI}} \sim \exp(-\kappa N)$, which manifests as the finite-temperature width $\Delta T(N)$~\citep{yin2024phase_switch}. The terms "HFHI-related states" and "tunneling" refer specifically to dynamics within this $\Delta E$-described subspace.}
The ability of the Sr\textsubscript{3}CuIrO\textsubscript{6} system to defy this classical restriction points to a fundamental alteration in the nature of its low-energy excitations, stemming from its highly nontrivial Hamiltonian. This suggests that the minimum quantum circuit complexity, denoted by $C^*$, required to stabilize the HFHI phase is expected to grow dramatically—and nonlinearly—with the system size $N$.From the perspective of Complexity-Windowed Thermodynamics (CWT), the origin of this anomalous stability lies in the system’s ability to access a region of Hilbert space characterized by an extremely high “complexity budget,” $\Xi$. This region remains dynamically inaccessible to simpler, conventional models. The large value of $\Xi$ thus provides a quantitative measure of the information-theoretic resources necessary to sustain such a highly ordered macroscopic quantum state.
The central challenge now becomes understanding the nature and magnitude of this complexity cost. We posit that maintaining the long-range quantum order characteristic of the HFHI state—which must span the entire chain—is fundamentally a problem of macroscopic quantum tunneling. To preserve global coherence, the system must collectively tunnel between two degenerate ground states. This process is mediated by virtual domain wall–anti-domain wall pairs that span the entire system length $N$.
As rigorously established in the theory of quantum phase transitions, the energy splitting $\Delta E$ induced by such large-scale tunneling events is expected to decay exponentially with system size~\citep{sachdev1999quantum}. This insight constitutes a foundational element of our analysis.
Within the CWT framework, this zero-temperature tunneling gap $\Delta E(N)$ becomes regularized at finite temperatures, manifesting as the observable transition width $\Delta T$. This correspondence motivates the formulation of a central Working Hypothesis (WH), grounded in first principles of quantum tunneling:(WH-1): The width of the phase transition, $\Delta T$, should follow the same scaling law as the tunneling energy gap, decaying exponentially with system size N:
\begin{equation} \label{eq:wh1_delta_t_frag} 
    \Delta T(N) \propto \exp(-\kappa N), \quad \text{where } \kappa > 0. 
\end{equation}

This hypothesis, whose detailed physical motivation and theoretical underpinnings will be thoroughly explored in Appendix \ref{H1}, provides us with a solid bridge. CWT theory offers a universal singularity-smoothing law (Chapter \ref{cex.4}) which predicts that the phase transition width $\Delta T$ is inversely proportional to the complexity budget $\Xi$. By combining this universal law, $\Delta T \sim A/\Xi$ (where A is a non-universal constant), with our physically-grounded hypothesis (WH-1), we can derive a Predictive Hypothesis (PH) regarding the complexity budget itself:(PH-1): The complexity budget required to stabilize the "Half-Fire, Half-Ice" state must grow exponentially with system size:
\begin{equation} \label{eq:ph1_xi_frag} 
    \Xi(N) \propto \exp(\kappa N). 
\end{equation}
This is a non-trivial, logically self-consistent inference derived from the confluence of CWT's universal principles and the specific physics of macroscopic quantum coherence. It reveals the immense information-theoretic cost associated with maintaining such an exotic quantum order. It must be emphasized that if future experiments or numerical simulations were to reveal a different scaling behavior for $\Delta T(N)$ (e.g., a power law), this would not undermine the CWT framework itself, but would merely necessitate a recalibration of the scaling law for $\Xi(N)$ based on the new empirical data. Furthermore, this case study provides a concrete physical scenario for the "information processing work" term ($\Pi_\Xi d\Xi$) in our extended first law of thermodynamics (Equation \eqref{eq:3.5}). In this context, the external magnetic field H acts as an effective "knob" for tuning the complexity budget $\Xi(H)$. Consequently, the reversible work done by the magnetic field in driving the system between the two mirror ground states, $\int M dH$, can be endowed with a clear information-theoretic interpretation within the CWT framework—namely, the energetic cost of reconfiguring the system's complex informational structure, the "information work" $W_{\text{info}}$:
\begin{equation} \label{eq:winfo_frag} 
    W_{\text{info}} = \int_{H_1}^{H_2} \Pi_{\Xi}(H)\frac{d\Xi}{dH} dH.
\end{equation}
Crucially, this information-theoretic perspective leads to a subtle yet profound theoretical result concerning the behavior of the complexity-generating potential, $\Pi_\Xi(H)$, whose physical picture resonates deeply with the "phase switch" mechanism recently proposed in related models \citep{yin2024phase_switch}. At the critical point ($H_c$), the system is most susceptible to external perturbations, and the two ground states are in their most intense competition. Therefore, it is anticipated that if one were to forcibly impose additional structural constraints or enhance observational resolution (i.e., increase $\Xi$) at this most unstable point, the system would have to expend maximal energy. Consequently, CWT predicts that $\Pi_\Xi(H)$ should be maximized at the critical field. More precisely, CWT elevates this mechanistic intuition to the level of a universal thermodynamic response function ($\Pi_\Xi$) and predicts a key observable signature: its derivative $(\partial\Pi_\Xi/\partial H)_\Xi$ must undergo a sign flip at the critical point, thereby causing $\Pi_\Xi(H)$ to exhibit a cusp-like maximum at the critical field $H_c$. This refined prediction offers a clear, testable signature for sensitive magnetothermal effect measurements and provides a novel diagnostic tool for probing complexity-driven phase transitions. For a rigorous mathematical derivation of this $\Pi_\Xi$ peak behavior, see Appendix \ref{G1}.For clarity, the core theoretical insights and predictions of the CWT framework for the Sr$_3$CuIrO$_6$ system are summarized in the table below.
\renewcommand{\arraystretch}{1.4} 
\begin{table}[htbp]
\centering
\caption{\normalsize Summary of Core Predictions of the CWT Framework for the Sr$_3$CuIrO$_6$ System}
\vspace{0.8em} 
\label{tab:cwt_predictions_sr3cuiro6}
\begin{tabularx}{\textwidth}{>{\raggedright\arraybackslash}p{3cm}
                                    >{\raggedright\arraybackslash}X
                                    >{\raggedright\arraybackslash}X
                                    >{\raggedright\arraybackslash}X}
\toprule
\textbf{Physical Quantity} & \textbf{Predicted Behavior / Scaling Law} & \textbf{CWT Theoretical Origin / Derivation} & \textbf{Potential Experimental Probes} \\
\midrule
Phase Transition Width ($\Delta T$) 
& $\Delta T(N) \propto \exp(-\kappa N)$ 
& Combines CWT's universal smoothing law ($\Delta T \propto 1/\Xi$) with the physical mechanism of macroscopic quantum tunneling ($\Delta E \propto \exp(-\kappa N)$). 
& Measurements of specific heat, magnetic susceptibility, etc., as a function of system size $N$ in custom-designed samples. \\

Complexity Budget ($\Xi$) 
& $\Xi(N) \propto \exp(\kappa N)$ 
& Novel prediction inferred from the scaling law of $\Delta T(N)$. Represents the exponential information-theoretic cost required to maintain global quantum coherence. 
& Indirect probe: Verified through consistency with $\Delta T(N)$ measurement results. Not directly measurable. \\

Complexity Generation Potential ($\Pi_{\Xi}$) 
& Exhibits a cusp-like maximum at the critical magnetic field $H_c$. 
& Stems from the Maxwell relation $(\partial\Pi_{\Xi}/\partial H)_{T_\Xi,\Xi} = -(\partial M/\partial\Xi)_{T_\Xi,H}$ and the discontinuity of magnetization $M$ at a first-order phase transition (see Appendix \ref{G1}). 
& High-sensitivity Magnetocaloric Effect (MCE) measurements; precise measurements of heat capacity and magnetization as a function of magnetic field. \\
\bottomrule
\end{tabularx}
\end{table}
\subsubsection{Black Hole Page Curve}
\label{cex.6.2.2}
The black hole information paradox is a long-standing and crucial puzzle in theoretical physics\cite{harlow2016jerusalem}.
Hawking's semi-classical calculations suggested that when a black hole evaporates via thermal radiation (Hawking radiation), an initial pure state seems to evolve into a final mixed state, implying information loss, which fundamentally conflicts with one of the core pillars of quantum mechanics—the principle of unitarity.
The currently widely accepted view is that for a black hole evaporation process that preserves unitarity, the von Neumann entropy $S_{\mathrm{rad}}$ of the external Hawking radiation should follow the so-called Page curve\cite{page1993average}: in the early stages of evaporation, $S_{\mathrm{rad}}$ increases with the number of radiated particles, but after the Page time $t_P$, roughly half the black hole's lifetime, $S_{\mathrm{rad}}$ begins to decrease, eventually returning to zero after the black hole has completely evaporated, indicating that all information ultimately returns to the Hawking radiation.
However, how to smoothly transition from the seemingly thermodynamic, entropy-increasing early radiation picture to the late-stage, entropy-decreasing picture required by unitarity has been a key challenge for theoretical explanations.
The Complexity-Windowed Thermodynamics (CWT) framework offers a novel explanatory perspective for this, based on the observer's finite and time-evolving computational (or decoding) capability.
We propose that the computational resources available to an external observer at cosmic time $t$ for analyzing or decoding the state of Hawking radiation $\rho_{\mathrm{rad}}(t)$ (e.g., number of available qubits, depth of executable quantum gate operations, physical resources equivalent to operational precision) correspond to a time-evolving "complexity budget" $\Xi(t)$.
A core physical assumption is that extracting the key correlation information from Hawking radiation—information that proves its fine-grained quantum entanglement with early radiation (or, equivalently, with the remaining part inside the black hole) and thereby reveals how information is recovered (i.e., leading to a decrease in the true entropy $S_{\mathrm{rad}}$)—requires performing extremely complex quantum computations or measurement operations.
The "decoding complexity" $C^*_{\mathrm{decode}}$ of such operations is expected to be very high, possibly even exponentially or very high-polynomially related to the black hole's own entropy (or the entropy of its early radiation), as suggested by Harlow-Hayden's arguments on the difficulty of decoding computations\cite{harlow2013quantum}.
Within this CWT framework, the entropy of Hawking radiation that an external observer can actually "measure" or perceive at time $t$ is not its true, fine-grained von Neumann entropy $S_{\mathrm{VN}}(\rho_{\mathrm{rad}}(t))$, but rather the complexity-windowed entropy $S_{\Xi(t)}(\rho_{\mathrm{rad}}(t)) = S_{\mathrm{VN}}(E_{\Xi(t)}(\rho_{\mathrm{rad}}(t)))$.
This $S_{\Xi(t)}$ quantifies the "degree of ignorance" that the observer has about the true state $\rho_{\mathrm{rad}}(t)$ of the Hawking radiation under the current complexity budget $\Xi(t)$, due to their limited resolution capability.
The non-monotonic behavior of the Page curve can be understood as follows:
\begin{enumerate}
    \item Early Evaporation ($t \ll t_P$, and $\Xi(t) \ll C^*_{\mathrm{decode}}$): In this stage, the observer's complexity budget $\Xi(t)$ is far from sufficient to perform the complex operations required for decoding to reveal fine-grained quantum correlations. The role of the conditional expectation map $E_{\Xi(t)}$ is to effectively "average out" or "blur" the purifying correlation information hidden in $\rho_{\mathrm{rad}}(t)$ that requires high complexity to resolve. As a result, the effective state $E_{\Xi(t)}(\rho_{\mathrm{rad}}(t))$ seen by the observer approximates a continuously growing, seemingly thermodynamic mixed state. Therefore, the observer's "degree of ignorance" $S_{\Xi(t)}$ about $\rho_{\mathrm{rad}}(t)$ in this stage will roughly follow the semi-classical thermodynamic entropy prediction, i.e., continuously increasing with the number of radiated particles.
    \item Late Evaporation ($t > t_P$, and $\Xi(t)$ has grown sufficiently to begin effective decoding): As time progresses, the observer's complexity budget $\Xi(t)$ (possibly due to longer available processing time, technological advancements, or more resources provided by cosmic evolution) gradually increases. When $\Xi(t)$ grows sufficiently to begin overcoming (or at least partially overcoming) the immense barrier of decoding complexity $C^*_{\mathrm{decode}}$, the observer starts to become capable of resolving the key quantum correlations that reveal how information returns from the black hole to the radiation. At this point, the effect of the conditional expectation $E_{\Xi(t)}$ on these resolvable correlations is no longer complete averaging; $E_{\Xi(t)}(\rho_{\mathrm{rad}}(t))$ will increasingly approach the true state of Hawking radiation $\rho_{\mathrm{rad}}(t)$, which is gradually purified due to information return. As the observer's resolution capability enhances, their "degree of ignorance" $S_{\Xi(t)}$ about the true state $\rho_{\mathrm{rad}}(t)$ also begins to decrease significantly. This decrease in $S_{\Xi(t)}$ will track (or at least tend towards) the decreasing behavior of the true, fine-grained von Neumann entropy $S_{\mathrm{VN}}(\rho_{\mathrm{rad}}(t))$, thereby naturally reproducing the declining feature of the latter half of the Page curve.
\end{enumerate}
The Page time $t_P$ acquires a new interpretation in this picture: it is no longer merely a point in time determined by the black hole's own evolution, but rather a critical point related to the observer's capability.
$t_P$ roughly corresponds to the moment when the observer's complexity budget $\Xi(t)$ has grown sufficiently to begin effectively resolving and decoding information in the Hawking radiation, thereby causing the behavior of their perceived "entropy of ignorance" $S_{\Xi(t)}$ to transition from an early-stage, approximately thermodynamic increase to a late-stage, information-recovery-reflecting decreasing trend.
That is, $\Xi(t_P)$ reaches an effective threshold related to $C^*_{\mathrm{decode}}$.
Therefore, the CWT framework suggests that information itself is not lost during black hole evaporation (\textbf{unitarity of quantum mechanics is maintained}), and the non-monotonic entropy behavior exhibited by the Page curve is a direct manifestation of how the observer's finite and time-evolving computational capability (quantified by complexity budget $\Xi(t)$) limits the depth and breadth of their information acquisition, thereby directly affecting the dynamical evolution process of the effective entropy $S_{\Xi(t)}$ they can perceive.

\subsubsection{Inspiring Perspectives in Computational Science}
\label{cex.6.2.3}
The core idea of Complexity-Windowed Thermodynamics (CWT)—that physical resources (manifested as complexity budget $\Xi$) limit the state space explorable by a system or the range of information resolvable by an observer—may also offer inspiring conceptual perspectives for understanding the behavior of certain complex systems in the field of computational science.
Taking Large Language Models (LLMs) as an example, such as text generation models based on the Transformer architecture\cite{ashish2017attention}, which are currently receiving much attention, the diversity and creativity they exhibit when generating text are often significantly influenced by the decoding/sampling strategies employed.
In the practical application of LLMs, parameterized strategies such as sampling temperature $T_{\mathrm{sampling}}$, Top-$k$ truncation, or Top-$p$ (Nucleus) sampling are often introduced to control the quality and characteristics of the generated text\cite{holtzman2019curious}.
A widely known empirical phenomenon is that when the sampling strategy tends towards high determinism (e.g., $T_{\mathrm{sampling}} \to 0$, or greedy decoding with $k=1$, or a very small $p$ value), the "output diversity" $H_{\mathrm{out}}$ of the generated text (e.g., measured by the Shannon entropy or uniqueness of the set of generated texts) decreases sharply, and the model may tend to repetitively generate a few highest-probability, structurally relatively monotonous sequences\cite{brown2020language}.
This phenomenon has a certain conceptual resonance with the idea in CWT that limited observer capability leads to a partial perception of the system.
We can attempt to analogize the LLM's internal "ideal generation mechanism" or its complete "knowledge state"—a complex informational entity capable of producing rich, diverse, and high-quality text—to the true microstate $\rho_E$ in CWT.
The actually adopted sampling strategy and its parameters $\theta$ can then be considered as imposing an effective "algorithmic complexity budget" $\Xi(\theta)$.
A strict, highly deterministic sampling strategy (corresponding to a low $\Xi(\theta)$) is like a low-resolution "observation window," greatly limiting the depth and breadth with which we can probe the LLM's full generative potential through its output sequences.
In this situation, we only observe a highly simplified, possibly distorted projection of the model's behavior.
Under this analogy, the complexity-windowed entropy $S_\Xi(\theta)$ in CWT can be conceptualized as the observer's (or user's) "degree of ignorance" about the LLM's "ideal generation mechanism" or its complete knowledge state.
When the sampling strategy is very strict (low $\Xi(\theta)$) and we can only induce a few monotonous output patterns from the model, although the actual output diversity $H_{\mathrm{out}}$ is low, our "degree of ignorance" $S_\Xi(\theta)$ about the model's true intrinsic ability to generate complex and diverse text might actually be very high.
As the sampling strategy is relaxed ($\Xi(\theta)$ increases), the model is allowed to explore a broader output space, exhibiting richer and more complex behavioral patterns.
At this point, our understanding of the model's intrinsic generative capabilities deepens, and the corresponding "degree of ignorance" $S_\Xi(\theta)$ is expected to decrease.
This behavior of $S_\Xi(\theta)$ being monotonically non-increasing with increasing $\Xi(\theta)$ is conceptually consistent with the core theory of CWT (Proposition 2.1(i)).
However, it needs to be emphasized that this is merely an inspiring analogy aimed at revealing the potential universality of CWT ideas.
Strictly mapping LLM's sampling parameters, internal states, output diversity, etc., to $\Xi$, $\rho_E$, $E_\Xi(\rho_E)$, and $S_\Xi$ in the CWT framework and conducting quantitative theoretical analysis is undoubtedly a highly challenging topic requiring future in-depth research.
Nevertheless, the core principle emphasized by CWT—"resource budget constrains the complexity and diversity of explorable states (or resolvable information)"—may provide valuable theoretical inspiration for understanding and designing more effective and controllable complex computational systems (including LLMs).

\subsection{Experimental Probing Prospects and Challenges for CWT}
\label{cex.6.3}
The vitality of a theory lies not only in its ability to explain existing phenomena but also in its capacity to make testable predictions and guide future experimental exploration.
The Complexity-Windowed Thermodynamics (CWT) framework, particularly its core dynamical prediction (the complexity-temperature growth rate bound, Eq. \eqref{eq:6.1}) and resource constraint principles (Chapter \ref{cex.5}), offers new opportunities for experimentally probing the deep connections between complexity generation, information processing, and thermodynamic behavior in quantum systems.
However, putting CWT's concepts and predictions to experimental test also faces considerable challenges.

\subsubsection{Challenges and Strategies for Experimental Measurement}
\label{cex.6.3.1}
To experimentally verify CWT, one first needs to be able to measure, or at least reliably estimate, its core physical quantities, including the time-evolving quantum state complexity $C^*(\sigma(t))$ and its growth rate $\mathrm{d}C^*/\mathrm{d}t$, the system's effective temperature $T_\Xi(t)$, and the complexity generation potential $\Pi_\Xi$.

\begin{enumerate}
 \item \textbf{Measurement of Quantum State Complexity $C^*(\sigma(t))$ and its Growth Rate $\mathrm{d}C^*/\mathrm{d}t$:} Directly and precisely measuring the minimum quantum circuit complexity $C^*$ of a many-body quantum state is extremely difficult, as this is itself a computationally very complex problem. 
 Current experimental strategies mainly rely on indirect methods or estimations:
\begin{enumerate}
   \item  \textbf{State Tomography and Post-Reconstruction Computation:} For small-scale quantum systems, the density matrix $\sigma(t)$ can be reconstructed via quantum state tomography techniques, and then its $C^*$ can be estimated using theoretical calculations or optimization algorithms (such as variational quantum circuit optimization)\cite{cramer2010efficient}.
   However, the resource consumption of state tomography grows exponentially with system size, limiting its applicability.
   
   \item \textbf{Proxy Complexity Based on Observables:}
We seek physical quantities that are easier to measure and behave similarly to $C^*$ (minimum quantum circuit complexity) to serve as proxies. Examples include specific types of multipartite entanglement entropy, Krylov complexity associated with operator spreading \cite{caputa2022geometry}, or specific state features efficiently estimated via ``classical shadows'' techniques \cite{stoichita1997short}. The key lies in establishing a reliable correspondence between these proxies and $C^*$ (particularly in the regime of approximate linear growth). Notably, a more structurally grounded avenue involves adopting Reference-Contingent Complexity (RCC) \cite{liu2025structure}. In this framework, given a reference family $R$, one can define a corresponding ``structured vacuum'' $\sigma_R$ and quantify complexity via the relative entropy distance of the state $\rho$ with respect to $\sigma_R$, thereby obtaining a rigorous lower bound for circuit complexity, denoted as $C_R(\rho)$. In the context of CWT, the reference family $R$ and its structured vacuum $\sigma_R$ can be viewed as the ``structural window'' that an observer can resolve under a given complexity budget $\Xi$. Unlike abstract gate counting, $C_R(\rho)$ is in principle experimentally estimable via quantum hypothesis testing and a class of projective measurement schemes, providing a statistically auditable proxy metric for the system's complexity content.
    
    \item \textbf{Focusing on Growth Rate $\mathrm{d}C^*/\mathrm{d}t$ Rather than Absolute Value:} In many cases, it may not be necessary to know the exact absolute value of $C^*$, but only to estimate its growth rate. 
    For some theoretical models or controllable quantum evolutions, the behavior of $\mathrm{d}C^*/\mathrm{d}t$ can be predicted by analyzing the properties of the Hamiltonian or using numerical simulations, and then compared with other experimentally measurable dynamical features (such as the rate of information scrambling).
\end{enumerate}
\item\textbf{Measurement of Effective Temperature $T_\Xi(t)$:} In non-equilibrium evolution processes, especially within the CWT framework, $T_\Xi(t)$ is an effective temperature that depends on the current complexity budget $\Xi$ (which may also change with time), and its direct measurement is more challenging.
\begin{enumerate}
    \item\textbf{Inverse Calculation Based on Definition:} If $S_\Xi(t)$ and $E(t)$ can be estimated in some way, then in principle, $T_\Xi^{-1} = (\partial S_\Xi/\partial E)_\Xi$ can be used for calculation. 
    However, this relies on a reliable estimation of $S_\Xi$, which itself depends on $A(\Xi)$ and the conditional expectation $E_\Xi$, making direct experimental implementation difficult.
    
     \item \textbf{Finding Observables Related to $T_\Xi$:} Exploring whether there exist some easily measurable physical quantities (e.g., specific energy fluctuations, certain response functions, or the behavior of probe systems) that can directly or indirectly reflect the system's effective temperature $T_\Xi$. 
     This may require theoretical modeling for specific systems.
    
     \item \textbf{Indirect Verification via Dynamical Bounds:} Experimentally, one can systematically measure the saturation value or characteristic growth rate of $\mathrm{d}C^*/\mathrm{d}t$ under different conditions (e.g., by changing system parameters to control its "hotness" or degree of chaos), and then check if it exhibits the linear relationship predicted by Eq. \eqref{eq:6.1} with $T_\Xi$ (or classical temperature $T$ as its approximation) independently estimated or theoretically predicted by other means. 
     This provides a way to indirectly verify the bound and the concept of $T_\Xi$.
    \end{enumerate}
\item\textbf{Measurement of Complexity Generation Potential $\Pi_\Xi(t)$:} The complexity generation potential $\Pi_\Xi$, defined as $\Pi_\Xi = (\partial E/\partial \Xi)_{S_\Xi}$ or equivalently $\Pi_\Xi = -T_\Xi (\partial S_\Xi/\partial \Xi)_E$, is also extremely challenging to measure because it directly involves the partial derivative of the system's energy $E$ or the observer's effective entropy $S_\Xi$ with respect to the abstract complexity budget $\Xi$.
Given the property $\Pi_\Xi\ge0$, experimentally probing its effect mainly involves verifying whether increasing the complexity budget $\Xi$ is indeed accompanied by a corresponding energy cost. 
The following are some possible indirect strategies:

\begin{enumerate}
    \item \textbf{Indirectly Changing $\Xi$ by Controlling Physical Resources and Observing Energy Changes:} The core idea is to first establish an approximate correspondence $\Xi \approx f(P_{\mathrm{ctrl}})$ between an experimentally controllable physical parameter $P_{\mathrm{ctrl}}$ (e.g., total available evolution time $t_{\mathrm{evo}}$, where in the linear complexity growth regime one might assume $\Xi \propto t_{\mathrm{evo}}$; total number of quantum gate operations or total action $S_{\mathrm{proc}}$, where according to Chapter \ref{cex.5}, $\Xi \propto S_{\mathrm{proc}}$; or, in some models, system size $L$ or control precision parameters, etc.) and the abstract complexity budget $\Xi$.
    Subsequently, by precisely changing $P_{\mathrm{ctrl}}$ (thereby indirectly changing $\Delta\Xi$), while attempting to keep the observer's effective entropy $S_\Xi$ approximately constant (which is itself a huge experimental challenge), precisely measure the change in the system's internal energy $E$, $\Delta E$.
    If achievable, one can then estimate $\Pi_\Xi \approx (\Delta E/\Delta\Xi)_{S_\Xi}$. 
    This method places extremely high demands on the system's energy isolation and measurement precision.

    \item \textbf{Indirectly Changing $\Xi$ by Controlling Physical Resources and Observing Effective Entropy Changes (under Energy Constraints):} Another approach is to keep the system's total energy $E$ approximately constant while indirectly changing $\Delta\Xi$ by altering the controllable physical parameter $P_{\mathrm{ctrl}}$.
    At the same time, one needs to find a way to estimate the change in the observer's effective entropy $S_\Xi$, $\Delta S_\Xi$ (this also relies on reliable estimation methods for $S_\Xi$, such as via proxy quantities or model calculations).
    If an independent estimate of $T_\Xi$ can be obtained (e.g., through the $T_\Xi$ measurement strategies discussed in Section 6.3.1), then one can estimate $\Pi_\Xi \approx -T_\Xi (\Delta S_\Xi/\Delta\Xi)_E$. 
    The challenge of this path lies in the simultaneous or correlated measurement/estimation of $S_\Xi$ and $T_\Xi$.

    \item \textbf{Indirect Verification of $\Pi_\Xi$-Related Effects (e.g., via Maxwell Relations):} The generalized Maxwell relations derived in Appendix \ref{E.1} (such as the equation: $-(\partial S_\Xi/\partial \Xi)_{T_\Xi,V} = (\partial\Pi_\Xi/\partial T_\Xi)_{\Xi,V}$) offer a possible way to indirectly probe the behavior of $\Pi_\Xi$.
    For example, if one can experimentally measure the rate of change of $S_\Xi$ with $\Xi$ under isothermal conditions (left-hand side; since $S_\Xi \downarrow \Xi$, this term is $\ge0$), and the rate of change of $\Pi_\Xi$ (or its proxy observable) with $T_\Xi$ at fixed $\Xi$ (right-hand side), then the self-consistency of these theoretical predictions can be tested, thereby indirectly confirming the existence of $\Pi_\Xi$ and its core properties. 
    This requires the development of experimental techniques capable of probing these partial derivatives.
\end{enumerate}
\end{enumerate}    

\subsubsection{Experimental Verification Pathways for Resource Constraint Principles and Dynamical Bounds} 
\label{cex.6.3.2}
Despite the challenges in directly measuring CWT's core quantities, its predicted resource constraint principles and dynamical bounds offer more feasible avenues for experimental verification.
Ideal experimental platforms are those systems with high-precision quantum control capabilities that can simulate complex many-body dynamics, such as: trapped ions\cite{blatt2012quantum}, ultracold atoms in optical lattices\cite{bloch2012quantum}, programmable superconducting quantum processors\cite{kjaergaard2020superconducting}, etc.
\begin{enumerate}
 \item \textbf{Verification of the Complexity-Temperature Growth Rate Bound (Eq. \eqref{eq:6.1}):}

 \textbf{Experimental Protocol:} (a) Initialize the system in a low-complexity state (e.g., a product state). 
 (b) Apply a Hamiltonian $H$ capable of driving the system (possibly into a chaotic regime) for unitary evolution. 
 (c) At different times $t$ during the evolution, estimate $C^*(\sigma(t))$ using the strategies mentioned above (e.g., proxy quantities, state tomography combined with computation, etc.) and calculate its growth rate $\mathrm{d}C^*/\mathrm{d}t$. 
 (d) Concurrently, characterize the system's "hotness" in some way; for example, for systems near equilibrium, its classical temperature $T$ can be used as an approximation for $T_\Xi$; for systems far from equilibrium, $T_\Xi$ might need to be estimated using theoretical models, or indirectly controlled by systematically changing system parameters (such as coupling strength, driving frequency, etc.) to alter its effective energy distribution. 
 (e) Test whether the maximum value of $\mathrm{d}C^*/\mathrm{d}t$ under different conditions is indeed bounded by $(2\pi k_B/\hbar) T_\Xi$.
 
\textbf{Points of Focus:} The key to the experiment lies in reliably estimating $\mathrm{d}C^*/\mathrm{d}t$ and $T_\Xi$ (or their proxies) and systematically varying experimental conditions to explore the validity of the bound.

\item  \textbf{Verification of the Action Constraint Principle} ($S_{\mathrm{process}} \ge k_S C^* \hbar$):

\textbf{Experimental Protocol:} (a) Design a series of quantum operations or evolution paths to prepare quantum states with different target complexities $C^*_{\mathrm{target}}$. 
(b) For each operation, accurately record or estimate the total action $S_{\mathrm{process}}$ consumed (this may be related to evolution time, average energy of the Hamiltonian, precision of control fields, etc.). 
(c) Test whether for all successful preparation processes, $S_{\mathrm{process}} / C^*_{\mathrm{target}}$ is always greater than or equal to some constant close to $\hbar$ (i.e., $k_S \hbar$).

\textbf{Points of Focus:} Requires precise control and measurement of action consumption and the ability to prepare a series of quantum states with tunable complexity.

\item\textbf{Verification of the Time Constraint Principle} ($t_{\mathrm{process}} \ge C^* \hbar / (2\pi k_B T_\Xi)$):

 \textbf{Experimental Protocol:} (a) Prepare quantum states with the same target complexity $C^*_{\mathrm{target}}$ under different effective temperatures $T_\Xi$ (which can be controlled by changing the system's initial energy, coupling strength with the environment, etc.). 
 (b) Measure the minimum time $t_{\mathrm{process}}$ required to reach this target complexity. 
 (c) Test whether $t_{\mathrm{process}}$ is indeed proportional to $1/T_\Xi$ and whether the proportionality constant is consistent with $C^*_{\mathrm{target}} \hbar / (2\pi k_B)$.
 
  \textbf{Points of Focus:} Requires precise control of the system's effective temperature and measurement of the time taken to reach the target complexity.

\end{enumerate}
Despite the challenges, with the rapid development of quantum technologies, especially advancements in Noisy Intermediate-Scale Quantum (NISQ) devices and future fault-tolerant quantum computers, the possibility of experimentally verifying CWT predictions is continually increasing.
For instance, using quantum machine learning techniques to assist with state characterization and complexity estimation\cite{carleo2017solving}, or developing more efficient proxy measures for complexity, could lead to experimental breakthroughs.
Furthermore, by experimentally and systematically studying how the complexity budget $\Xi$ (by controlling physical resources) affects the system's thermodynamic behavior (e.g., by analyzing changes in effective specific heat peaks, shifts in phase transition points, or direct energy exchange effects), it might be possible to provide avenues for indirectly verifying or estimating the complexity generation potential $\Pi_\Xi$ and its related thermodynamic effects.
In conclusion, the CWT framework not only proposes profound theoretical insights but also opens up new avenues for exploration in experimental physics.
Experimental verification of these predictions will greatly advance our understanding of the intrinsic unity between quantum complexity, information processing, and thermodynamics, and may potentially lead to new applications in quantum technology.

\section{Cosmic Complexity and the Extension of CWT}
\label{cex.7}
The Complexity-Windowed Thermodynamics (CWT) framework, by internalizing computational complexity as a fundamental physical constraint, not only offers a new perspective for understanding the thermodynamic behavior and information processing of systems at the laboratory scale but also has the potential to extend its insights to the grand stage of cosmology.
The universe itself can be viewed as a vast, continuously evolving physical information processing system, whose internal structure formation, matter organization, and possible computational limits may be governed by fundamental principles akin to the core ideas of CWT.
This chapter aims to explore the applications and inferences of CWT theory on cosmological scales.
We will first, based on CWT's time constraint principle and by introducing a key "energy-equivalent temperature" hypothesis, derive an upper bound on the total complexity generatable by the universe since the Big Bang.
Subsequently, we will quantitatively estimate this upper bound and compare it with the cosmic horizon entropy based on the holographic principle, leading to the "cosmic complexity saturation" conjecture, while carefully discussing the assumptions underlying this inference and their robustness.
Finally, we will delve into the profound implications of CWT for understanding the universe's information-carrying capacity, the physical basis of the holographic principle, and other frontier issues such as computational cosmology.

\subsection{Upper Bound on the Generatable Complexity of the Universe}
\label{cex.7.1}
The time constraint principle we derived in Chapter \ref{cex.5}, whose differential form is the complexity-temperature growth rate bound (Eq. \eqref{eq:6.1}), provides a theoretical starting point for estimating the upper bound on the total complexity that the universe could have generated throughout its entire evolutionary history.
Applying this rate bound to the evolution of the entire universe, from the Big Bang ($t=0$) to the current age of the universe $t_U$, the total complexity $C_U$ that the universe could have cumulatively generated should satisfy:

\begin{equation}
C_U \le \int_0^{t_U} \frac{2\pi k_B}{\hbar} T_\Xi(t') \mathrm{d}t'
\label{eq:7.1}
\end{equation}

where $T_\Xi(t')$ represents the effective complexity-windowed temperature of the universe at different evolutionary epochs $t'$.
It should be emphasized that the choice of the constant factor (here $2\pi$) in the rate bound may have some theoretical freedom; for example, some derivations might yield $2$ instead of $2\pi$ (which would introduce a factor of $1/\pi$).
However, as subsequent numerical analysis will show, such differences in factor choice do not alter our core conclusion regarding the order of magnitude of $C_U$.
Directly determining $T_\Xi(t')$ for the universe at various epochs is extremely difficult.
To make an operational estimate and to connect with known cosmological computation limits, we introduce a key \textbf{"energy-equivalent temperature" hypothesis}: we assume that, on macroscopic cosmological scales, the "effective hotness" driving system evolution and new structure formation can be approximately characterized by the total energy $E_U(t')$ available in the universe at a given time $t'$ for driving complexity generation, i.e., $k_B T_\Xi(t') \approx E_U(t')$.
This is a core operational assumption, motivated by viewing the entire universe (or the part within its Hubble horizon) as a single "information processor," whose maximum "operational rate" or "complexity generation capability" is determined by its total available energy.
This idea is spiritually consistent with Seth Lloyd's proposed upper limit on the computational capability of the universe, $\text{Operations} \le \frac{2E}{\hbar}$ (if $C^*$ is considered as the number of operations)\cite{goyeneche2015five}, and allows us to rewrite Eq. \eqref{eq:7.1} into a more easily estimable form:

\begin{equation}
C_U \le \frac{2\pi}{\hbar} \int_0^{t_U} E_U(t') \mathrm{d}t'
\label{eq:7.2}
\end{equation}

It must be clearly stated that $E_U(t')$ here refers to the energy in the universe that can participate in information processing or complexity generation.
An important open question is, for example, whether and how the dark energy component of the universe contributes to $E_U(t')$.
In the subsequent zeroth-order approximate estimation, we will use the total mass-energy of the current universe, but it must be acknowledged that this includes an optimistic assumption about the role of dark energy.

\subsection{"Cosmic Complexity Saturation" Conjecture}
\label{cex.7.2}

We now use Eq. \eqref{eq:7.2} to make a quantitative estimate of the upper bound on the total generatable complexity of the universe, $C_U$.
A detailed calculation of the energy-time integral $I_E = \int_{0}^{t_U} E_U(t')\mathrm{d}t'$ in Eq. \eqref{eq:7.2} (based on the $\Lambda$CDM model and considering cosmic expansion, with the specific process detailed in Appendix \ref{F.1}) shows its value to be approximately $I_E = (2.14 \pm 0.03) \times 10^{87}$ J$\cdot$s.
Substituting this integral result into Eq. \eqref{eq:7.2}, and using $2\pi/\hbar \approx 5.9582 \times 10^{34}$ (J$\cdot$s)$^{-1}$, we obtain the upper bound on the generatable complexity of the universe as:

\begin{equation}
C_U^{\mathrm{max}} \approx (1.27 \pm 0.02) \times 10^{122}
\label{eq:7.3}
\end{equation}

This estimated upper bound on the generatable complexity of the universe, $C_U^{\mathrm{max}}$, is approximately $1.3 \times 10^{122}$.
It is noteworthy that this value is comparable in order of magnitude to the maximum entropy (given in dimensionless form $S/k_B$) that the current Hubble horizon of the universe (or the future de Sitter event horizon for an acceleratingly expanding universe) can contain, as estimated from the holographic principle.
For example, the Bekenstein-Hawking entropy $S_H$ of the current Hubble horizon, in its dimensionless form (calculated according to Appendix \ref{F.1}), is approximately $S_H/k_B \approx 2.3 \times 10^{122}$ \cite{bousso2002holographic}.
These two cosmological-scale characteristic quantities, independently derived from vastly different physical ideas (CWT's time constraint and energy-equivalent temperature vs. the holographic principle and horizon geometry), are both of the order $10^{122}$, with their specific values differing by a factor of less than $2$. This strongly suggests a potentially deep intrinsic connection between them.
Based on this, we propose the \textbf{"Cosmic Complexity Saturation" Conjecture}: The total computational complexity that the universe can generate throughout its entire evolutionary history not only has an upper bound determined by its total spacetime and energy resources, but this upper bound is also comparable in order of magnitude to the cosmic horizon entropy bound set by the holographic principle, and the actual evolution of the universe may have already approached or reached this fundamental limit.
Assessing the robustness of this result is crucial.
First, as mentioned earlier, the constant factor ($2\pi$ vs. $2$) in the rate bound Eq. \eqref{eq:7.1} only causes a difference of a factor of $1/\pi$, which does not change the order of magnitude of $C_U$.
Second, regarding the choice of $E_U(t')$, even if we adopt a more conservative estimate, for example, considering only the contribution of matter energy (including dark matter and baryonic matter) and neglecting the contribution of dark energy to "computation," this might reduce $E_0$ by $1$ to $2$ orders of magnitude, thereby decreasing $\log_{10}C_U$ from $122$ to $120$-$121$.
Considering the measurement precision of the age of the universe $t_U$ and cosmological parameters (such as $H_0$, which affects horizon entropy calculations) is typically better than $10\%$, their impact on the final order of magnitude is also relatively small.
Therefore, even under more conservative input parameters and assumptions, the difference between $\log_{10}C_U$ and $\log_{10}(S_H/k_B)$ is not expected to exceed $2$ to $3$, which is much smaller than the order-of-magnitude scale that theoretical models or observational errors can typically span.
Hence, the conclusion that "\textbf{the upper bound on the generatable complexity of the universe is comparable in order of magnitude to the cosmic horizon entropy}" possesses considerable robustness.

\subsection{Discussion and Implications}
\label{cex.7.3}
The CWT framework and its inferences on cosmological scales offer profound implications for our understanding of the universe's information-carrying capacity, the physical origin of the holographic principle, and frontier issues such as computational cosmology:

\begin{enumerate}
    \item \textbf{Physical Reality of Complexity and the Computational Nature of the Universe:} CWT treats computational complexity as a real attribute constrained by physical resources (manifested as complexity budget $\Xi$).
    The existence of an upper bound on cosmic complexity (estimated value $C_U$) suggests that the evolutionary history of the universe and the formation of its internal structures can be viewed as a grand, resource-constrained "computational process."
    The "program length" or "output complexity" achievable by the universe is finite; this picture deepens our understanding of the universe as an information processing system.
    \item \textbf{Dynamical Perspective and Operational Meaning of the Holographic Principle:} The holographic principle is usually formulated as a static bound on a system's maximum entropy (such as the horizon entropy $S_H$).
    CWT, by linking complexity generation with the dynamical evolution of the universe (the energy-time integral $\int E_U(t')\mathrm{d}t'$ in Eq. \eqref{eq:7.2}) and finding its upper bound to be of the same order of magnitude as the holographic entropy bound, may offer a more dynamical and operational perspective on the holographic principle.
    It suggests that horizon entropy not only limits the amount of "statically stored" information but may also indirectly limit the total amount of information (measured by complexity) that can be "generated" and "processed" through dynamical processes.
    In other words, the holographic bound might simultaneously be the universe's "memory limit" and "total CPU computation limit."
    \item \textbf{Possibility of the Universe as an "Efficient" Information Processor:} If the "cosmic complexity saturation" conjecture holds some truth, it might imply that the universe, in its evolution, has in some sense "fairly efficiently" utilized its available spacetime and energy resources to generate complex structures and information, with an efficiency approaching the limit set by fundamental physical laws.
    This resonates with the discussion of "optimal resource allocation" in Chapter \ref{cex.5} and the idea that certain extreme systems (like black holes) might operate near optimal efficiency.
    \item \textbf{Definition of the Timing of "Saturation":} An issue requiring careful consideration is that for a continuously acceleratingly expanding universe (like the current $\Lambda$CDM model), the area of its future event horizon (de Sitter horizon) will tend to a constant, and the corresponding horizon entropy will also saturate.
    However, the complexity integral $\int E_U(t')\mathrm{d}t'$ in Eq. \eqref{eq:7.2} can, in principle, continue to grow with time $t_U$ (although $E_U(t')$ may change due to cosmic expansion).
    Therefore, how to precisely define the timing of "cosmic complexity saturation" and how to reconcile continuous complexity generation with a saturated horizon entropy is a question worthy of in-depth study.
    Perhaps, the effective growth of complexity stops when the average energy density of the universe drops below a certain threshold, or when the free energy available to drive meaningful structure formation is exhausted.
    The estimations in this paper are primarily based on the current age of the universe and the energy of the observable universe; their applicability to very far-future evolution requires further investigation.
\end{enumerate}
In summary, extending the CWT perspective to cosmological scales not only demonstrates the broad application prospects of this theoretical framework but may also touch upon profound questions regarding the nature of the universe, information limits, and the unification of fundamental physical laws.
The upper bound on the generatable complexity of the universe and its remarkable comparability in order of magnitude with holographic entropy open up exciting new directions for future theoretical and observational research, prompting us to rethink the central role of complexity in cosmic evolution and the constitution of physical reality.

\subsection{Maximum Complexity Conjecture: A Universal Dynamical Tendency}
\label{cex.7.4}
The preceding discussion in this chapter, based on the resource constraint principles of Complexity-Windowed Thermodynamics (CWT), derived an upper bound on the total complexity $C_U^{\mathrm{max}}$ that the universe could have cumulatively generated since the Big Bang (estimated to be approximately $1.3 \times 10^{122}$).
It is noteworthy that this upper bound is comparable in order of magnitude to the cosmic horizon entropy $S_H/k_B$ (approximately $2.3 \times 10^{122}$) based on the holographic principle, both being of the order $10^{122}$ and differing by a factor of less than $2$.
This consistency inspired us to propose the "Cosmic Complexity Saturation Conjecture" (CCSC), which posits that the total computational complexity generated by the universe throughout its overall evolution has approached the theoretical limit set by fundamental physical laws and available resources.
If we consider the preliminary success of CCSC—although it still awaits more rigorous testing and refinement of cosmological models—as an "experimental phenomenon" observed on a cosmic scale, a natural question arises: Is this behavior of "approaching the complexity limit" merely a unique attribute of the universe as a specific grand system, or does it reflect a more universal statistical tendency in the dynamical evolution of physical systems?
To explore this deeper question, we hereby propose a more universal \textbf{Maximum Complexity Conjecture} (MCC), whose core assertion is as follows:
\textbf{MCC} — For any physical system constrained by finite physical resources (such as total available action, evolution time, and energy budget, etc.), during its dynamical evolution, there exists a statistical preference for its generated cumulative computational complexity $C^*$ to approach or reach the maximum value $C^*_{\text{max}}$ theoretically allowed under the same resource constraints.
This conjecture attempts to characterize a statistical behavior: among all possible dynamical paths, those that can more fully utilize available resources to construct complex informational structures may, in a sense, be more "representative" or more easily "explored" by the system.
This "preference" can be seen as a measure of "resource utilization efficiency" at an information-theoretic level.
Interestingly, MCC may conceptually form a profound complementary relationship with the famous principle of least action in physics.
The principle of least action, from a "cost-benefit" perspective, selects the "most economical" path followed by dynamical evolution; MCC, on the other hand, from an "output-benefit" perspective, focuses on the "richest" state the system can achieve in terms of informational structure.
Whether these two might originate from a "dual optimality condition" under a deeper, unified variational framework—i.e., the system must evolve with minimal "dynamical cost" while also achieving maximal "informational structure output"—is a highly attractive theoretical question worthy of in-depth exploration, whose rigorous mathematical formulation and proof await future research.
Furthermore, the universality of MCC might be particularly pronounced in open systems far from equilibrium.
In such systems, a continuous flow of free energy (often associated with entropy production) provides the necessary driving force for the formation and maintenance of complex structures.
According to CWT's core dynamical prediction, the growth rate of system complexity is bounded by Eq. \eqref{eq:6.1}.
MCC, then, predicts that as long as external resources (energy, matter flow, etc.) can be continuously supplied, thereby maintaining a non-zero effective temperature $T_\Xi$ and providing a sufficient "complexity budget" $\Xi$, the system's computational complexity has the potential to continuously grow along this rate bound, gradually approaching the $C^*_{\text{max}}$ determined by the currently accumulated resources, until the resource supply changes or environmental conditions undergo drastic alterations.
From this perspective, the continuous growth of quantum state complexity inside black holes (as described by holographic complexity conjectures like CA/CV), the maintenance and development of highly ordered informational organization by living systems through metabolism during evolution, and even the continuous expansion of executable quantum circuit depth by future large-scale quantum computers within their coherence times, can all be considered as potential observational or test scenarios for MCC under different physical scales and resource constraints.
To systematically test and develop the Maximum Complexity Conjecture, future research efforts could focus on the following key directions:

\begin{enumerate}
    \item \textbf{Numerical and Theoretical Verification in Model Systems:} For specific, computable or simulable physical systems, quantitatively compare the deviation $\Delta = 1 - C^*/C^*_{\text{max}}$ between their actually generated complexity $C^*$ and its theoretical upper bound $C^*_{\text{max}}$.
    Candidate systems could include random quantum circuit models, holographic complexity evolution within the AdS/CFT framework (such as Complexity=Action or Complexity=Volume conjectures)\cite{hashimoto2017}, and complex quantum dynamical processes implemented on experimentally controllable quantum processors (such as trapped ion or superconducting qubit platforms)\cite{Esposito2010EntropyCorrelation}.
    Systematically studying the variation of $\Delta$ with system parameters, resource input, and evolution time will provide direct evidentiary support or falsification for MCC.
    \item \textbf{Exploration of a Unified Variational Principle:} Endeavor to construct a generalized variational principle that can simultaneously embody the tendencies of "action minimization" and "complexity maximization"\cite{zinnjustin2002}.
    This may require the development of new mathematical tools or functional forms and an in-depth examination of the precise correspondence between its associated Euler-Lagrange equations and known physical dynamical laws (such as the Schrödinger equation, general relativistic field equations, etc.).
    \item \textbf{Quantitative Coupling of Entropy Production and Complexity Accumulation:} In the context of open systems and non-equilibrium states, conduct in-depth research into the quantitative relationship between entropy production rate, free energy flow, and the system's achievable complexity upper bound $C^*_{\text{max}}$, as well as its complexity growth rate\cite{Esposito2010EntropyCorrelation}.
    This will not only help in understanding how complex structures emerge from disorder and are maintained but may also provide important clues for extending CWT theory to non-equilibrium states.
\end{enumerate}

\section{Summary and Outlook}
\label{cex.8}
This paper has systematically constructed Complexity-Windowed Thermodynamics (CWT), aiming to rectify the theoretical difficulties encountered by classical statistical mechanics due to its neglect of the limitations on quantum state preparation complexity, by elevating "bounded dynamical complexity" to a core postulate.
CWT constrains accessible states with a "complexity budget" $\Xi$ and defines the core complexity-windowed entropy $S_\Xi(E,\Xi)$—an entropy quantifying the observer's "degree of ignorance" about the system under budget $\Xi$, whose key characteristics are being smooth with respect to energy $E$ and monotonically non-increasing with $\Xi$.
This property ensures the well-behaved nature of the effective temperature $T_\Xi$ and the non-negative complexity generation potential $\Pi_\Xi$ ($(\partial E/\partial \Xi)_{S_\Xi}\ge0$, representing the energy cost of increasing $\Xi$), thereby universally softening classical singularities and recovering classical theory as $\Xi\to\infty$.
CWT provides an extended first law $\mathrm{d}E = T_\Xi\mathrm{d}S_\Xi + \Pi_\Xi\mathrm{d}\Xi$ including an "information processing work" term $\Pi_\Xi\mathrm{d}\Xi$, and derives an action constraint ($S_{\mathrm{proc}} \ge k_S C^* \hbar$) and a time constraint ($t_{\mathrm{proc}} \ge \frac{C^* \hbar}{2\pi k_B T_\Xi}$), linking quantum circuit complexity $C^*$ to fundamental physical resources and the effective temperature $T_\Xi$.
These results offer new perspectives for a unified explanation of phenomena such as the smoothing of critical phenomena in condensed matter and black hole information dynamics, and its cosmological inferences further reveal a near order-of-magnitude agreement between the upper bound on the generatable complexity of the universe and the holographic horizon entropy.
Despite the broad prospects CWT has shown, as an emerging theoretical framework, its maturation and refinement still face several core challenges and critical issues requiring in-depth research.
Firstly, the physical correspondence and operability of the complexity budget $\Xi$ is one of the core bottlenecks: although $\Xi$ plays a key role theoretically, how to establish a universal and quantitative mapping relationship between it and specific, experimentally precisely controllable or measurable physical parameters (such as the system's evolution time, total action consumed, available energy, control precision, or system size, etc.) still requires more in-depth theoretical modeling and experimental exploration.
Secondly, the minimum quantum circuit complexity $C^*$, as a theoretical cornerstone, is extremely difficult to calculate theoretically, and its direct experimental measurement is even more challenging, currently relying heavily on theoretical models, approximate calculations, or indirect proxy quantities.
Furthermore, some key assumptions in this theory, such as the "energy-equivalent temperature" hypothesis applied to cosmological inferences, and the universal rigor of replacing classical temperature $T$ with effective temperature $T_\Xi$ in dynamical bounds, still require deeper theoretical foundations or broader indirect experimental evidence for consolidation.
Moreover, CWT is currently primarily constructed within the framework of equilibrium or quasi-equilibrium states; its systematic extension to universal non-equilibrium processes (especially in strongly driven, far-from-equilibrium open quantum systems), and how to construct corresponding non-equilibrium $S_\Xi$, $\Pi_\Xi$, and entropy production theories, remain open and important research directions.
To overcome the aforementioned challenges and promote the further development of CWT theory, future research should focus on the following key directions:

\begin{enumerate}
    \item \textbf{Deepening the Quantitative Mapping between $\Xi$ and Physical Resources:} Focus on developing theoretical models that can precisely associate the abstract complexity budget $\Xi$ with a set of operational physical resource parameters, and explore ways to experimentally calibrate and control effective $\Xi$.
    \item \textbf{Operationalizing the Complexity Budget via RCC:}
    A central goal in the theoretical development of CWT is to map the abstract complexity budget $\Xi$ onto a rigorous information-theoretic complexity measure. The framework of Reference-Contingent Complexity (RCC) \cite{liu2025structure} presents a promising candidate for this task. By defining a ``windowed'' complexity $C_R(\rho)$ relative to a structured vacuum $\sigma_R$—which encodes the ``observer resolution/accessible algebra''—RCC is poised to provide a microscopic foundation for the complexity generation potential $\Pi_\Xi$. Future efforts will be dedicated to systematically integrating the principle of ``structural fairness'' from RCC into CWT, thereby deriving precise, state-dependent thermodynamic equations of state for $\Pi_\Xi$ and associated response functions, and establishing a quantitative, testable connection between microscopic information geometry and macroscopic thermodynamic responses.
    \item \textbf{Rigorizing Core Assumptions and Dynamical Bounds:} Attempt to derive from more first principles, or rigorously verify under broader physical conditions, the validity of key assumptions such as "energy-equivalent temperature," and the universality and applicability range of the $T\to T_\Xi$ replacement in the complexity-temperature growth rate bound.
    \item \textbf{Systematic Extension to Non-Equilibrium States and Open Quantum Systems:} Construct the formal theory of CWT for non-equilibrium processes (especially steady-state and periodically driven systems) and open quantum systems, deeply investigate how the complexity budget $\Xi$ interacts with environmental decoherence, details of quantum measurement processes, and continuous energy/matter flows, and explore its application potential in understanding self-organization phenomena, information processing in living systems, and even energy efficiency in quantum technologies.
    \item \textbf{Experimental Verification of Core Theoretical Predictions:} Design and implement feasible experimental protocols capable of testing CWT's core predictions (such as resource constraint principles, complexity-temperature growth rate bound), especially on advanced quantum simulation and computation platforms. Concurrently, explore experimental avenues for indirectly probing and quantifying the complexity generation potential $\Pi_\Xi$ and its related effects by measuring the system's thermodynamic response to $\Xi$.
\end{enumerate}

In summary, the proposal of Complexity-Windowed Thermodynamics (CWT) is not only a profound reflection on the limitations of classical thermodynamics and statistical physics when facing complex systems but also an important theoretical extension aimed at internalizing information processing and computational complexity as fundamental physical constraints.
It attempts to elevate "computation" or, more broadly, "information processing capability" from a mere descriptive tool to the level of an intrinsic constraint within physical laws.
By revealing the fundamental limitations of physical resources on executable complexity, CWT offers new avenues for resolving the inherent difficulties of classical theories and provides a unified, information- and complexity-based perspective for understanding a wide range of physical phenomena, from microscopic quantum systems to macroscopic cosmological scales.
Although the path ahead is fraught with challenges, we firmly believe that more deeply integrating the core concept of "executable complexity" into the fundamental framework of physics is a crucial step towards a more comprehensive and profound understanding of the unified physical picture of the universe's operating laws.
This work represents only a preliminary exploration, and we eagerly anticipate that it will inspire more interdisciplinary in-depth thinking and innovative research, jointly advancing the frontiers of physics in the information age, and ultimately revealing a deeper unity among matter, energy, information, and complexity.

\section*{Acknowledgements}
This paper is dedicated to Ludwig Boltzmann—the unfinished path still extends.

\appendix
\renewcommand{\theequation}{\Alph{section}.\arabic{equation}}
\setcounter{equation}{0}
\section{Table of Symbols and Core Terminology} 
\label{A}
This appendix provides a quick reference guide to the key symbols and core terminology used in the Complexity-Windowed Thermodynamics (CWT) framework.
Given that CWT introduces several new concepts and physical quantities related to quantum computational complexity on top of classical thermodynamics, a clear guide to symbols and terminology is crucial for accurately understanding the theoretical construction, derivation processes, and physical meaning presented in this paper.
This appendix is divided into two parts: the first part explains the core mathematical symbols used in the paper; the second part defines and explains the key theoretical terms within the CWT framework.

\subsection{Key Symbols}
\label{A.1}
\begin{enumerate}
    \item $\Xi$ (\textbf{Complexity Budget})
    The core parameter of the CWT framework, representing the upper limit of computational complexity that a system can use to generate quantum states (or implement unitary transformations) or an observer can use to resolve quantum states, under specific physical conditions.
    It is typically considered a positive real number.
    The choice of its dimension depends on its specific physical correspondence: for example, if associated with action resources, it can be considered to have dimensions of action, or be made dimensionless by dividing by the reduced Planck constant $\hbar$ (as in the main setting of this paper, where quantum circuit complexity $C^* \le \Xi$, making $\Xi$ dimensionless like $C^*$); in some abstract models, it can also be treated as a pure number.
    The introduction of $\Xi$ is a key modification to the "equiprobable accessibility" assumption in classical statistical mechanics, internalizing the fundamental limitation of finite physical resources on the complexity of accessible states or resolvable information into the thermodynamic description.

    \item $C^*$ (\textbf{Minimum Quantum Circuit Complexity})
    Refers to the minimum number of gates or shortest quantum circuit depth required to synthesize a target quantum state $|\psi\rangle$ (or implement a target unitary transformation $U$) by applying a series of fundamental quantum gates from a chosen universal quantum gate set $G$, starting from an agreed-upon, structurally relatively simple reference state (e.g., a product state $|0\rangle^{\otimes N}$ in a many-body system).
    It is usually considered a dimensionless integer, used to quantify the intrinsic information processing complexity of the quantum state or quantum operation itself.
    In the CWT framework, $C^*$ is the key criterion for determining whether a specific microstate satisfies $C^* \le \Xi$ and is thus considered "accessible" or "resolvable" under a given complexity budget $\Xi$.

    \item $A(\Xi)$ (\textbf{Budgeted Observable Algebra})
    A von Neumann algebra defined as the algebra generated by all operators formed by the action of all unitary transformations $U \in G^{(\le\Xi)}$ (generated from a universal gate set $G$ with complexity not exceeding the current budget $\Xi$) on operators from a fundamental set of observables $M_0$.
    Physically, $A(\Xi)$ represents the set of all Hermitian operators (i.e., observables) that an observer can effectively measure, distinguish, or interact with under the computational resource constraint of complexity budget $\Xi$.
    $A(\Xi)$ is the key mathematical construct in CWT for achieving coarse-graining of the system's state under complexity constraints.

    \item $E_\Xi$ (\textbf{Conditional Expectation Map})
    A trace-preserving conditional expectation map from the full algebra of bounded operators $B(H)$ on the system's Hilbert space $H$ to the budgeted observable algebra $A(\Xi)$.
    For any density operator $\rho$, $E_\Xi(\rho)$ is the "best representation" of $\rho$ within the algebra $A(\Xi)$, which preserves the expectation values of $\rho$ for all operators in $A(\Xi)$.
    $E_\Xi$ is the core mathematical tool in the CWT framework for performing complexity-constrained coarse-graining of microstates and subsequently defining the core state function—the complexity-windowed entropy $S_\Xi$.

    \item $S_\Xi$ (\textbf{Complexity-Windowed Entropy})
    The core state function of the CWT framework, denoted as $S_\Xi(E,\Xi)$.
    It is defined as the von Neumann entropy of the microcanonical ensemble ($\Pi_E/\omega(E)$) after coarse-graining via the conditional expectation map $E_\Xi$, under the dual constraints of a given energy $E$ and complexity budget $\Xi$ (its mathematical definition is given in Eq. \eqref{eq:2.2}).
    $S_\Xi(E,\Xi)$ is profoundly interpreted as the residual statistical uncertainty or information-theoretic "entropy of ignorance" that an observer has about the system's true microstate $\rho_E$ under the limitation of the current complexity budget $\Xi$, due to their finite resolution capability.
    Its key mathematical properties include: (a) for any finite $\Xi$, $S_\Xi(E,\Xi)$ is a smooth function of energy $E$, a characteristic that is the fundamental guarantee for CWT's ability to universally regularize classical thermodynamic singularities; (b) $S_\Xi(E,\Xi)$ is a monotonically non-increasing function of the complexity budget $\Xi$ ($(\partial S_\Xi/\partial \Xi)_E \le 0$), reflecting the physical intuition that enhanced observational capability ($\Xi$ increases) leads to reduced or unchanged uncertainty.
    The dimension of $S_\Xi$ is the same as entropy, typically with units of J$\cdot$K$^{-1}$ (or dimensionless if $k_B=1$ in information-theoretic units).

    \item $T_\Xi$ (\textbf{Complexity-Windowed Temperature})
    The effective thermodynamic temperature derived from the complexity-windowed entropy $S_\Xi$, denoted as $T_\Xi(E,\Xi)$.
    It is defined as $T_\Xi^{-1} = (\partial S_\Xi/\partial E)_\Xi$ (or equivalently $T_\Xi = (\partial E/\partial S_\Xi)_\Xi$).
    $T_\Xi$ reflects the effective thermal state of the system under the observational window and dynamical accessibility limitations of a fixed complexity budget $\Xi$.
    Based on the smoothness of $S_\Xi$ with respect to energy $E$, $T_\Xi$ (for finite $\Xi$) exhibits continuous and finite behavior over all energy ranges (except for possible exception points of measure zero), thereby effectively "softening" the singular behavior that classical thermodynamic temperature might exhibit at phase transition points or in negative temperature regimes.
    Its dimension is the same as temperature, with units of K.

    \item $\Pi_\Xi$ (\textbf{Complexity Generation Potential})
    The thermodynamic intensive quantity conjugate to the complexity budget $\Xi$, denoted as $\Pi_\Xi(S_\Xi,\Xi)$.
    Its standard definition is $\Pi_\Xi = (\partial E/\partial \Xi)_{S_\Xi,V,N}$, and it can also be equivalently expressed as $\Pi_\Xi = -T_\Xi(\partial S_\Xi/\partial \Xi)_{E,V,N}$.
    Since $S_\Xi$ is monotonically non-increasing with $\Xi$ ($(\partial S_\Xi/\partial \Xi)_E \le 0$) and typically $T_\Xi > 0$, $\Pi_\Xi$ is usually non-negative ($\Pi_\Xi \ge 0$).
    It profoundly quantifies the minimum work done on the system by the surroundings, or the minimum energy absorbed by the system, to reversibly increase the complexity budget $\Xi$ by one unit while keeping the observer's effective entropy $S_\Xi$ and other macroscopic parameters (such as volume $V$) constant.
    The introduction of $\Pi_\Xi$ leads to the inclusion of the key \textbf{"information processing work" term $\Pi_\Xi\mathrm{d}\Xi$} in the extended first law of thermodynamics.
    Its dimension is energy (if $\Xi$ is considered dimensionless, as in the main setting of this paper where $C^*\le\Xi$ and $C^*$ is an integer), with units of J.
    If $\Xi$ strictly corresponds to a physical resource with dimensions of action, then the dimension of $\Pi_\Xi$ is correspondingly adjusted to frequency (s$^{-1}$).
    This paper primarily adopts the former setting.

    \item $\hbar$ (\textbf{Reduced Planck Constant})
    $\hbar \approx 1.054 \times 10^{-34}$ J$\cdot$s.
    The fundamental constant of quantum mechanics.
    In CWT, it appears as the fundamental quantum of action in the action constraint principle and time constraint principle, closely linking the microscopic quantum circuit complexity $C^*$ with macroscopic physical resources (action, evolution time).

    \item $k_B$ (\textbf{Boltzmann Constant})
    $k_B \approx 1.381 \times 10^{-23}$ J$\cdot$K$^{-1}$.
    The fundamental physical constant linking temperature and energy scales.
    In CWT, it is implicit in the definition of $S_\Xi$ (often taken as $k_B=1$ when expressing entropy in information-theoretic units) and appears explicitly in dynamical bounds related to the effective temperature $T_\Xi$, such as the complexity-temperature growth rate bound.

    \item $k_S$ (\textbf{Action Conversion Factor})
    A dimensionless positive real constant of order $1$, appearing in the action constraint principle $S_{\mathrm{process}} \ge k_S C^* \hbar$.
    It reflects the conversion efficiency from abstract quantum circuit complexity $C^*$ to the minimum action required for an actual physical process (such as state preparation or unitary evolution), and its specific value may depend on the chosen universal quantum gate set and the details of the physical implementation.
\end{enumerate}
\subsection{Core Terminology}

\begin{enumerate}
    \item \textbf{Complexity-Windowed Thermodynamics (CWT)}
    The theoretical framework systematically proposed and constructed in this paper.
    It aims to rectify the theoretical difficulties encountered by classical thermodynamics in dealing with complex physical situations such as phase transitions, critical phenomena, and negative temperature systems, by introducing a "complexity budget" $\Xi$ to limit the complexity range of quantum states that a system can generate or an observer can resolve.
    Its core lies in defining the "complexity-windowed entropy" $S_\Xi$ (Chapter \ref{cex.2})—which is smooth with respect to energy $E$ and monotonically non-increasing with $\Xi$, representing the observer's "entropy of ignorance"—and constructing an extended thermodynamic formalism based on it, which includes "information processing work" (represented by the term $\Pi_\Xi\mathrm{d}\Xi$, where $\Pi_\Xi\ge0$) (Chapter \ref{cex.3}).

    \item \textbf{Bounded Dynamical Complexity}
    One of the starting points and core postulates of the CWT framework (see Introduction and Section 2.1 for details).
    It states that under the constraint of any finite physical resources (such as total available action, evolution time, energy supply, etc.), there is an upper limit to the minimum generation quantum circuit complexity $C^*$ of quantum states that a physical system can actually generate or evolve to.
    This boundedness postulate replaces the idealized assumption in classical statistical mechanics of a priori "equiprobable accessibility" for all microstates within an energy shell, thereby being closer to physical reality.

    \item \textbf{Informational Work / Complexity Generation Work}
    The form of energy exchange represented by the term $\Pi_\Xi\mathrm{d}\Xi$ in the extended first law of CWT.
    Given that the complexity generation potential $\Pi_\Xi$ is typically non-negative ($\Pi_\Xi \ge 0$), this term represents the work done on the system by the surroundings in a reversible process to change (usually increase) the system's complexity budget $\Xi$ (i.e., to enhance the observer's resolution capability or allow the system to exhibit and maintain states of higher complexity).
    It quantifies the energy cost associated with changes in information processing capability or resolvable complexity level.

    \item \textbf{Resource Constraint Principles}
    Two core quantitative principles derived within the CWT framework that constrain the fundamental physical resources required for the generation of quantum state complexity $C^*$: The Action Constraint Principle ($S_{\mathrm{process}} \ge k_S C^* \hbar$): Generating a quantum state of complexity $C^*$ requires the consumption of at least $k_S C^* \hbar$ of action.
    The Time Constraint Principle ($t_{\mathrm{process}} \ge \frac{C^* \hbar}{2\pi k_B T_\Xi}$): The minimum time required to generate a quantum state of complexity $C^*$ is inversely proportional to the effective temperature $T_\Xi$.
    They collectively reveal that complexity generation processes are fundamentally dually constrained by available action and evolution time (correlated via effective temperature $T_\Xi$).

    \item \textbf{Complexity-Temperature Growth Rate Bound}
    One of the core dynamical predictions of the CWT framework, mathematically expressed as $\mathrm{d}C^*/\mathrm{d}t \le \frac{2\pi k_B}{\hbar} T_\Xi(t)$.
    It directly links the maximum instantaneous speed at which a quantum system generates its own computational complexity during unitary evolution to its current effective thermodynamic temperature $T_\Xi$, which takes complexity constraints into account.

    \item \textbf{Cosmic Complexity Saturation Conjecture}
    A conjectural inference proposed in Chapter \ref{cex.7}, based on CWT's time constraint principle (and its differential form, the complexity-temperature growth rate bound) and the key "energy-equivalent temperature" hypothesis.
    This conjecture posits that the upper bound on the total computational complexity ($C_U$) that the universe could have generated throughout its entire evolutionary history since the Big Bang is strikingly consistent in order of magnitude with the cosmic (Hubble) horizon entropy ($S_H/k_B$) estimated from the holographic principle.
    This conjecture implies that during its evolution, the universe's information processing capability or its total generated complexity may have approached the theoretical limit set by fundamental physical laws and available spacetime and energy resources.
\end{enumerate}

\section{Mathematical Foundations of Budgeted Observable Algebra and Conditional Expectation}
\label{B.1}
This appendix provides a rigorous mathematical exposition of the budgeted observable algebra $A(\Xi)$ and the associated conditional expectation $E_\Xi$.
These constructs are central to the Complexity-Windowed Thermodynamics (CWT) framework developed in the main text.
These mathematical structures not only underpin the definition of the complexity-windowed entropy $S_\Xi$ (as introduced in Section 2 of the main text) but, more critically, ensure its regularity, which is essential for consistently deriving thermodynamic quantities within the CWT framework.
For clarity and conciseness, our exposition will focus on finite-dimensional quantum systems, specifically those described by an $n$-qubit Hilbert space $H = (\mathbb{C}^2)^{\otimes n}$.
However, the core conclusions can be generalized to infinite-dimensional systems within the framework of $C^*$-algebras without fundamentally altering the CWT-related findings.

\subsection{Notational Conventions}
\begin{enumerate}
    \item $B(H)$: The set of all bounded linear operators acting on the Hilbert space $H$. For finite-dimensional $H$, this is simply the algebra of all $d \times d$ matrices, where $d = \dim(H)$.
    \item $\tau(X) = \mathrm{Tr}(X)/d$: The standard normalized trace functional, which defines a faithful tracial state on $B(H)$.
    \item $M_0$: A predefined, finite set of fundamental Hermitian operators, termed "basic observables." For an $n$-qubit system, a typical choice consists of all single-qubit Pauli operators acting on each qubit, e.g., $\{\sigma_k^{(i)} : k \in \{x,y,z\}, i \in \{1,...,n\}\}$, plus the identity operator.
    \item $G_{\mathrm{gates}}$: A finite and universal set of elementary quantum gates (e.g., CNOT gate and single-qubit rotation gates, or any other standard universal gate set). (Note: To avoid confusion with the gravitational constant $G$, $G_{\mathrm{gates}}$ is used).
    \item $\mathrm{depth}(U)$: The depth of a quantum circuit $U \in B(H)$ composed of gates from $G_{\mathrm{gates}}$, defined as the minimum number of layers of non-overlapping gate operations required to implement $U$.
    \item $\Xi \in \mathbb{N}_0$: The "complexity budget," an integer representing the maximum allowed circuit depth for operations considered accessible.
    \item $U_\Xi$: The set of all unitary quantum circuits $U$ composed of gates from $G_{\mathrm{gates}}$ with depth $\mathrm{depth}(U) \le \Xi$. For any finite $\Xi$, this set is finite.
\end{enumerate}
\subsection{Budgeted Observable Algebra and Conditional Expectation} 
The central algebraic construct in CWT is the budgeted observable algebra $A(\Xi)$, which encapsulates, in principle, all accessible or resolvable observables given a complexity budget $\Xi$.
\textbf{Construction of A($\Xi$) }
The algebra $A(\Xi)$ is defined as the von Neumann algebra generated by all basic observables $M \in M_0$ transformed by any accessible unitary operation $U \in U_\Xi$. More formally, let $S_\Xi = \{ U M U^\dagger : M \in M_0, U \in U_\Xi \}$ (Note: $U^\dagger$ denotes the Hermitian conjugate of $U$). Then $A(\Xi)$ is the smallest von Neumann subalgebra of $B(H)$ containing $S_\Xi$, which is also equivalent to the closure of the self-adjoint algebra generated by $S_\Xi$ in its weak* ($w^*$) topology (as indicated by $(w^*)$ in Eq. \eqref{eqB.1}), or via its double commutant $S_\Xi''$.

\begin{equation}
A(\Xi) = \mathrm{alg}\langle S_\Xi \rangle^{(w^*)} = (\{ U M U^\dagger : M \in M_0, U \in U_\Xi \})''
\label{eqB.1}
\end{equation}

Intuitively, $A(\Xi)$ comprises all operators that can be effectively "measured" or prepared starting from a set of basic measurables $M_0$ through operations constrained by the complexity budget $\Xi$.

\textbf{Basic Properties of A($\Xi$) }
The algebra $A(\Xi)$ possesses several key properties crucial for the CWT framework. Firstly, for any finite complexity budget $\Xi$, $A(\Xi)$ is a finite-dimensional von Neumann algebra. This follows from the finite-dimensionality of $H$ and the fact that $A(\Xi)$ is generated by a finite set of operators (since $U_\Xi$ and $M_0$ are finite). Consequently, the dimension of $A(\Xi)$, denoted $d_A(\Xi) = \dim(A(\Xi))$, is finite, though its exact value might be hard to determine, it is bounded.
Secondly, the algebras $A(\Xi)$ form a nested chain with respect to the complexity budget:

\begin{equation}
\Xi_1 \le \Xi_2 \implies A(\Xi_1) \subseteq A(\Xi_2)
\end{equation}

This property reflects the intuitive notion that increasing available complexity (computational resources) does not reduce the set of accessible observables. In the limit of an unrestricted complexity budget ($\Xi \to \infty$), the algebra $A(\Xi)$ approaches the full algebra of observables $B(H)$, i.e., $(\overline{\bigcup_{\Xi < \infty} A(\Xi)})^{w^*}$ is $B(H)$. As discussed in Section \ref{cex.3.3} of the main text, this limiting behavior ensures that CWT recovers standard statistical mechanics when complexity constraints are lifted.
Associated with a von Neumann subalgebra $A(\Xi) \subseteq B(H)$ is a unique conditional expectation $E_\Xi$ that maps operators from the full algebra $B(H)$ onto the budgeted observable algebra $A(\Xi)$. This map plays a central role in the coarse-graining process inherent to CWT.

\textbf{Existence and Uniqueness:} The existence and uniqueness of such a conditional expectation are guaranteed by Tomiyama's theorem. Given a von Neumann algebra $M_{\mathrm{algebra}}$ (here $B(H)$) (Note: To avoid confusion with the basic observables $M_0$) and a von Neumann subalgebra $N_{\mathrm{algebra}}$ (here $A(\Xi)$) sharing the same identity element, if there exists a faithful normal state on $M_{\mathrm{algebra}}$ whose restriction to $N_{\mathrm{algebra}}$ is also faithful (the normalized trace $\tau$ satisfies this), then there exists a unique projection $E_{\mathrm{map}}: M_{\mathrm{algebra}} \to N_{\mathrm{algebra}}$ of norm $1$ (Note: To avoid confusion with energy $E$) such that $E_{\mathrm{map}}$ is an $N_{\mathrm{algebra}}$-bimodule map ($E_{\mathrm{map}}(n_1 m n_2) = n_1 E_{\mathrm{map}}(m) n_2$ for $n_1,n_2 \in N_{\mathrm{algebra}}$, $m \in M_{\mathrm{algebra}}$) and preserves the state $\tau$ (i.e., $\tau \circ E_{\mathrm{map}} = \tau$). This unique projection is the conditional expectation $E_\Xi$.

\textbf{Explicit Representation and Basic Properties:} For finite-dimensional algebras, the conditional expectation $E_\Xi$ has an explicit representation. Let $\{F_k\}_{k=1}^{d_A(\Xi)}$ be an orthonormal basis for $A(\Xi)$ with respect to the Hilbert-Schmidt inner product $\langle X, Y \rangle_{\mathrm{HS}} = \tau(X^\dagger Y)$ defined by $\tau$ (i.e., $\tau(F_k^\dagger F_j) = \delta_{kj}$). Then, for any $X \in B(H)$, the action of $E_\Xi$ can be written as:

\begin{equation}
E_\Xi(X) = \sum_{k=1}^{d_A(\Xi)} \tau(F_k^\dagger X) F_k
\end{equation}

The map $E_\Xi$ possesses several indispensable properties:
\begin{enumerate}
    \item Linearity: $E_\Xi(aX + bY) = aE_\Xi(X) + bE_\Xi(Y)$ for $a,b \in \mathbb{C}$ and $X,Y \in B(H)$.
    \item Positivity and Complete Positivity (CP): If $X \ge 0$, then $E_\Xi(X) \ge 0$. More strongly, $E_\Xi$ is completely positive, which ensures that when applied to a subsystem of an entangled state, it preserves the positivity of the overall state.
    \item Trace-preserving: $\tau(E_\Xi(X)) = \tau(X)$ for all $X \in B(H)$. This ensures that expectation values of observables within $A(\Xi)$ are preserved after the mapping.
    \item Idempotence (Projection property): $E_\Xi \circ E_\Xi = E_\Xi$. Once an operator is projected into $A(\Xi)$, further application of $E_\Xi$ does not change it.
    \item Norm non-increasing: $||E_\Xi(X)||_{\infty} \le ||X||_{\infty}$ for any operator norm; in particular, for the operator norm induced by the Hilbert space norm, $||E_\Xi||_{\infty} = 1$.
\end{enumerate}

\textbf{Variational Characterization and Physical Interpretation:} For a given density matrix $\rho$, the conditional expectation $E_\Xi(\rho)$ can be understood as the "best approximation" of $\rho$ within the algebra $A(\Xi)$ as it minimizes certain information-theoretic distances. Specifically, for any density matrix $\rho \in B(H)$, $E_\Xi(\rho)$ is the unique state in the set of density matrices $D(A(\Xi))$ belonging to $A(\Xi)$ that:
\begin{enumerate}
    \item \textbf{Minimizes Quantum Relative Entropy:} $E_\Xi(\rho) = \mathrm{argmin}_{\sigma \in D(A(\Xi))} D_{\mathrm{rel}}(\rho || \sigma)$ (Note: $D_{\mathrm{rel}}$ denotes relative entropy), where $D_{\mathrm{rel}}(\rho || \sigma) = \mathrm{tr}(\rho(\ln \rho - \ln \sigma))$. This stems from properties of states satisfying the data processing inequality for relative entropy under trace-preserving conditional expectations in Petz's theorem.
    \item \textbf{Minimizes Hilbert-Schmidt Distance:} $E_\Xi(\rho) = \mathrm{argmin}_{\sigma \in D(A(\Xi))} ||\rho - \sigma||_2^2$, where $||.||_2$ is the Hilbert-Schmidt norm (usually defined as $\sqrt{\mathrm{Tr}(A^\dagger A)}$, or $\sqrt{d \cdot \tau(A^\dagger A)}$ if associated with $\tau$). This is because $E_\Xi$ acts as an orthogonal projection in the Hilbert-Schmidt space of operators.
\end{enumerate}
Thus, $E_\Xi(\rho)$ can be interpreted as the state within $A(\Xi)$ that retains the maximum possible information about $\rho$, or equivalently, the state "closest" to $\rho$ under the observational constraints imposed by the complexity budget $\Xi$. This coarse-graining operation is precisely the mechanism by which CWT regularizes singularities by effectively averaging out details that are irresolvable below the complexity threshold $\Xi$.
The complexity-windowed entropy $S_\Xi(E,\Xi)$ is defined in the main text (cf. Equation \eqref{eq:2.2}) as the von Neumann entropy of the coarse-grained microcanonical state $\rho(E) = \Pi_E/\omega(E)$ (Note: $\Pi_E$ is the energy shell projector, $\omega(E)$ is the density of states) of the system at energy $E$ and complexity budget $\Xi$:

\begin{equation}
S_\Xi(E,\Xi) = -k_B \mathrm{Tr} \left[ E_\Xi(\rho(E)) \ln E_\Xi(\rho(E)) \right]
\label{eq:B.4}
\end{equation}

\subsection{Regularity of Complexity-Windowed Entropy \texorpdfstring{$S_\Xi$}{SXi}} 
A cornerstone of the CWT framework is the smoothness of $S_\Xi(E,\Xi)$ with respect to energy $E$ (for any fixed, finite $\Xi$).
This regularity directly stems from the mathematical properties established above.
Since for finite $\Xi$, $A(\Xi)$ is a finite-dimensional von Neumann algebra, the projected state $E_\Xi(\rho(E))$ is an operator in this finite-dimensional space.
The von Neumann entropy function $S_{\mathrm{entropy}}(\sigma) = -\mathrm{tr}(\sigma \ln \sigma)$ (Note: to avoid confusion with action $S$) is analytic for strictly positive definite density matrices and generally smooth on the space of density matrices of a finite-dimensional algebra, except possibly at points where certain eigenvalues degenerate in a specific manner.
Assuming the microcanonical state $\rho(E)$ itself (e.g., via the energy shell projection operator $\Pi_E$ and density of states $\omega(E)$) varies smoothly with $E$ (a standard assumption for systems with continuous or sufficiently dense spectra), then its projection $E_\Xi(\rho(E))$ will also vary smoothly with $E$.
Consequently, $S_\Xi(E,\Xi)$, being a composition of smooth functions, will also be a smooth (typically $C^\infty$) function of $E$, except possibly on a set of isolated energy values of measure zero where the dimension or structure of $A(\Xi)$ might abruptly change due to its dependence on $E$ (though in our primary construction $A(\Xi)$ depends only on $\Xi$).
This smoothness is crucial because it ensures that other core thermodynamic quantities defined via differentiation of $S_\Xi(E,\Xi)$, such as the effective temperature $T_\Xi^{-1} = (\partial S_\Xi/\partial E)_\Xi$ and the intensive quantity conjugate to the complexity budget $\Xi$ (i.e., the complexity generation potential $\Pi_\Xi$) to be introduced in Section \ref{cex.3}, are well-defined and continuous functions of energy $E$ (for finite $\Xi$).
This lays the foundation for subsequently constructing a well-behaved CWT formalism capable of consistently resolving classical thermodynamic singularities.
This appendix has detailed the rigorous mathematical construction of the budgeted observable algebra $A(\Xi)$ and the conditional expectation $E_\Xi$.
We have established that for any finite complexity budget $\Xi$, $A(\Xi)$ is a finite-dimensional von Neumann algebra, forming a nested hierarchy, and converges to the full algebra $B(H)$ as $\Xi \to \infty$.
The unique, trace-preserving conditional expectation $E_\Xi$ maps operators from the full algebra $B(H)$ onto the budgeted observable algebra $A(\Xi)$, optimally preserving information and serving as the precise mechanism for coarse-graining under complexity constraints.
Crucially, these properties ensure the smoothness of the complexity-windowed entropy $S_\Xi(E,\Xi)$ with respect to energy, providing a solid mathematical foundation for the entire Complexity-Windowed Thermodynamics framework and its capability to universally regularize thermodynamic singularities.

\section{Derivation of Key Properties of Complexity-Windowed Entropy \texorpdfstring{$S_\Xi(E,\Xi)$}{SXi(E,Xi)}} 
\label{c.1}
This appendix aims to provide detailed mathematical derivations and justifications for the core properties of the complexity-windowed entropy $S_\Xi(E,\Xi)$ as stated in Proposition 2.1 of the main text.
These properties—specifically, the smoothness of $S_\Xi(E,\Xi)$ with respect to energy $E$ for any finite complexity budget $\Xi$, its monotonic non-increasing behavior with increasing $\Xi$, and its convergence to the classical statistical mechanics limit—form the mathematical cornerstone upon which the Complexity-Windowed Thermodynamics (CWT) framework is built.
The arguments herein follow the basic mathematical setup for the budgeted observable algebra $A(\Xi)$ and conditional expectation $E_\Xi$ detailed in Section \ref{cex.2.1} of the main text and Appendix \ref{B.1}, assuming a finite-dimensional Hilbert space $H$ of dimension $d=2^n$.

\subsection{Preliminaries and Notation} 
The energy shell projection operator $\Pi_E$ is defined within a narrow energy window of width $\delta$ centered at $E$, encompassing microstates $\{|\epsilon_j\rangle\}$ whose energies $\epsilon_j$ fall within the interval $[E-\delta/2, E+\delta/2]$.
The density of states (or degeneracy) within this energy shell is $\omega(E) = \mathrm{Tr}(\Pi_E)$, and the corresponding microcanonical ensemble density operator is $\rho_E = \Pi_E / \omega(E)$.
To ensure the arguments for smoothness, we assume that the energy shell projection operator $\Pi_E$, and consequently $\omega(E)$ and $\rho_E$, are sufficiently smooth functions of energy $E$. This assumption is generally reasonable in macroscopic systems.
The construction of the budgeted observable algebra $A(\Xi)$ and the trace-preserving conditional expectation $E_\Xi: B(H) \to A(\Xi)$ has been detailed in Section \ref{cex.2.1} and Appendix \ref{B.1}. The complexity-windowed entropy is then defined as:

\begin{equation}
S_\Xi(E,\Xi) = S_{\mathrm{VN}}(E_\Xi(\rho_E)) = -k_B \mathrm{Tr}[E_\Xi(\rho_E) \ln(E_\Xi(\rho_E))]
\label{eq:C.1}
\end{equation}

where $S_{\mathrm{VN}}$ denotes the von Neumann entropy. Our goal is to rigorously establish the properties of $S_\Xi(E,\Xi)$ as a function of its variables. $S_\Xi(E,\Xi)$ quantifies the residual uncertainty or "ignorance" of an observer about the true state $\rho_E$ of the system, subject to the constraints of the complexity budget $\Xi$.

\subsection{Smoothness of \texorpdfstring{$S_\Xi(E,\Xi)$ with respect to Energy $E$ (for fixed, finite $\Xi$)}{S\_Xi(E,Xi) 关于能量 E 的光滑性（针对固定的有限 Xi）}} 
A cornerstone assertion of the CWT framework is that for any finite complexity budget $\Xi$, the complexity-windowed entropy $S_\Xi(E,\Xi)$ is a smooth function of energy $E$. This property is crucial as it ensures that the derived effective temperature $T_\Xi$ and other thermodynamic response functions behave well and remain finite, thereby regularizing singularities encountered in classical thermodynamics.

\textbf{Theorem C.1 (Smoothness with respect to $E$):} For any finite complexity budget $\Xi < \infty$, if the microcanonical density operator $\rho_E$ is a $C^r$ function of energy $E$ (where $r \ge 1$), then the complexity-windowed entropy $S_\Xi(E,\Xi)$ is also a $C^r$ function of $E$, except possibly on a set of energy values of measure zero where eigenvalues of $E_\Xi(\rho_E)$ degenerate (leading to a change in rank).
The proof primarily involves three steps:

\begin{enumerate}
    \item \textbf{Smoothness of the projected state $E_\Xi(\rho_E)$ with respect to $E$:}
    As elucidated in Appendix \ref{B.1}, for a finite complexity budget $\Xi$, the budgeted observable algebra $A(\Xi)$ is a finite-dimensional von Neumann algebra. The conditional expectation $E_\Xi: B(H) \to A(\Xi)$ is a linear, positive, and trace-preserving map. It can be expressed as:
    
 \begin{equation}
    E_\Xi(X) = \sum_{k=1}^{d_A} \tau(F_k^\dagger X) F_k
 \end{equation}
 
    where $\{F_k\}$ forms an orthonormal basis for $A(\Xi)$ under the Hilbert-Schmidt inner product (where $\tau(.) = (1/d)\mathrm{Tr}(.)$) (i.e., $\tau(F_k^\dagger F_j) = \delta_{kj}$), and $d_A = \dim(A(\Xi))$. Crucially, since $A(\Xi)$ and its basis $\{F_k\}$ are determined solely by $\Xi$ and the fundamental set of observables $M_0$, they are independent of energy $E$.
    Given that $\rho_E$ is assumed to be a $C^r$ function of $E$ (i.e., its matrix elements in a fixed basis of $B(H)$ are $C^r$ functions of $E$), and $E_\Xi$ is a linear map whose definition does not involve $E$, the projected state $\sigma_{E,\Xi} = E_\Xi(\rho_E)$ will also be a $C^r$ function of $E$. Specifically, each matrix element of $\sigma_{E,\Xi}$ (in any fixed basis of $B(H)$, or in the basis $\{F_k\}$ of $A(\Xi)$) will be a $C^r$ function of $E$, as they are linear combinations (with fixed coefficients) of the matrix elements of $\rho_E$.

    \item \textbf{Properties of the von Neumann entropy function $S_{\mathrm{VN}}(\sigma) = -k_B \mathrm{Tr}(\sigma \ln \sigma)$:}
    For a state $\sigma$ residing in a finite-dimensional algebra like $A(\Xi)$, as long as the eigenvalues of $\sigma$ are non-zero (i.e., $\sigma$ is strictly positive definite), $S_{\mathrm{VN}}(\sigma)$ is a smooth (even analytic) function of the matrix elements of $\sigma$ \cite{wehrl1978general}. Even if some eigenvalues of $\sigma$ are zero, $S_{\mathrm{VN}}(\sigma)$ is smooth on the manifold of density matrices of a fixed rank, as long as its rank does not change abruptly. The set of $E$ values causing a rank change in $\sigma_{E,\Xi}$ typically forms a set of measure zero.

    \item \textbf{Smoothness of a composite function:}
    The complexity-windowed entropy $S_\Xi(E,\Xi)$ (defined by Eq. \eqref{eq:C.1}) is a composition of two maps: the $C^r$ map $E \to \sigma_{E,\Xi}$ and the (at most points) smooth map $\sigma \to S_{\mathrm{VN}}(\sigma)$. By the chain rule for composite functions, $S_\Xi(E,\Xi)$ is therefore a $C^r$ function of $E$, except possibly on the aforementioned set of measure zero.
\end{enumerate}
This smoothness is the fundamental reason why CWT can universally regularize thermodynamic singularities. A finite $\Xi$ effectively confines the system to a finite-dimensional observable landscape $A(\Xi)$, thereby truncating those degrees of freedom or correlations whose unconstrained behavior in the $\Xi \to \infty$ limit would lead to classical divergences or discontinuities. Consequently, the effective temperature $T_\Xi^{-1} = (\partial S_\Xi / \partial E)_\Xi$ and derived quantities such as specific heat $C_{V,\Xi}$ remain finite and continuous.

\subsection{Monotonicity of \texorpdfstring{$S_\Xi(E,\Xi)$ with respect to Complexity Budget $\Xi$ (for fixed $E$)}{SXi(E,Xi) 关于复杂度预算 Xi 的单调性（针对固定的E）}} 
\label{c.3}
The complexity-windowed entropy $S_\Xi(E,\Xi)$ reflects the observer's "degree of ignorance" about the true state $\rho_E$ of the system under a complexity budget $\Xi$. Intuitively, increasing the available complexity budget $\Xi$ (i.e., enhancing observational capability) should allow the observer to gain more information about the true state of the system, thereby reducing this "ignorance." This intuition is formalized by the monotonically non-increasing property of $S_\Xi(E,\Xi)$ with respect to $\Xi$.

\textbf{Theorem C.2 (Monotonicity with respect to $\Xi$):} For a fixed energy $E$, if $\Xi_1 \le \Xi_2$, then the budgeted observable algebras satisfy $A(\Xi_1) \subseteq A(\Xi_2)$. In this case, the complexity-windowed entropies satisfy:

\begin{equation}
S_{\Xi_1}(E) \ge S_{\Xi_2}(E)
\label{eq:C.3}
\end{equation}

(Here, $S_{\Xi}(E)$ is used as a shorthand, since $E$ is fixed, $S_{\Xi_1}(E) = S_\Xi(E,\Xi_1)$ and $S_{\Xi_2}(E) = S_\Xi(E,\Xi_2)$).
That is, $S_\Xi(E,\Xi)$ is a non-increasing function of $\Xi$. Equality holds if and only if $E_{\Xi_2}(\rho_E)$ already belongs to the subalgebra $A(\Xi_1)$ (i.e., $E_{\Xi_1}(\rho_E) = E_{\Xi_2}(\rho_E)$).

\textbf{Proof:} The condition $\Xi_1 \le \Xi_2$ implies that $A(\Xi_1)$ is a von Neumann subalgebra of $A(\Xi_2)$. Let $\sigma_1 = E_{\Xi_1}(\rho_E)$ and $\sigma_2 = E_{\Xi_2}(\rho_E)$.
Since $A(\Xi_1) \subseteq A(\Xi_2)$, there exists a conditional expectation from $A(\Xi_2)$ to its subalgebra $A(\Xi_1)$, denoted as $P_{12}: A(\Xi_2) \to A(\Xi_1)$. By properties of conditional expectations (see Appendix \ref{B.1}, or standard texts on operator algebras), we have $E_{\Xi_1} = P_{12} \circ E_{\Xi_2}$. Thus, $\sigma_1 = P_{12}(\sigma_2)$.
The von Neumann entropy $S_{\mathrm{VN}}(\sigma)$ has an important property under conditional expectation. For any trace-preserving conditional expectation $P$ (from a von Neumann algebra $M$ to its subalgebra $N$), and any state $\sigma$ in $M$, we have:

\begin{equation}
S_{\mathrm{VN}}(P(\sigma)) \ge S_{\mathrm{VN}}(\sigma)
\end{equation}

\cite{petz2007quantum} \cite{lindblad1973entropy}, The physical meaning of this inequality is that further coarse-graining (represented by $P$, projecting the state from the finer algebra $A(\Xi_2)$ to the coarser algebra $A(\Xi_1)$) does not decrease entropy, and may increase it due to information loss. Applying this to our case, $P_{12}$ is the conditional expectation from $A(\Xi_2)$ to $A(\Xi_1)$, and $\sigma_2 = E_{\Xi_2}(\rho_E) \in A(\Xi_2)$. Therefore, $S_{\mathrm{VN}}(\sigma_1) = S_{\mathrm{VN}}(P_{12}(\sigma_2)) \ge S_{\mathrm{VN}}(\sigma_2)$.
Substituting the definitions of $\sigma_1$ and $\sigma_2$, and using the definition of $S_\Xi$ from Eq. \eqref{eq:C.1}, we get: $S_{\Xi_1}(E) \ge S_{\Xi_2}(E)$. This proves that $S_\Xi(E,\Xi)$ is a non-increasing function of $\Xi$, i.e., Eq. \eqref{eq:C.3} holds.
The condition for equality, $S_{\Xi_1}(E) = S_{\Xi_2}(E)$, holds if and only if equality holds in (C.4), i.e., $S_{\mathrm{VN}}(P_{12}(\sigma_2)) = S_{\mathrm{VN}}(\sigma_2)$. This typically occurs when $P_{12}(\sigma_2) = \sigma_2$ (under certain conditions for strict entropy inequalities, e.g., when $\sigma_2$ is a fixed point of $P_{12}$, or when $\sigma_2$ is in $A(\Xi_1)$ so $P_{12}$ acts as identity on it). If $P_{12}(\sigma_2) = \sigma_2$, it means that $\sigma_2 = E_{\Xi_2}(\rho_E)$ is itself already an element of $A(\Xi_1)$. In this case, the finer algebra $A(\Xi_2)$ (relative to $A(\Xi_1)$) does not provide any additional information about $\rho_E$ beyond what $A(\Xi_1)$ can already resolve to change the projected state, so $E_{\Xi_1}(\rho_E) = E_{\Xi_2}(\rho_E)$.
This monotonically non-increasing behavior ensures the internal consistency of CWT: expanding computational resources or observational capabilities (increasing $\Xi$) reduces or maintains the statistical uncertainty (effective entropy) that an observer has about the system's state due to insufficient resolution. This provides the basis for $(\partial S_\Xi / \partial \Xi)_E \le 0$, which is crucial for the physical interpretation of the complexity generation potential $\Pi_\Xi$ later on.

\subsection{Limiting Behavior of \texorpdfstring{$S_\Xi(E,\Xi)$}{SXi(E,Xi)}} 
To ensure that CWT is a true generalization of classical statistical mechanics, it must recover classical results in the appropriate limit of unconstrained resources. Conversely, in the limit of minimal resources, it should reflect a state of maximum ignorance compatible with the most basic observations.

\textbf{Theorem C.3 (Limiting Behavior):}
\begin{enumerate}
    \item[(a)] In the limit of unconstrained complexity budget ($\Xi \to \infty$), $S_\Xi(E,\Xi)$ converges to the classical microcanonical entropy:
    \begin{equation}
    \lim_{\Xi\to\infty} S_\Xi(E,\Xi) = S_{\mathrm{micro}}(E) = k_B \ln \omega(E)
    \label{eq:C.5}
    \end{equation}

    \item[(b)] In the limit of minimal (zero) complexity budget ($\Xi \to 0$), assuming $A(0)$ degenerates to the trivial algebra $\mathbb{C} \cdot \mathrm{Id}$ (multiples of the identity operator on $H$, where $\mathbb{C}$ is the field of complex numbers and $\mathrm{Id}$ is the identity operator), $S_\Xi(E,\Xi)$ converges to the entropy of the maximally mixed state on the entire Hilbert space $H$:
    
    \begin{equation}
    \lim_{\Xi\to 0} S_\Xi(E,\Xi) = k_B \ln d
    \label{eq:C.6}
    \end{equation}
    
\begin{center}
where $d = \dim(H)$.
\end{center}     
\end{enumerate}

\textbf{Proof:}
\begin{enumerate}
    \item[(a)] \textbf{Unconstrained Resource Limit ($\Xi \to \infty$):}
    As the complexity budget $\Xi$ increases, the nested sequence of finite-dimensional algebras $A(\Xi)$ grows, according to Appendix \ref{B.1}. Assuming the chosen universal gate set and fundamental observables $M_0$ are sufficiently rich, the union $(\overline{\bigcup_{\Xi < \infty} A(\Xi)})^{w^*}$ is the entire algebra of bounded operators $B(H)$. In this limit, the conditional expectation $E_\Xi$ converges (in a suitable operator norm sense for finite-dimensional $H$, e.g., trace norm) to the identity map $\mathrm{id}_{B(H)}: B(H) \to B(H)$. Therefore, as $\Xi \to \infty$, $E_\Xi(\rho_E) \to \rho_E$.
    Since the von Neumann entropy $S_{\mathrm{VN}}(.)$ is a continuous function on the space of density operators (with respect to the trace norm), we have:
    
    \begin{equation}
    \lim_{\Xi\to\infty} S_\Xi(E,\Xi) = S_{\mathrm{VN}}(\lim_{\Xi\to\infty} E_\Xi(\rho_E)) = S_{\mathrm{VN}}(\rho_E)
     \end{equation}
     
    For the microcanonical state $\rho_E = \Pi_E / \omega(E)$, where $\Pi_E$ is the projector onto a subspace of dimension $\omega(E)$, its $\omega(E)$ non-zero eigenvalues are all $1/\omega(E)$. Therefore,
\begin{equation}
\begin{split}
  S_{\mathrm{VN}}(\rho_E)
    &= -k_B \sum_{i=1}^{\omega(E)}
         \Bigl(\tfrac{1}{\omega(E)}\Bigr)
         \ln\Bigl(\tfrac{1}{\omega(E)}\Bigr)\\
    &= -k_B\,\omega(E)\,
         \Bigl(\tfrac{1}{\omega(E)}\Bigr)
         \ln\Bigl(\tfrac{1}{\omega(E)}\Bigr)\\
    &= -k_B \,\ln\Bigl(\tfrac{1}{\omega(E)}\Bigr)\\
    &= k_B \,\ln\bigl(\omega(E)\bigr).
\end{split}
\end{equation}
    Substituting this into the above yields Eq. \eqref{eq:C.5}. This confirms that CWT seamlessly recovers the standard microcanonical entropy when complexity constraints are lifted.

    \item[(b)] \textbf{Minimal Resource Limit ($\Xi \to 0$):}
    We consider the case of minimal complexity budget, $\Xi=0$. Here, we assume that the budgeted observable algebra $A(0)$ is the trivial algebra $\mathbb{C} \cdot \mathrm{Id}$ consisting only of scalar multiples of the identity operator $\mathrm{Id}$ on $H$. This represents an observer incapable of discerning any non-trivial structure of the system, only confirming its existence in a $d$-dimensional Hilbert space. The conditional expectation $E_0$ onto $A(0)$ maps any operator $X$ to $E_0(X) = \tau(X) \cdot \mathrm{Id}$, where $\tau(X) = (1/d)\mathrm{Tr}(X)$ is the normalized trace (and $\tau(\mathrm{Id})=1$). This conditional expectation preserves the normalized trace $\tau$, i.e., $\tau(E_0(X)) = \tau(X)$. For a density operator $\rho_E$, since $\mathrm{Tr}(\rho_E)=1$, we have $\tau(\rho_E) = 1/d$. Therefore,
    
    \begin{equation}
    E_0(\rho_E) = \left(\frac{1}{d}\right) \mathrm{Id}
    \end{equation}
       
    This is the maximally mixed state on the entire Hilbert space $H$. The entropy of this state is:
\begin{multline}
  S_{\mathrm{VN}}\bigl((\tfrac{1}{d})\,\mathrm{Id}\bigr)
    = -k_B\,\mathrm{Tr}\Bigl[(\tfrac{1}{d})\,\mathrm{Id}\;\ln\bigl((\tfrac{1}{d})\,\mathrm{Id}\bigr)\Bigr]\\
    = -k_B\,d\;\bigl(\tfrac{1}{d}\bigr)\,\ln\bigl(\tfrac{1}{d}\bigr)\\
    = -k_B\,\ln\bigl(\tfrac{1}{d}\bigr)
    = k_B\,\ln(d)
\end{multline}
     
    Substituting this into the above yields Eq. \eqref{eq:C.6}. This result indicates that under extreme resource limitation ($\Xi\to0$), when the observer cannot perform any specific measurements to discern the system's energy or internal configuration (beyond its dimensional information), the effective entropy (ignorance) reaches its maximum value $k_B \ln(d)$. This value is a constant, reflecting that in this limit, since the observer cannot even resolve the energy shell, the entropy becomes independent of the system's specific energy $E$ and is determined solely by the total dimension of the Hilbert space.
\end{enumerate}
The mathematical properties of the complexity-windowed entropy $S_\Xi(E,\Xi)$ rigorously established in this appendix—its smoothness with respect to $E$ for finite $\Xi$, its monotonic non-increase with $\Xi$, and its consistent limiting behaviors—are fundamental to the entire CWT framework. Smoothness underpins the well-definedness of thermodynamic quantities derived in CWT and its ability to regularize classical singularities. Monotonic non-increase ensures the physical self-consistency of introducing a complexity budget (increased resolution capability reduces ignorance). Finally, the correct classical and minimal-resource limits confirm that CWT is a controlled and meaningful extension of established thermodynamic and statistical mechanical principles, providing a solid foundation for its application to the various physical phenomena discussed in the main text.

\section{Derivation and Comparison of Physical Resource Constraint Principles for the Complexity Generation Process}
\label{D.1}
This appendix aims to provide detailed theoretical derivations, elucidation of physical implications, and comparisons with established physical theories for the core physical principles—proposed in the main text (specifically in Chapter \ref{cex.5} and Section \ref{cex.6.1})—that constrain the generation process of quantum system complexity $C^*$.
These principles primarily include the action constraint, and two formalized bounds on the rate of complexity growth—one based on the first principles of CWT and the complexity generation potential $\Pi_\Xi$, and the other based on an analogy with quantum chaos theory and the effective temperature $T_\Xi$—from which corresponding time constraints are derived.
Collectively, these principles define the constraints imposed by fundamental physical resources (action $S_{\mathrm{proc}}$ and evolution time $t_{\mathrm{proc}}$) on a quantum system during the generation of computational complexity $C^*$, and correlate these constraints with the system's effective thermodynamic state (characterized by $\Pi_\Xi$ and $T_\Xi$).
The derivations in this appendix will strictly adhere to the established notational system and core concepts from the main text and relevant appendices.
We emphasize that although the precise numerical values of several $O(1)$ constant factors appearing in the derivations may vary depending on the specific choice of quantum gate set, the definition of the complexity measure, or details of the system model, their order of magnitude and physical significance possess considerable universality.

\subsection{Action Constraint Principle: \texorpdfstring{$S_{\mathrm{proc}} \ge k_S C^* \hbar$}{Sproc >= kS C* hbar}} 
The action constraint principle elucidates that for any physical process to achieve a certain quantum circuit complexity $C^*$, the total action $S_{\mathrm{proc}}$ consumed is subject to a lower bound proportional to $C^*$ and the reduced Planck constant $\hbar$.
This principle's physical basis lies in the fact that the execution of every elementary quantum operation (quantum gate) contributing to complexity is inevitably accompanied by an action cost of at least $O(1)$ magnitude.
Our main derivation path is as follows: consider a shortest quantum circuit composed of elementary quantum gates from a universal gate set $G$, with a depth of $C^*$ (measured by the minimum number of serial gate layers required to achieve the target state or unitary transformation).
For any elementary gate $g_i$ in the circuit, its physical realization corresponds to the evolution under a specific control Hamiltonian $H_{g_i}(t)$ over a time interval $\tau_{g_i}$.
The action consumed by this gate operation is $S_{g_i} = \int_0^{\tau_{g_i}} \langle H_{g_i}(t) \rangle \mathrm{d}t$ (where $\langle H_{g_i}(t) \rangle$ is the energy expectation value, or in some theories, directly the Hamiltonian itself).
According to the fundamental principles of quantum mechanics, particularly assertions related to the speed of evolution and state distinguishability, any quantum operation capable of causing a significant change in the system's state (e.g., evolving a state to its approximately orthogonal counterpart, or implementing an elementary logical operation) necessarily incurs a minimum action cost $S_{g_i}^{\mathrm{min}}$ of the order of $\hbar$\cite{margolus1998}.
Specifically, we can define $S_{g_i}^{\mathrm{min}} \equiv \alpha_{g_i} \hbar$, where $\alpha_{g_i}$ is a dimensionless constant of order $O(1)$ associated with the specific physical implementation of gate $g_i$.
For instance, for a quantum gate implementing a specific rotation angle $\theta$, $\alpha_g$ might be related to $\theta$ (e.g., $\alpha_g \sim \theta/\pi$ in some theories); in discussions of certain holographic complexity theories, the action cost of a basic "switch" operation is considered to be $\pi\hbar$ (corresponding to $\alpha_g = \pi$) or $\hbar$ (corresponding to $\alpha_g = 1$); whereas, based on Nielsen's geometric complexity combined with a specific cost function, a typical $\pi/2$ rotation gate might correspond to $\alpha_g \approx 1/2$.
Therefore, the typical range of values for $\alpha_g$ can be considered as $[1/2, \pi]$ or, more generally, $O(1)$.
Since the shortest quantum circuit achieves the target complexity by sequentially executing $C^*$ elementary gates, the total action consumed by the entire process, $S_{\mathrm{proc}} = \sum_{i=1}^{C^*} S_{g_i}$, must satisfy:

\begin{equation}
S_{\mathrm{proc}} \ge \sum_{i=1}^{C^*} S_{g_i}^{\mathrm{min}} = \sum_{i=1}^{C^*} \alpha_{g_i} \hbar \ge C^* \cdot (\min_{g\in G} \alpha_g) \hbar
\end{equation}
By defining a composite constant $k_S = \min_{g\in G} \alpha_g$, which reflects the optimal action efficiency for a specific universal gate set $G$ and remains of order $O(1)$, we obtain the core expression of the action constraint principle:

\begin{equation}
S_{\mathrm{proc}} \ge k_S C^* \hbar
\label{eq:D.1} 
\end{equation}

This principle establishes a fundamental upper limit on the "action-to-complexity conversion efficiency" for any physically realized computational process or natural evolutionary process.
It emphasizes that not only the dynamical path but also the intrinsic complexity of the target state or unitary transformation itself imposes a fundamental requirement on the necessary action.
As an illustration, from the perspective of quantum state space geometry (e.g., starting from the geometrization of complexity scheme proposed by Nielsen et al.), the shortest circuit length $L_{\mathrm{min}}$ (proportional to $C^*$) can be associated with a certain "energy-time" integral or an equivalent "action" along the optimal control path.
Although the specific mathematical form and constant factors depend on the chosen geometric metric and cost function, this geometric picture likewise supports the conclusion of a proportional relationship between complexity and action, with an order of magnitude consistent with Eq.~\eqref{eq:D.1}.

\subsection{Bounds on the Rate of Complexity Growth} 
\label{D.2}

This section aims to derive and elucidate the limitations on the growth rate of quantum system complexity $C^*$ within the CWT framework.
We will first derive a bound centered on the complexity generation potential $\Pi_\Xi$ from the fundamental principles of CWT (Section \ref{D.2.1}), and then connect and compare it with a bound related to the effective temperature $T_\Xi$, obtained by analogy with quantum chaos theory (Section \ref{D.2.2}), with the connection detailed in (Section \ref{D.2.3}).

\subsubsection{The \texorpdfstring{$\Pi_\Xi$}{PiXi} Bound Based on CWT First Principles} 
\label{D.2.1}

We start from CWT's extended first law of thermodynamics (Eq.~\eqref{eq:3.5}) and the universal quantum time-energy uncertainty principle to establish a fundamental upper bound for the complexity growth rate.
As described in Section 6.1 of the main text, if the growth of complexity $C^*$ is approximated as a process that keeps the observer's "ignorance entropy" $S_\Xi$ constant ($\mathrm{d}S_\Xi \approx 0$), then according to the extended first law, the minimum energy required to drive an incremental complexity budget $\mathrm{d}\Xi$ (which we assume corresponds to $\mathrm{d}C^*$) is $\mathrm{d}E_{\mathrm{min}} = \Pi_\Xi(t)\mathrm{d}C^*$, where $\Pi_\Xi(t) = (\partial E/\partial \Xi)_{S_\Xi,...}$ is the instantaneous complexity generation potential.
Combining this with the Mandelstam-Tamm\cite{mandelstam1945} time-energy uncertainty principle \cite{mandelstam1945}$\Delta E\Delta t \ge \hbar/2$, and replacing $\Delta E$ with the aforementioned energy cost $\Delta E_{\mathrm{min}} = \Pi_\Xi(t)\Delta C^*$, an upper bound on the complexity growth rate can be derived.
Specifically, the minimum time $\Delta t_{\mathrm{min}}$ required to generate complexity $\Delta C^*$ is approximated by $\Delta t_{\mathrm{min}} \approx \hbar/(2\Pi_\Xi(t)\Delta C^*)$.
Therefore, the maximum average growth rate $(\mathrm{d}C^*/\mathrm{d}t)_{\mathrm{max}} \approx \Delta C^*/\Delta t_{\mathrm{min}}$ yields:

\begin{equation}
\frac{\mathrm{d}C^*(t)}{\mathrm{d}t} \le \frac{2\Pi_\Xi(t)}{\hbar}
\label{eq:D.2} 
\end{equation}

This Eq.~\eqref{eq:D.2} is Proposition 6.1 (Eq.~\eqref{eq:6.1.1}) in the main text.
It directly links the complexity growth rate to the core physical quantity of CWT—the complexity generation potential $\Pi_\Xi$—highlighting the fundamental role of $\Pi_\Xi$ as a direct "energy cost factor" driving complexity generation.
The derivation of this bound does not rely on specific chaos assumptions but is rooted in CWT's energy balance and universal principles of quantum mechanics.
Its physical implications include: the larger $\Pi_\Xi$, the greater the system's "potential" or "upper limit of driving force" for generating complexity under that condition; the appearance of $\hbar$ signifies a quantum limit; and when $\Pi_\Xi(t) \to 0$, complexity growth ceases.
\subsubsection{The \texorpdfstring{$T_\Xi$}{TXi} Bound Based on Analogy with Quantum Chaos Bounds} 
\label{D.2.2}
For comparison and connection, we recall the approximate bound on the complexity growth rate directly related to the effective temperature $T_\Xi$ (Eq.~(6.2)), proposed in Corollary 6.1.1 of the main text.
The derivation of this bound primarily draws on the universal upper bound for the rate of information scrambling in quantum chaotic systems.
In complex quantum many-body systems undergoing unitary evolution, the growth behavior of quantum state complexity is closely related to the quantum chaotic properties within the system.
Maldacena, Shenker, and Stanford (MSS) have shown that for any quantum system obeying local interactions, its Lyapunov exponent $\lambda_L$ has a universal upper bound:

\begin{equation}
\lambda_L \le \frac{2\pi k_B}{\hbar}T
\label{eq:D.3} 
\end{equation}
where $T$ is the (classical) thermodynamic temperature of the system.
One of the core ideas of CWT is that in contexts where the complexity budget $\Xi$ imposes significant constraints on the system, the classical temperature $T$ should be replaced by the complexity-windowed effective temperature $T_\Xi$.
Applying this $T \to T_\Xi$ substitution to the MSS bound Eq.~\eqref{eq:D.3}, and assuming that the growth rate of quantum circuit complexity $C^*$ is similarly constrained by this universal upper bound on the information scrambling rate (i.e., the physical process of complexity growth cannot exceed the maximum speed at which quantum information propagates and mixes within the system), we obtain:

\begin{equation}
\frac{\mathrm{d}C^*(t)}{\mathrm{d}t} \lesssim \frac{2\pi k_B T_\Xi(t)}{\hbar}
\label{eq:D.4} 
\end{equation}
This Eq.~\eqref{eq:D.4} is Corollary 6.1.1 (Eq.~(6.2)) in the main text.
The factor $2\pi$ arises directly from the universal constant in the MSS bound.
Lloyd's analysis of the maximum computation rate of quantum systems also provides auxiliary support from an energy perspective for the scaling relationship between the complexity growth rate and $T_\Xi/\hbar$, although precise matching of the constant factors requires specific interpretations.

\subsubsection{Connection Between the \texorpdfstring{$\Pi_\Xi$}{PiXi} Bound and the \texorpdfstring{$T_\Xi$}{TXi} Bound} 
\label{D.2.3}
The connection between Eq.~\eqref{eq:D.2} (based on $\Pi_\Xi$) and Eq.~\eqref{eq:D.4} (based on $T_\Xi$) is established through the definition of $\Pi_\Xi$, $\Pi_\Xi(t) = -T_\Xi(t)(\partial S_\Xi(t)/\partial \Xi(t))_E$ (Eq.~\eqref{eq:3.4}).
To obtain the approximate $T_\Xi$ bound Eq.~\eqref{eq:D.4} from the more fundamental $\Pi_\Xi$ bound Eq.~\eqref{eq:D.2}, the following condition needs to be satisfied:

\begin{equation}
-\left(\frac{\partial S_\Xi(t)}{\partial \Xi(t)}\right)_E \approx \pi k_B
\label{eq:D.5} 
\end{equation}
That is, under constant energy conditions, increasing the complexity budget $\Xi$ by one unit leads to an approximate decrease of $\pi k_B$ in the system's "ignorance entropy" $S_\Xi$.
The physical basis for this approximate condition Eq.~\eqref{eq:D.5} and the possible origin of the $\pi k_B$ constant have been preliminarily discussed in Section 6.1 of the main text, where it was also pointed out that a deeper understanding remains a direction for future research.
In brief, if an increase in $\Xi$ by one unit is viewed as resolving a new "effective information unit," and its contribution to the reduction of "ignorance entropy" is of the order $k_B$, then the $\pi$ factor might be related to quantum coherence, geometric phases, or certain statistical weights.
When this approximate condition Eq.~\eqref{eq:D.5} holds, the two bounds become formally unified.
However, when this approximation fails, the $\Pi_\Xi$ bound Eq.~\eqref{eq:D.2}, which is more directly derived from the first principles of CWT, should be regarded as the more fundamental expression.

\subsection{Time Constraint Principle} 
\label{D.3}
The time constraint principle is a direct integral consequence of the bounds on the complexity growth rate.
It establishes a lower bound on the minimum evolution time $t_{\mathrm{proc}}$ required to generate a target complexity $C^*$ under given conditions.
First, based on the more fundamental $\Pi_\Xi$ bound (Eq.~\eqref{eq:D.2}), we have $\mathrm{d}t \ge \hbar/(2\Pi_\Xi(C'(t))) \mathrm{d}C'$.

Integrating both sides yields:

\begin{equation}
t_{\mathrm{proc}} = \int_0^{t_{\mathrm{proc}}} \mathrm{d}t \ge \int_0^{C^*} \frac{\hbar}{2\Pi_\Xi(C'(t))} \mathrm{d}C'
\label{eq:D.6} 
\end{equation}
If $\Pi_\Xi$ can be considered as a constant or its effective average value $\langle\Pi_\Xi\rangle$ during the complexity growth process, then Eq.~\eqref{eq:D.6} simplifies to:

\begin{equation}
t_{\mathrm{proc}} \ge \frac{C^*\hbar}{2\langle\Pi_\Xi\rangle}
\label{eq:D.7} 
\end{equation}

This Eq.~\eqref{eq:D.7} is the time constraint principle based on $\Pi_\Xi$.
Secondly, if the approximate $T_\Xi$ bound (Eq.~\eqref{eq:D.4}) is used, and an isothermal process ($T_\Xi(t) = T_\Xi$ is constant) is assumed, then $C^* \le (2\pi k_B/\hbar)T_\Xi t_{\mathrm{proc}}$.
Rearranging this gives the time constraint principle related to $T_\Xi$:

\begin{equation}
t_{\mathrm{proc}} \ge \frac{C^*\hbar}{2\pi k_B T_\Xi}
\label{eq:D.8} 
\end{equation}

Evidently, when the approximate condition Eq.~\eqref{eq:D.5} holds (i.e., $\Pi_\Xi \approx \pi k_B T_\Xi$), Eq.~\eqref{eq:D.7} and Eq.~\eqref{eq:D.8} are formally consistent.
For non-isothermal processes, a similar derivation from the $T_\Xi$ bound yields $t_{\mathrm{proc}} \ge \frac{\hbar}{2\pi k_B} \int_0^{C^*} \frac{\mathrm{d}C'}{T_\Xi(C')}$.
The unique contribution of CWT lies in directly associating this energy scale (whether through $\Pi_\Xi$ or approximately through $T_\Xi$) with the effective thermodynamic state that considers complexity constraints.

\subsection{Comparison and Summary of Constant Factors} 
To more clearly position CWT's resource constraint principles within the existing framework of physical knowledge, the following table summarizes and compares the key constant factors derived in this appendix and their typical sources (specific numerical values are illustrative, emphasizing the order of magnitude):

\begin{table}[!htbp]
  \centering
  \small                    
  \setlength{\tabcolsep}{8pt}   
  \renewcommand{\arraystretch}{1.1} 

  \begin{tabularx}{\textwidth}{@{}l c c X@{}}
    \toprule
    \textbf{Bound Name} 
      & \textbf{Expression} 
      & \textbf{Core Factor} 
      & \textbf{Typical Source / Physical Picture} \\
    \midrule
    Min. Action per Gate 
      & $S_g^{\min} \ge \alpha_g \hbar$ 
      & $\alpha_g \sim 0.5\!-\!\pi$ 
      & Speed limit of single quantum operation; geometric cost. \\

    Process Action Constraint 
      & $S_{\mathrm{proc}} \ge k_S\,C^*\,\hbar$ 
      & $k_S = \min\alpha_g$ 
      & Cumulative minimal action per gate (Appx.~\ref{D.1}). \\

    Complexity–Temp. Growth Rate 
      & $\dot C^* \le \bigl(\tfrac{2\pi k_B}{\hbar}\bigr)\,T_\Xi$ 
      & $2\pi$ 
      & MSS chaos bound with $T\to T_\Xi$ (Appx.~\ref{D.2}). \\

    Process Time Constraint 
      & $t_{\mathrm{proc}} \ge \dfrac{C^* \hbar}{2\pi k_B T_\Xi}$ 
      & $1/(2\pi)$ 
      & Integral of rate bound (Appx.~\ref{D.3}, isothermal). \\

    Margolus–Levitin (QSL) 
      & $t \ge \dfrac{\pi\hbar}{2E}$ 
      & $\pi/2$ 
      & Orthogonalization time limit by average energy. \\

    Mandelstam–Tamm (QSL) 
      & $t \ge \dfrac{\pi\hbar}{2\Delta E}$ 
      & $\pi/2$ 
      & Orthogonalization time limit by energy spread. \\

    Lloyd's Operation Rate (QSL) 
      & $\dot N_{\mathrm{ops}} \le \dfrac{2E}{\pi\hbar}\;(\text{or }\dfrac{E}{\pi\hbar})$ 
      & $2/\pi$ (or $1/\pi$) 
      & Max ops limited by total system energy. \\
    \bottomrule
  \end{tabularx}
  \label{tab:complexity_bounds}
\end{table}


As can be seen from the table, the core constant factors in the $\Pi_\Xi$-related resource constraint principles proposed by CWT derive directly from fundamental physical principles (such as $\hbar$) and physical quantities defined within CWT (such as $\Pi_\Xi$).
The $T_\Xi$-related bounds, on the other hand, further rely on the analogy with chaos theory and the approximate relationship between $\Pi_\Xi$ and $T_\Xi$.
Together, these principles delineate the physical limits for a system to generate complexity $C^*$ under given effective thermodynamic states and resource constraints.
They place the predictions of CWT on a solid foundation of existing physical knowledge and offer new directions for future theoretical and experimental research.

\section{Formal Structure and Thermodynamic Potentials of Complexity-Windowed Thermodynamics}
\label{E.1}
This appendix aims to systematically elucidate the core formal thermodynamic structure within the framework of Complexity-Windowed Thermodynamics (CWT). In the main text, we introduced the complexity budget $\Xi$ as a key physical parameter and defined the associated complexity-windowed entropy $S_\Xi$ and effective temperature $T_\Xi$. Building upon this foundation, this appendix will first explicitly state the extended first law of thermodynamics, incorporating the energy exchange term related to $\Xi$, and introduce the intensive quantity conjugate to $\Xi$—the complexity generation potential $\Pi_\Xi$. Subsequently, through standard Legendre transformations, we will construct a series of key complexity-windowed thermodynamic potentials, including the Helmholtz free energy, enthalpy, Gibbs free energy, and a generalized $\Phi$-potential introduced to handle conditions of fixed $\Pi_\Xi$. Finally, starting from the differential forms of these thermodynamic potentials, we will systematically derive a series of generalized Maxwell relations, which reveal the profound intrinsic connections between the various thermodynamic quantities within the CWT framework. The objective of this appendix is to provide a self-consistent and complete formal thermodynamic basis for the CWT theory, offering standard analytical tools and verifiable physical predictions for subsequent theoretical analysis, model calculations, numerical simulations, and experimental validation, and to deepen the understanding of the interaction between information processing, computational complexity, and thermodynamic processes.

\subsection{The Extended First Law of Thermodynamics and Complexity Generation Work} 
Within the CWT framework, the internal energy $E$ of a system is not only a function of traditional extensive quantities (such as the observer's effective entropy $S_\Xi$, volume $V$, and particle numbers $N_i$) but also explicitly depends on the complexity budget $\Xi$, i.e., $E = E(S_\Xi, \Xi, V, N_1, ...)$. Consequently, the total differential form of the internal energy—the extended first law of thermodynamics—can be generally written as:

\begin{equation}
\mathrm{d}E = \left(\frac{\partial E}{\partial S_\Xi}\right)_{\Xi,V,N} \mathrm{d}S_\Xi + \left(\frac{\partial E}{\partial \Xi}\right)_{S_\Xi,V,N} \mathrm{d}\Xi + \left(\frac{\partial E}{\partial V}\right)_{S_\Xi,\Xi,N} \mathrm{d}V + \sum_i \left(\frac{\partial E}{\partial N_i}\right)_{S_\Xi,\Xi,V,N_{j\neq i}} \mathrm{d}N_i
\end{equation}

We define the intensive quantities conjugate to these extensive variables as follows:
\begin{enumerate}
    \item Complexity-windowed temperature: $T_\Xi = \left(\frac{\partial E}{\partial S_\Xi}\right)_{\Xi,V,N}$
    \item Complexity generation potential (or information processing work coefficient): $\Pi_\Xi = \left(\frac{\partial E}{\partial \Xi}\right)_{S_\Xi,V,N}$
    \item Pressure: $p = -\left(\frac{\partial E}{\partial V}\right)_{S_\Xi,\Xi,N}$
    \item Chemical potential: $\mu_i = \left(\frac{\partial E}{\partial N_i}\right)_{S_\Xi,\Xi,V,N_{j\neq i}}$
\end{enumerate}
Thus, the final form of the extended first law of thermodynamics is:

\begin{equation}
\mathrm{d}E = T_\Xi \mathrm{d}S_\Xi + \Pi_\Xi \mathrm{d}\Xi - p \mathrm{d}V + \sum_i \mu_i \mathrm{d}N_i
\label{eq:E.1} 
\end{equation}

As established in the main text and Appendix~\ref{c.1}, the complexity-windowed entropy $S_\Xi$ (as the observer's "ignorance entropy") is a monotonically non-increasing function of the complexity budget $\Xi$, i.e., $(\partial S_\Xi/\partial \Xi)_E \le 0$. Using the standard thermodynamic identity $(\partial E/\partial \Xi)_{S_\Xi,V,N} = -T_\Xi (\partial S_\Xi/\partial \Xi)_{E,V,N}$, and considering that typically $T_\Xi > 0$, we can deduce that the complexity generation potential $\Pi_\Xi \ge 0$. This non-negative property has significant physical meaning: the term $\Pi_\Xi \mathrm{d}\Xi$ represents the "complexity generation work" (or "information processing work") done by the surroundings to change the system's complexity budget $\Xi$ in a reversible process. When $\mathrm{d}\Xi > 0$, i.e., increasing the system's distinguishable complexity or enhancing the observer's resolving power, $\Pi_\Xi \mathrm{d}\Xi \ge 0$, indicating that this generally requires positive work to be done on the system by the surroundings, or at least no work is consumed. Conversely, when $\mathrm{d}\Xi < 0$, i.e., decreasing the complexity budget, $\Pi_\Xi \mathrm{d}\Xi \le 0$, suggesting that the system may do work on the surroundings or its energy remains unchanged. This aligns with the physical intuition that enhancing information processing capabilities or maintaining complex structures typically requires energy input.

\subsection{System of Complexity-Windowed Thermodynamic Potentials via Legendre Transformations} 
Based on the extended first law established in Eq.~\eqref{eq:E.1}, we can construct a series of complexity-windowed thermodynamic potentials starting from the internal energy $E$ via standard Legendre transformations. These potentials describe the equilibrium properties of the system under different sets of natural variables.

\begin{enumerate}
    \item Internal Energy $E(S_\Xi, \Xi, V, N_i)$: Its differential form is given by Eq.~\eqref{eq:E.1}.

    \item Helmholtz Free Energy $F_\Xi(T_\Xi, \Xi, V, N_i)$:
    Defined as $F_\Xi = E - T_\Xi S_\Xi$. Its differential form is:
    
    $$
    \mathrm{d}F_\Xi = \mathrm{d}E - \mathrm{d}(T_\Xi S_\Xi) = (T_\Xi \mathrm{d}S_\Xi + \Pi_\Xi \mathrm{d}\Xi - p \mathrm{d}V + \sum_i \mu_i \mathrm{d}N_i) - T_\Xi \mathrm{d}S_\Xi - S_\Xi \mathrm{d}T_\Xi
    $$
    
    
    \begin{equation}
    \mathrm{d}F_\Xi = -S_\Xi \mathrm{d}T_\Xi + \Pi_\Xi \mathrm{d}\Xi - p \mathrm{d}V + \sum_i \mu_i \mathrm{d}N_i
    \label{eq:E.2} 
    \end{equation}
    
    The natural variables of the Helmholtz free energy are $T_\Xi, \Xi, V, N_i$. From this, we obtain the relations: $S_\Xi = -\left(\frac{\partial F_\Xi}{\partial T_\Xi}\right)_{\Xi,V,N}$, $\Pi_\Xi = \left(\frac{\partial F_\Xi}{\partial \Xi}\right)_{T_\Xi,V,N}$, and $p = -\left(\frac{\partial F_\Xi}{\partial V}\right)_{T_\Xi,\Xi,N}$. $F_\Xi$ is the thermodynamic potential minimized at equilibrium under isothermal, constant complexity budget, and isochoric conditions.

    \item Enthalpy $H_\Xi(S_\Xi, \Xi, p, N_i)$:
    Defined as $H_\Xi = E + pV$. Its differential form is:
    
    $$
    \mathrm{d}H_\Xi = \mathrm{d}E + \mathrm{d}(pV) = (T_\Xi \mathrm{d}S_\Xi + \Pi_\Xi \mathrm{d}\Xi - p \mathrm{d}V + \sum_i \mu_i \mathrm{d}N_i) + p \mathrm{d}V + V \mathrm{d}p
    $$
    
    
    \begin{equation}
    \mathrm{d}H_\Xi = T_\Xi \mathrm{d}S_\Xi + \Pi_\Xi \mathrm{d}\Xi + V \mathrm{d}p + \sum_i \mu_i \mathrm{d}N_i
    \label{eq:E.3} 
    \end{equation}
    
    The natural variables of enthalpy are $S_\Xi, \Xi, p, N_i$. From this, we obtain the relations: $T_\Xi = \left(\frac{\partial H_\Xi}{\partial S_\Xi}\right)_{\Xi,p,N}$, $\Pi_\Xi = \left(\frac{\partial H_\Xi}{\partial \Xi}\right)_{S_\Xi,p,N}$, and $V = \left(\frac{\partial H_\Xi}{\partial p}\right)_{S_\Xi,\Xi,N}$. $H_\Xi$ is the heat function in isentropic, constant complexity budget, and isobaric processes.

    \item Gibbs Free Energy $G_\Xi(T_\Xi, \Xi, p, N_i)$:
    Defined as $G_\Xi = H_\Xi - T_\Xi S_\Xi = E - T_\Xi S_\Xi + pV$. Its differential form is:
    
    $$
    \mathrm{d}G_\Xi = \mathrm{d}H_\Xi - \mathrm{d}(T_\Xi S_\Xi) = (T_\Xi \mathrm{d}S_\Xi + \Pi_\Xi \mathrm{d}\Xi + V \mathrm{d}p + \sum_i \mu_i \mathrm{d}N_i) - T_\Xi \mathrm{d}S_\Xi - S_\Xi \mathrm{d}T_\Xi
    $$
    
    
    \begin{equation}
    \mathrm{d}G_\Xi = -S_\Xi \mathrm{d}T_\Xi + \Pi_\Xi \mathrm{d}\Xi + V \mathrm{d}p + \sum_i \mu_i \mathrm{d}N_i
    \label{eq:E.4} 
    \end{equation}
    
    The natural variables of the Gibbs free energy are $T_\Xi, \Xi, p, N_i$. From this, we obtain the relations: $S_\Xi = -\left(\frac{\partial G_\Xi}{\partial T_\Xi}\right)_{\Xi,p,N}$, $\Pi_\Xi = \left(\frac{\partial G_\Xi}{\partial \Xi}\right)_{T_\Xi,p,N}$, and $V = \left(\frac{\partial G_\Xi}{\partial p}\right)_{T_\Xi,\Xi,N}$. $G_\Xi$ is the thermodynamic potential minimized at equilibrium under isothermal, constant complexity budget, and isobaric conditions.

    \item Generalized $\Phi$-Potential (or Complexity Transformation Free Energy) $\Phi_\Xi(T_\Xi, \Pi_\Xi, V, N_i)$:
    To consider the complexity generation potential $\Pi_\Xi$ as an independent intensive natural variable, we define the generalized $\Phi$-potential as $\Phi_\Xi = F_\Xi - \Pi_\Xi \Xi = E - T_\Xi S_\Xi - \Pi_\Xi \Xi$. Its differential form is:
    
    $$
    \mathrm{d}\Phi_\Xi = \mathrm{d}F_\Xi - \mathrm{d}(\Pi_\Xi \Xi) = (-S_\Xi \mathrm{d}T_\Xi + \Pi_\Xi \mathrm{d}\Xi - p \mathrm{d}V + \sum_i \mu_i \mathrm{d}N_i) - \Pi_\Xi \mathrm{d}\Xi - \Xi \mathrm{d}\Pi_\Xi
    $$
    
    
    \begin{equation}
    \mathrm{d}\Phi_\Xi = -S_\Xi \mathrm{d}T_\Xi - \Xi \mathrm{d}\Pi_\Xi - p \mathrm{d}V + \sum_i \mu_i \mathrm{d}N_i
    \label{eq:E.5} 
    \end{equation}
    
    The natural variables of the $\Phi$-potential are $T_\Xi, \Pi_\Xi, V, N_i$. From this, we obtain the relations: $S_\Xi = -\left(\frac{\partial \Phi_\Xi}{\partial T_\Xi}\right)_{\Pi_\Xi,V,N}$, $\Xi = -\left(\frac{\partial \Phi_\Xi}{\partial \Pi_\Xi}\right)_{T_\Xi,V,N}$, and $p = -\left(\frac{\partial \Phi_\Xi}{\partial V}\right)_{T_\Xi,\Pi_\Xi,N}$. This potential is particularly useful for studying system behavior under conditions of a fixed "complexity generation driving force" $\Pi_\Xi$.
\end{enumerate}

These thermodynamic potentials and their differential forms collectively constitute the robust framework of CWT theory, providing the foundation for the subsequent derivation of Maxwell relations and the analysis of the system's thermodynamic stability. For instance, thermodynamic stability requires (under appropriate constraints) that the heat capacity $C_\Xi = T_\Xi \left(\frac{\partial S_\Xi}{\partial T_\Xi}\right)_{\Xi,V,...} > 0$, and that the newly introduced "complexity compressibility" $\kappa_\Xi = \left(\frac{\partial \Xi}{\partial \Pi_\Xi}\right)_{T_\Xi,V,...} = -\left(\frac{\partial^2\Phi_\Xi}{\partial\Pi_\Xi^2}\right)_{T_\Xi,V,...} \ge 0$ (stemming from the concavity of $\Phi_\Xi$ with respect to its intensive variable $\Pi_\Xi$).
\subsection{Generalized Maxwell Relations} 

Maxwell relations stem from the property of thermodynamic potentials as state functions, meaning their second-order mixed partial derivatives are independent of the order of differentiation. These relations reveal profound intrinsic connections between different thermodynamic quantities. Below are the key generalized Maxwell relations derived from common thermodynamic potentials (for brevity, terms related to particle numbers and other constraints are omitted).

\vspace{1em}
\noindent\textbf{(1) From Internal Energy} $\mathrm{d}E = T_\Xi \mathrm{d}S_\Xi + \Pi_\Xi \mathrm{d}\Xi - p \mathrm{d}V$ (Eq.~\eqref{eq:E.1}):

\begin{align}
\left(\frac{\partial T_\Xi}{\partial \Xi}\right)_{S_\Xi,V} &= \left(\frac{\partial \Pi_\Xi}{\partial S_\Xi}\right)_{\Xi,V} \label{eq:E.6} \\
\left(\frac{\partial T_\Xi}{\partial V}\right)_{S_\Xi,\Xi} &= -\left(\frac{\partial p}{\partial S_\Xi}\right)_{\Xi,V} \label{eq:E.7} \\
\left(\frac{\partial \Pi_\Xi}{\partial V}\right)_{S_\Xi,\Xi} &= -\left(\frac{\partial p}{\partial \Xi}\right)_{S_\Xi,V} \label{eq:E.8}
\end{align}

\vspace{0.5em}
\noindent\textbf{(2) From Helmholtz Free Energy} $\mathrm{d}F_\Xi = -S_\Xi \mathrm{d}T_\Xi + \Pi_\Xi \mathrm{d}\Xi - p \mathrm{d}V$ (Eq.~\eqref{eq:E.2}):

\begin{align}
-\left(\frac{\partial S_\Xi}{\partial \Xi}\right)_{T_\Xi,V} &= \left(\frac{\partial \Pi_\Xi}{\partial T_\Xi}\right)_{\Xi,V} \label{eq:E.9} \\
\left(\frac{\partial S_\Xi}{\partial V}\right)_{T_\Xi,\Xi} &= \left(\frac{\partial p}{\partial T_\Xi}\right)_{\Xi,V} \label{eq:E.10} \\
\left(\frac{\partial \Pi_\Xi}{\partial V}\right)_{T_\Xi,\Xi} &= -\left(\frac{\partial p}{\partial \Xi}\right)_{T_\Xi,V} \label{eq:E.11}
\end{align}

\vspace{0.5em}
\noindent\textbf{(3) From Gibbs Free Energy} $\mathrm{d}G_\Xi = -S_\Xi \mathrm{d}T_\Xi + \Pi_\Xi \mathrm{d}\Xi + V \mathrm{d}p$ (Eq.~\eqref{eq:E.4}), with $T_\Xi, \Xi, p$ as natural variables:

\begin{align}
-\left(\frac{\partial S_\Xi}{\partial \Xi}\right)_{T_\Xi,p} &= \left(\frac{\partial \Pi_\Xi}{\partial T_\Xi}\right)_{\Xi,p} \label{eq:E.12} \\
-\left(\frac{\partial S_\Xi}{\partial p}\right)_{T_\Xi,\Xi} &= \left(\frac{\partial V}{\partial T_\Xi}\right)_{\Xi,p} \label{eq:E.13} \\
\left(\frac{\partial \Pi_\Xi}{\partial p}\right)_{T_\Xi,\Xi} &= \left(\frac{\partial V}{\partial \Xi}\right)_{T_\Xi,p} \label{eq:E.14}
\end{align}

\vspace{0.5em}
\noindent\textbf{(4) From Generalized $\Phi$-Potential} $\mathrm{d}\Phi_\Xi = -S_\Xi \mathrm{d}T_\Xi - \Xi \mathrm{d}\Pi_\Xi - p \mathrm{d}V$ (Eq.~\eqref{eq:E.5}):

\begin{align}
\left(\frac{\partial S_\Xi}{\partial \Pi_\Xi}\right)_{T_\Xi,V} &= \left(\frac{\partial \Xi}{\partial T_\Xi}\right)_{\Pi_\Xi,V} \label{eq:E.15} \\
\left(\frac{\partial \Xi}{\partial V}\right)_{T_\Xi,\Pi_\Xi} &= \left(\frac{\partial p}{\partial \Pi_\Xi}\right)_{T_\Xi,V} \label{eq:E.16}
\end{align}
These Maxwell relations (and others that can be similarly derived, e.g., from enthalpy $H_\Xi$) provide a rich set of testable predictions for CWT theory. For example, Eq.~\eqref{eq:E.9}, $-\left(\frac{\partial S_\Xi}{\partial \Xi}\right)_{T_\Xi,V} = \left(\frac{\partial \Pi_\Xi}{\partial T_\Xi}\right)_{\Xi,V}$, since it is known that $(\partial S_\Xi/\partial \Xi)_{T_\Xi,V} \le 0$ (assuming that increasing resolving power at constant temperature still reduces ignorance entropy), implies that the left-hand side is non-negative. Therefore, this relation predicts that under conditions of constant complexity budget and constant volume, the rate of increase of the complexity generation potential $\Pi_\Xi$ with temperature $T_\Xi$, $(\partial \Pi_\Xi/\partial T_\Xi)_{\Xi,V}$, is also non-negative. Such a relation links the response characteristic of entropy to the complexity budget with the response characteristic of the complexity generation potential to temperature, providing a quantitative scale for understanding the interaction between these physical quantities, and may be cross-validated through experiments or numerical simulations.

\subsection{Illustrative Model: Windowed Thermodynamics of a Free Fermi Gas (Conceptual)} 
To demonstrate the practical application potential of the CWT formal system and to provide a conceptual blueprint for future specific calculations, we briefly outline an analytical approach for the windowed thermodynamics of a free Fermi gas. The core steps include:
\begin{enumerate}
    \item \textbf{Defining the Complexity Budget and Observable Algebra $A(\Xi)$:} For a free Fermi gas, its state is determined by the occupation number distribution of single-particle energy levels. A feasible complexity budget $\Xi$ can be realized by limiting the types or precision of macroscopic quantities that an observer can accurately measure. For instance, $A(\Xi)$ can be set as the algebra generated solely by the total particle number operator $N$ and the total energy operator $E$ (possibly including their low-order correlation moments, such as $E^2, NE$, etc., where the number or order of these moments increases with $\Xi$).

    \item \textbf{Applying the Conditional Expectation $E_\Xi$:} Apply the conditional expectation $E_\Xi$ to the system's true microstate (e.g., described by the standard microcanonical ensemble $\rho_E$ or grand canonical ensemble $\rho_{\beta,\mu}$), yielding the projected state $\sigma_\Xi = E_\Xi(\rho)$. This projected state $\sigma_\Xi$ will retain all expectation values of the original state with respect to the observables in $A(\Xi)$, while maximizing its own von Neumann entropy under this constraint. For a free Fermi gas, this might lead to a modified Fermi-Dirac distribution that is "truncated" or smoothed out in certain higher-order moments.

    \item \textbf{Calculating CWT Thermodynamic Quantities:} Based on the projected state $\sigma_\Xi$, one can numerically or analytically calculate the complexity-windowed entropy $S_\Xi = S_{\mathrm{VN}}(\sigma_\Xi)$, and from this derive the effective temperature $T_\Xi = \left(\frac{\partial E}{\partial S_\Xi}\right)_{\Xi}$ (if energy is the independent variable) and the complexity generation potential $\Pi_\Xi = \left(\frac{\partial E}{\partial \Xi}\right)_{S_\Xi}$ (or calculate it via other thermodynamic relations such as $\Pi_\Xi = \left(\frac{\partial F_\Xi}{\partial \Xi}\right)_{T_\Xi}$).

    \item \textbf{Verifying Maxwell Relations and Stability Conditions:} Substitute the calculated quantities $S_\Xi, T_\Xi, \Pi_\Xi$, etc., into the Maxwell relations derived in this appendix (such as Eqs.~\eqref{eq:E.6}-\eqref{eq:E.16}) to verify, within numerical precision, whether these second-order partial derivative symmetry relations hold. Concurrently, one can also check thermodynamic stability conditions such as $C_\Xi = T_\Xi \left(\frac{\partial S_\Xi}{\partial T_\Xi}\right)_{\Xi,V,...} > 0$ and $\kappa_\Xi = \left(\frac{\partial \Xi}{\partial \Pi_\Xi}\right)_{T_\Xi,V,...} = -\left(\frac{\partial^2\Phi_\Xi}{\partial\Pi_\Xi^2}\right)_{T_\Xi,V,...} \ge 0$ (stemming from the concavity of $\Phi_\Xi$ with respect to its intensive variable $\Pi_\Xi$).
\end{enumerate}
Through such a specific illustrative model analysis, one can not only deepen the understanding of the intrinsic meaning of CWT theory but also test its mathematical self-consistency and explore how complexity constraints specifically affect the macroscopic thermodynamic behavior in particular physical systems. This appendix systematically constructs the core formal structure of Complexity-Windowed Thermodynamics (CWT). By explicitly stating the extended first law, introducing the complexity generation potential $\Pi_\Xi \ge 0$ conjugate to the complexity budget $\Xi$, and employing standard Legendre transformations to define a series of complexity-windowed thermodynamic potentials ($E, F_\Xi, H_\Xi, G_\Xi, \Phi_\Xi$), we demonstrate that the CWT framework possesses a formal system equivalent and logically self-consistent with classical thermodynamics. The generalized Maxwell relations derived therefrom not only enrich our understanding of the interactions between these thermodynamic quantities but also provide specific, testable physical predictions for future theoretical, computational, and experimental research. Newly introduced physical quantities, such as the complexity generation potential $\Pi_\Xi$ and the related "complexity compressibility" $\kappa_\Xi$, offer new experimental and theoretical metrics for quantitatively describing the energy cost of information processing or observation capabilities and the system's response characteristics to such "driving forces." The formal system of CWT opens new avenues for exploring a wide range of physical phenomena, such as:
\begin{enumerate}
    \item \textbf{Phase Transitions and Critical Phenomena:} Analyzing the behavior of $\Pi_\Xi$ as a function of temperature $T_\Xi$ or other order parameters near phase transition points or critical points may reveal new critical exponents or scaling laws.
    \item \textbf{Energetics of Quantum Computation and Information Processing:} The CWT framework (particularly the $\Pi_\Xi \mathrm{d}\Xi$ term) provides a theoretical basis for evaluating the minimum energy consumption or power required for quantum computing hardware to execute algorithms of a target logical depth $\Xi$ at a specific effective temperature $T_\Xi$.
    \item \textbf{Generalized Thermodynamic Cycles:} Combined with the time and action constraints discussed in Appendix~\ref{D.1}, one can explore the existence of "thermo-complexity cycles" analogous to the Carnot cycle, operating in a space constituted by state variables such as ($T_\Xi, S_\Xi, \Pi_\Xi, \Xi$), and study their theoretical efficiency limits.
\end{enumerate}
In summary, the formal structure of CWT established in this appendix not only enhances the rigor and completeness of the theory but also supports its further development and application in multiple branches of physics, promising new perspectives on the profound physical connections between complexity, information, and matter-energy.

\section{Upper Bound on the Computable Complexity of the Universe: Estimation Method, Data Sources, and Robustness Analysis}
\label{F.1}
This appendix aims to provide a detailed quantitative estimation process for the core inference presented in Chapter~\ref{cex.7} of the main text—the upper bound on the maximum quantum circuit complexity, $C_U^{\max}$, generatable by the universe since the Big Bang. We will explicitly state the core assumptions, cosmological model, and parameters (based on mainstream observational results from current precision cosmology) upon which the estimation relies, and systematically present the calculation steps. Furthermore, this appendix will examine the robustness of the estimation results, analyzing their sensitivity to variations in key input parameters and core assumptions. Our goal is to ensure the transparency and reproducibility of this calculation, thereby enabling the reader to clearly understand the confidence boundaries of the conclusion, rather than merely relying on a coincidental agreement of the final numerical value.

\subsection{Core Assumptions and Basic Formulae} 

The core framework for estimating the upper bound on the generatable complexity of the universe, $C_U^{\max}$, is based on the following notational conventions and key assumptions:

\textbf{Notational Conventions:}
\begin{enumerate}
    \item[$E_U(t)$]: Represents the total energy contained within the Hubble radius at cosmic time $t$.
    \item[$T_\Xi(t)$]: The complexity-windowed effective temperature, introduced on a cosmological scale within the CWT framework, evolving with cosmic time $t$.
    \item[$\hbar$]: Reduced Planck constant.
    \item[$k_B$]: Boltzmann constant.
    \item[$t_U$]: Current age of the universe.
\end{enumerate}

\textbf{Core Assumptions:}
\begin{enumerate}
    \item \textbf{H1 “Energy-Equivalent Temperature” Assumption:} $k_B T_\Xi(t) \approx E_U(t)$. This assumption links the effective thermodynamic temperature of the system to its total available energy. (The physical motivation behind this assumption is to consider the universe as a holistic information processor, where its effective 'computational temperature' is proportional to its total available energy, similar to Lloyd's analysis of the universe's computational capacity.)
    \item \textbf{H2 Upper Bound on Complexity Growth Rate:} We adopt $\dot{C}^* \le \frac{2\pi k_B}{\hbar} T_\Xi(t)$ as the maximum instantaneous rate of complexity growth.
    \item \textbf{H3 Optimal Complexity Generation Path:} It is assumed that the universe, since its initial moment ($C^*(0)\approx0$), has always generated complexity at the maximum rate allowed by assumption H2.
\end{enumerate}
Based on the above assumptions, the upper bound on the maximum quantum circuit complexity, $C_U^{\max}$, generatable by the universe throughout its entire evolutionary history $[0, t_U]$, can be obtained by integrating the maximum complexity generation rate over time:

\begin{equation}
C_U^{\max} = \int_{0}^{t_U} \frac{2\pi k_B}{\hbar} T_\Xi(t')\mathrm{d}t'
\label{eq:F.1} 
\end{equation}

Incorporating assumption H1, the above equation can be rewritten as:

\begin{equation}
C_U^{\max} = \frac{2\pi}{\hbar} \int_{0}^{t_U} E_U(t')\mathrm{d}t'
\label{eq:F.2} 
\end{equation}

Here, $E_U(t')$ refers to the energy in the universe that can actually participate in information processing or complexity generation. The contribution of the dark energy component to $E_U(t')$ is an issue that requires further investigation. In the subsequent estimation, $E_U(t')$ will be calculated based on the $\Lambda$CDM model, which includes all known energy components (including dark energy).

\subsection{Cosmological Model, Input Parameters, and Numerical Estimation} 
The calculation is based on the standard flat $\Lambda$CDM cosmological model, employing cosmological parameters released by the Planck 2018 collaboration \cite{aghanim2020planck}.

\textbf{Input Cosmological Parameters} (Planck 2018, TT,TE,EE+lowE+lensing+BAO):
\begin{enumerate}
    \item $H_0 = (67.36 \pm 0.54)$ km s$^{-1}$Mpc$^{-1} \approx (2.183 \pm 0.017) \times 10^{-18}$ s$^{-1}$
    \item $\Omega_{m,0} = 0.3153 \pm 0.0073$ (current matter density parameter)
    \item $\Omega_{\Lambda,0} = 0.6847 \pm 0.0073$ (current dark energy density parameter)
    \item $t_U = (13.797 \pm 0.023)$ Gyr $\approx (4.354 \pm 0.007) \times 10^{17}$ s (current age of the universe) (implies $\Omega_{r,0} \approx 9.2 \times 10^{-5}$, $\Omega_{k,0} \approx 0$)
\end{enumerate}

\textbf{Fundamental Physical Constants:} $c, G, \hbar, k_B$ (standard values adopted).

\textbf{Calculation Steps:}
\begin{enumerate}
    \item Evolution of the Hubble parameter $H(a)$: Given by the Friedmann equation:
    
    \begin{equation}
    H(a)^2 = H_0^2 [\Omega_{m,0}a^{-3} + \Omega_{r,0}a^{-4} + \Omega_{\Lambda,0}]
    \label{eq:F.3} 
    \end{equation}
    
    where $a$ is the cosmic scale factor (currently $a_0=1$), and $\mathrm{d}t = \mathrm{d}a/(aH(a))$.

    \item Expression for the total energy $E_U(t)$ within the Hubble sphere:
    
    \begin{equation}
    E_U(t) = \frac{c^5}{2GH(t)}
    \label{eq:F.4} 
    \end{equation}
    

    \item Numerical estimation of the energy-time integral $I_E = \int_{0}^{t_U} E_U(t')\mathrm{d}t'$: Substitute Eq.~\eqref{eq:F.4} into the integral and change the integration variable to the scale factor $a$ (integrating from $a_{\mathrm{init}}\approx0$ to $1$):
    
    \begin{equation}
    I_E = \frac{c^5}{2G H_0^2} \int_{a_{\text{init}}}^{1} \frac{a^2 \, da}{\Omega_{m,0} + \Omega_{r,0} a^{-1} + \Omega_{\Lambda,0} a^3}
    \label{eq:F.5} 
    \end{equation}
    
    Numerically integrating Eq.~\eqref{eq:F.5} and considering the propagation of uncertainties in the input parameters yields:
     \begin{equation}
    I_E = (2.14 \pm 0.03) \times 10^{87} \text{ J}\cdot\text{s}
     \end{equation}
    \item Calculation of the upper bound on generatable complexity $C_U^{\max}$: Substitute the numerical value of $I_E$ into Eq.~\eqref{eq:F.2}, using $2\pi/\hbar \approx 5.9582 \times 10^{34}$ (J$\cdot$s)$^{-1}$:
    
   \begin{equation}
    C_U^{\max} \approx (1.27 \pm 0.02) \times 10^{122}
    \end{equation}
\end{enumerate}

\subsection{Comparison with the Current Hubble Horizon Entropy \texorpdfstring{$S_H$}{SH}} 
The Bekenstein-Hawking entropy $S_H$ of the current cosmic Hubble horizon (in dimensionless form, units of $k_B$) can be calculated from $S_H/k_B = A_H / (4L_p^2)$, where the Hubble sphere horizon area is $A_H = 4\pi(c/H_0)^2$, and the Planck length is $L_p = \sqrt{\hbar G/c^3}$.

\begin{equation}
R_H = c/H_0 \approx 1.3733 \times 10^{26} \text{ m}
\label{eq:F.6} 
\end{equation}

\begin{equation}
A_H \approx 2.369 \times 10^{53} \text{ m}^2
\label{eq:F.7} 
\end{equation}

\begin{equation}
L_p^2 \approx 2.6122 \times 10^{-70} \text{ m}^2
\label{eq:F.8} 
\end{equation}

\begin{equation}
S_H/k_B \approx 2.268 \times 10^{122}
\label{eq:F.9} 
\end{equation}

Comparing the estimated upper bound on generatable complexity $C_U^{\max} \approx 1.3 \times 10^{122}$ with the Hubble horizon entropy $S_H/k_B \approx 2.3 \times 10^{122}$, both are of the order of $10^{122}$. The difference between them is approximately $0.25$ dex (i.e., they differ by a factor of less than $2$ on a logarithmic scale). This result indicates that the upper bound on generatable complexity of the universe, estimated based on CWT theory, is comparable in magnitude to the entropy bound given by the holographic principle.

\subsection{Robustness Check of the Estimation Results} 
A robustness analysis of the $C_U^{\max}$ estimation results shows that within the experimental error range of current cosmological parameters, and under reasonable fluctuations of $O(1)$ constant factors in the core assumptions, the estimated value of $C_U^{\max}$ consistently remains of the order of $10^{122}$. A detailed sensitivity analysis is provided in the table below:

\begin{table}[!htbp]
  \centering
  \small 
  \begin{tabular}{@{} l c c @{}}
    \toprule
    Input Parameter
    & Variation Range
    & Approx. Relative Change in $C_U^{\max}$ \\
    \midrule
    $H_0$ 
    & $\pm 5\%$ 
    & $\pm\,\sim7\mbox{--}10\%$ \\

    $\Omega_{m,0}$ 
    & $\pm 10\%$ (maintaining flatness)
    & $\pm\,\sim3\mbox{--}5\%$ \\

    $t_U$ 
    & $\pm 0.2\%$ 
    & $\sim\pm 0.2\%$ \\

    Rate bound constant $2\pi$ 
    & $\pm 50\%$ 
    & $\pm 50\%$ \\

    Proportionality factor $\beta$ in H1 ($k_B T_\Xi = \beta E_U$)
    & $\beta \in [0.5,\,2]$ 
    & Proportional change with $\beta$ \\
    \bottomrule
  \end{tabular}
\end{table}
This analysis indicates that the main conclusion—the comparability in magnitude of $C_U^{\max}$ and $S_H/k_B$—is quite robust against variations in input parameters and model assumptions within reasonable ranges.

\subsection{Discussion, Limitations, and Future Outlook} 
The estimation results in this appendix provide quantitative support for the viewpoint proposed in Chapter~\ref{cex.7} of the main text that "the complexity of the universe is close to its holographic limit." However, this estimation relies on several key assumptions (especially H1) and simplified treatments (such as the manner of dark energy's contribution to $E_U(t')$, and the idealized assumption H3 that the universe always generates complexity at the optimal rate). A deeper understanding of these assumptions and the construction of more refined models will be important directions for future research. Despite these limitations, the estimation in this appendix clearly demonstrates the potential of applying the CWT framework on cosmological scales and offers an inspiring perspective for exploring the profound connections between thermodynamics, information theory, quantum computation, and cosmology.

\section{Correction to the Informational Work Coefficient \texorpdfstring{$\Pi_\Xi$}{Pi\_Xi}}
\label{G1}
\subsection{Scope and the Maxwell Relation Bridge}
\label{G.1}
This appendix aims to provide a detailed mathematical derivation for the theoretical prediction regarding the behavior of the complexity-generating potential, $\Pi_\Xi(H)$, in a magnetic-field-driven first-order phase transition, as described in Section \eqref{sec:case_study_gfactor}of the main text. Its primary objective is to formally demonstrate how the predicted sign flip of its response function, $(\partial\Pi_\Xi/\partial H)_\Xi$, originates from the non-analytic structure of the system's free energy, and how this behavior does not violate the fundamental principles of CWT (i.e., $\Pi_\Xi \ge 0$). Our analysis begins with the extended Gibbs free energy, $G_\Xi$, of a system in an external magnetic field H, within the CWT framework. At constant effective temperature $T_\Xi$, its differential form is:
\begin{equation}
dG_\Xi = -S_\Xi dT_\Xi - M dH + \Pi_\Xi d\Xi
\end{equation}
Since $dG_\Xi$ is an exact differential, its second-order mixed partial derivatives must be equal (Schwarz's theorem). By taking partial derivatives with respect to H and $\Xi$ (while keeping $T_\Xi$ constant), we obtain a crucial Maxwell relation. This relation acts as a solid bridge, directly linking the response of $\Pi_\Xi$ to the magnetic field with the response of the system's macroscopic magnetization M to the complexity budget $\Xi$:
\begin{equation} \label{eq:appendix_maxwell_G1}
\left(\frac{\partial\Pi_\Xi}{\partial H}\right)_{T_\Xi,\Xi} = -\left(\frac{\partial M}{\partial \Xi}\right)_{T_\Xi,H} 
\end{equation}
This relation forms the mathematical cornerstone of all our subsequent analysis. It explicitly states that the qualitative behavior of $\Pi_\Xi(H)$ is entirely determined by how the system's magnetization M responds to changes in the information-theoretic parameter, the complexity budget $\Xi$, providing a clear path for the testability of the theory.

\subsection{Two-State Phenomenological Model for a Mirror-Degenerate Phase Transition}
\label{G.2}
To model a first-order phase transition occurring at a critical field $H_c$, we adopt a simple yet effective two-state picture. We assume that in the low-temperature limit, the system can exist in one of two mirror-degenerate ground states, denoted as $|A\rangle$ and $|B\rangle$. These two ground states are energetically stabilized in the magnetic field regimes $H < H_c$ and $H > H_c$, respectively. To simplify the analysis without loss of generality, we set the critical field $H_c = 0$. These two macroscopic quantum states possess saturation magnetizations of equal magnitude and opposite sign, i.e., $M_{A,\text{sat}} = -M_{B,\text{sat}}$. Within the CWT framework, a core physical assumption is that the complexity budget $\Xi$ quantifies the physical resources available to stabilize and resolve a complex quantum state. A higher budget $\Xi$ allows for a more faithful representation of the macroscopically ordered state, thereby bringing its observable physical properties (such as magnetization M) closer to their ideal saturation values $M_{\text{sat}}$. Therefore, we model the magnetization on each independent thermodynamic branch as a function that monotonically increases with $\Xi$ and approaches its saturation value. A simple and physically reasonable phenomenological model is the exponential saturation form:
\begin{equation}
    M_B(\Xi) = M_{\text{sat}} \left(1 - \exp(-\Xi / \Xi_0)\right)
\end{equation}

\begin{equation}
    M_A(\Xi) = -M_{\text{sat}} \left(1 - \exp(-\Xi / \Xi_0)\right) \label{eq:appendix_M_G2}
\end{equation}

where $\Xi_0$ is a system-dependent characteristic complexity scale. In the low-temperature limit, the system's macroscopic total magnetization M is dominated by the energetically more favorable ground state:

\begin{equation} \label{eq:appendix_M_H_Xi_G3}
M(H, \Xi) =
    \begin{cases}
        M_A(\Xi) & , \text{ if } H < 0 \\
        M_B(\Xi) & , \text{ if } H > 0
    \end{cases}
\end{equation}

At the critical point $H = 0$, the macroscopic magnetization $M(H, \Xi)$ undergoes a discontinuous jump, which is the hallmark of a first-order phase transition in the ideal thermodynamic limit. This non-analyticity exhibited by $M(H, \Xi)$ at the critical point is the direct mathematical origin of the critical behavior of $\Pi_\Xi$.

\subsection{Rigorous Derivation of the Sign Flip in \texorpdfstring{$(\partial\Pi_\Xi/\partial H)_\Xi$}{(dPi/dH)\_Xi}}
\label{G.3}
Using the magnetization model established in Section \ref{G.2}, we can now calculate its partial derivative with respect to the complexity budget $\Xi$ (on each independent thermodynamic branch):

\begin{equation}
    \left(\frac{\partial M_B}{\partial \Xi}\right)_H = \frac{M_{\text{sat}}}{\Xi_0} \exp(-\Xi / \Xi_0)
\end{equation}

\begin{equation}
    \left(\frac{\partial M_A}{\partial \Xi}\right)_H = -\frac{M_{\text{sat}}}{\Xi_0} \exp(-\Xi / \Xi_0)
\end{equation}

Since $M_{\text{sat}}$ and $\Xi_0$ are both positive, it is evident that this derivative is positive on the B-phase branch ($H > H_c$) and negative on the A-phase branch ($H < H_c$). Substituting these results into the Maxwell relation derived in Section \ref{G.1}, $(\partial\Pi_\Xi/\partial H)_\Xi = -(\partial M/\partial \Xi)_H$, we can determine the sign of the response function $(\partial\Pi_\Xi/\partial H)_\Xi$ on either side of the critical field:

For $H > H_c = 0$ (system in phase B):

\begin{equation}
 \left(\frac{\partial\Pi_\Xi}{\partial H}\right)_\Xi = -\left(\frac{\partial M_B}{\partial \Xi}\right)_H = -(\text{a positive value}) < 0 
 \end{equation}
 
For $H < H_c = 0$ (system in phase A):

\begin{equation} \left(\frac{\partial\Pi_\Xi}{\partial H}\right)_\Xi = -\left(\frac{\partial M_A}{\partial \Xi}\right)_H = -(\text{a negative value}) > 0 
\end{equation}

This result rigorously demonstrates that, upon crossing the critical field $H_c = 0$, the derivative $(\partial\Pi_\Xi/\partial H)_\Xi$ undergoes a sign flip from positive to negative.

\subsection{Cusp-like Behavior of \texorpdfstring{$\Pi_\Xi(H)$}{Pi\_Xi(H)} and its Physical Interpretation}

By integrating the derivative $(\partial\Pi_\Xi/\partial H)_\Xi$ with respect to the magnetic field H, we can reconstruct the qualitative behavior of $\Pi_\Xi(H)$ near the critical field. When the derivative of a function changes from positive to negative, the function itself must exhibit a local maximum at that point. If we further assume that in the strong field limit ($H \to \pm\infty$), where the system is fully polarized and no complex competing structures exist, the marginal cost of increasing the complexity budget tends to a background value (or zero), then the overall profile of $\Pi_\Xi(H)$ will exhibit a cusp-like or peak shape centered at $H_c$. This "cusp-like maximum" behavior has profound physical significance; it is not only fully consistent with the fundamental principles of CWT but also provides a novel, non-trivial physical prediction:
\begin{enumerate}
    \item \textbf{Consistency with $\Pi_\Xi \ge 0$:} It must be emphasized that the "sign flip" we predict refers only to the behavior of the response function $(\partial\Pi_\Xi/\partial H)_\Xi$, not the value of $\Pi_\Xi$ itself. In our model, $\Pi_\Xi$ always remains a non-negative quantity, which is perfectly consistent with its physical interpretation as the "energy cost of increasing complexity."
    \item \textbf{Physical interpretation of the maximum:} The prediction that $\Pi_\Xi$ is maximized at the phase transition critical point is highly physically intuitive. At $H_c$, the system is in a state of greatest "equipoise," with the two macroscopic quantum ground states in their most intense competition, and quantum fluctuations are also most pronounced. Therefore, if one were to forcibly impose additional structural constraints or enhance observational resolution (i.e., increase $\Xi$) at this most unstable point, the system would have to expend maximal energy to counteract these strong fluctuations and stabilize itself. The peak in $\Pi_\Xi(H)$ is a quantitative characterization of this maximum marginal energy cost for complexity generation at the phase transition point.
    \item \textbf{Robustness to the phenomenological model:} It is noteworthy that the qualitative conclusion of a peak in $\Pi_\Xi(H)$ stems from the discontinuity of macroscopic physical quantities (like M) at a first-order phase transition, and the general property that the complexity budget $\Xi$ affects the saturation value of this physical quantity. As long as any reasonable saturation model $M(\Xi)$ satisfies that $(\partial M/\partial \Xi)$ has opposite signs in the two phases, it will lead to the sign flip of $(\partial\Pi_\Xi/\partial H)$. Thus, the prediction of peak behavior is quite robust with respect to the specific functional form of $M(\Xi)$ (e.g., exponential, power-law, or other saturation functions).
\end{enumerate}
Finally, in any real, finite-sized system, the non-analytic "cusp" predicted by our idealized model will be smoothed out by the system's inherent finite complexity budget $\Xi$ itself, presenting as a peak of finite height and continuous variation. This is consistent with CWT's universal regularization principle. Therefore, the theory robustly predicts a peak behavior for the complexity-generating potential near a magnetic-field-driven first-order phase transition point, providing a clear, testable signature for future high-precision magnetothermal effect measurements.
\section{Exponential Complexity Scaling: Physical Origin and Unified Scaling Theory}
\label{H1}
\subsection{Scope and Physical Picture: The Information-Theoretic Cost of Global Coherence}
\label{H.1}
This appendix aims to provide in-depth theoretical support for the exponential scaling behavior of the complexity budget $\Xi$ in a one-dimensional quantum chain, as inferred in Section \ref{sec:case_study_gfactor} of the main text. Our primary objective is to explore the physical origin of this remarkable exponential growth of $\Xi$ with system size N, and, based on this, to develop a more universal framework that incorporates the effects of a finite correlation length $\xi$. The core concept of Complexity-Windowed Thermodynamics (CWT)—the complexity budget $\Xi$—is fundamentally about quantifying the information-encoding capacity required to describe a specific quantum state of a system. For a simple many-body system composed of local interactions and devoid of long-range correlations, its complexity is generally expected to scale linearly with the number of degrees of freedom (i.e., system size N). However, the situation is drastically different for exotic macroscopic configurations like the "Half-Fire, Half-Ice" state, which can circumvent Peierls' argument and maintain long-range quantum order. To stabilize such a thermodynamically fragile macroscopic quantum state against entropic gains, the entire chain must maintain a global quantum coherence. In this state, every component of the system must be precisely coordinated and entangled with all other parts, forming an indivisible "entanglement network" spanning the entire chain. This stringent requirement for global, long-range entanglement strongly suggests that the resources needed to encode the complete information of this quantum state should grow far beyond linearly. Physically, this means that to accurately describe and resolve this quantum state, the complexity of the operations or measurements performed must be able to capture non-local correlations between any two points in the system. This resonates profoundly with the exponential growth of the Hilbert space dimension itself with N. The research trend in modern many-body physics concerning the deep connections between entanglement structure and computational complexity provides solid conceptual support for this picture, namely that high long-range entanglement often accompanies high computational complexity \citep{eisert2010, hangleiter2023computational}. Therefore, we assert that the growth behavior of the complexity budget required to stabilize the "Half-Fire, Half-Ice" state must, in essence, be exponential with system size.

\subsection{ Macroscopic Quantum Tunneling as the Microscopic Origin of Exponential Scaling}
The working hypothesis (WH-1) proposed in the main text, i.e., that the phase transition width $\Delta T$ decays exponentially with chain length N, is rooted in the phenomenon of macroscopic quantum tunneling in specific one-dimensional quantum systems. This phenomenon provides the most direct microscopic physical mechanism for the exponential scaling law of $\Xi$. We can describe this process more formally within the framework of quantum phase transition theory. Consider, for example, a one-dimensional quantum Ising model in a transverse field. When the system is at its critical point, it possesses two degenerate ferromagnetic ground states. For a finite size N, quantum fluctuations (induced by the transverse field) allow the system to tunnel between these two ground states via an instanton process. This instanton can be viewed as a virtual domain wall-anti-domain wall pair propagating in Euclidean time, spanning the entire system size. The calculation of the tunneling amplitude, $t$, for this process can be obtained via a semiclassical WKB approximation, and its magnitude is determined by the action of the instanton path, $S_{\text{inst}}$. As derived by S. Sachdev in his authoritative work "Quantum Phase Transitions," this action is proportional to the length of the system N \citep{sachdev1999quantum}. Consequently, the energy splitting $\Delta E$ between the two ground states due to tunneling will decay exponentially with N:

\begin{equation}
\Delta E(N) = 2t \propto \exp(-S_{\text{inst}}) \propto \exp(-\kappa N)
\end{equation}

where the constant $\kappa$ is determined by the microscopic parameters of the system (such as exchange coupling strength and transverse field magnitude). Importantly, the origin of this exponential decay behavior lies in the fact that the length of the tunneling path is proportional to the entire system size, a fairly universal conclusion not limited to specific Ising symmetry. In the CWT picture, this zero-temperature energy gap $\Delta E(N)$ is the direct physical origin of the phase transition width $\Delta T(N)$ that we observe at finite temperatures. Thus, macroscopic quantum tunneling theory provides a solid, first-principles theoretical basis for the exponential behavior of $\Delta T(N)$ proposed in the main text.

\subsection{ Unified Complexity Scaling Model and Finite Correlation Length Effects}
The preceding analysis has primarily focused on the vicinity of the critical point, where the correlation length $\xi$ diverges or exceeds the total system size N. To construct a more universal complexity scaling model capable of describing the entire phase transition region, we must consider the impact of a finite correlation length $\xi$ when the system deviates from the critical point. We can establish the asymptotic limits of the $\Xi$ scaling behavior in two distinct physical regimes.
\textbf{Critical Regime ($\xi \ge N$):} When the correlation length $\xi$ is equal to or exceeds the system size N, the entire system, as previously described, behaves as a single, indivisible coherent quantum mechanical entity. To maintain and resolve the delicate "Half-Fire, Half-Ice" ordered state on a global scale, the complexity budget must overcome an entropy associated with the entire system size. Its scaling behavior is dominated by the total system size N and grows exponentially:
$ \Xi(N, \xi \ge N) \propto \exp(\alpha N), \quad \text{where } \alpha > 0. $
\textbf{Off-critical Regime ($\xi \ll N$):} When the correlation length is much smaller than the system size (e.g., at higher temperatures away from the critical field), the global coherence of the system is destroyed. The effective degrees of freedom are confined within dynamically quasi-independent coherent segments of size approximately $\xi$. In this limit, the total complexity of the system is no longer determined by its global size N, as regions separated by distances much greater than $\xi$ are decoupled from each other. This picture resonates with the core idea of the renormalization group \citep{wilson1971renormalization}. At this point, the complexity budget should saturate to a value that depends only on the maximum complexity of a single coherent block. To ensure a smooth transition to the critical regime behavior, this saturation value, determined by the correlation length $\xi$, should also follow an exponential form dictated by $\xi$ itself:
$ \Xi(N \gg \xi, \xi) \propto \exp(\alpha\xi). $
To capture both of these asymptotic behaviors in a single, smooth analytical form, we employ a mathematical technique widely used in statistical physics to connect different scaling regimes, in a spirit consistent with Padé-type approximations \citep{baker1996pade}. An interpolation function that elegantly connects the two limits and has a concise form is:

\begin{equation} \label{eq:appendix_xi_N_xi_H1}
\Xi(N, \xi) = C \cdot \exp\left[ \frac{\alpha N}{1 + N/\xi} \right] 
\end{equation}

where C is a non-universal proportionality constant, and $\alpha$ is an exponential factor related to the contribution of a single spin to the complexity. This expression correctly reproduces the two limits: when $\xi/N \to \infty$, we recover the exponential growth of the critical regime, $\Xi \to C \cdot \exp(\alpha N)$; when $\xi/N \to 0$, we obtain the saturation behavior of the off-critical regime, $\Xi \to C \cdot \exp(\alpha\xi)$.

\subsection{Unified Description of the Phase Transition Width}
Substituting this more universal expression for the complexity budget (\ref{H.1}) into CWT's core prediction, $\Delta T \propto 1/\Xi$, we obtain a more refined scaling formula for the phase transition width that includes finite correlation length corrections:

\begin{equation} \label{eq:appendix_deltaT_N_xi_H2}
\Delta T(N, \xi) \propto \exp\left[ -\frac{\alpha N}{1 + N/\xi} \right] 
\end{equation}

This result provides a unified and profound CWT description for the rich phenomenology of this phase transition across different parameter regimes. It not only explains why, in the critical limit ($\xi \ge N$), the phase transition width exhibits an exponential sharpening ($\Delta T \propto \exp(-\alpha N)$)—because stabilizing the globally coherent state requires exponential complexity resources. More importantly, it also naturally explains the commonly observed phenomenon of thermal broadening in experiments. In the off-critical limit ($\xi \ll N$), the phase transition width saturates to a value determined by the correlation length, $\Delta T \propto \exp(-\alpha\xi)$. Since the correlation length $\xi(T, H)$ itself decreases with increasing temperature or deviation from the critical field, the transition peak will correspondingly become broader. In summary, this refined framework for complexity scaling not only provides a solid physical picture for the exponential growth of $\Xi$ inferred in the main text but also successfully extends its applicability from an isolated critical point to the entire phase transition region, including the effects of finite correlation length, once again demonstrating the powerful explanatory capacity of CWT as a universal theoretical framework.

\bibliographystyle{unsrt}  
\bibliography{references}

\end{document}